# Fundamental Phenomena and Applications of Swift Heavy Ion Irradiations

Maik Lang[1], Flyura Djurabekova[2], Nikita Medvedev[3,4], Marcel Toulemonde[5], and Christina Trautmann[6,7]

[1]Department of Nuclear Engineering, University of Tennessee, Knoxville, TN 37996, USA

[2]Helsinki Institute of Physics and Department of Physics, University of Helsinki, PO Box 43, Helsinki FI-00014, Finland

[3]Institute of Physics Czech Academy of Science Na Slovance 2, 182 21 Prague 8, Czech Republic

[4]Institute of Plasma Physics, Czech Academy of Science, Za Slovankou 4, 182 00 Prague 8, Czech Republic

[5]Centre de Recherche sur les Ions, les Matériaux et la Photonique, CIMAP-GANIL, CEA-CNRS-ENSICAEN-Univ. Caen, Bd., H. Becquerel, 14070 Caen, France

[6]GSI Helmholtzzentrum für Schwerionenforschung, Planckstr. 1, 64291 Darmstadt, Germany

[7]Technische Universität Darmstadt, Petersenstraße 23, 64287 Darmstadt, Germany

## ABSTRACT

This review concentrates on the specific properties and characteristics of damage structures generated with high-energy ions in the electronic energy loss regime. Irradiation experiments with so-called swift heavy ions (SHI) find applications in many different fields, with examples presented in ion-track nanotechnology, radiation hardness analysis of functional materials, and laboratory tests of cosmic radiation. The basics of the SHI-solid interaction are described with special attention to processes in the electronic subsystem. The broad spectrum of damage phenomena is exemplified for various materials and material classes, along with a description of typical characterization techniques. The review also presents state-of-the-art modeling efforts that try to account for the complexity of the coupled processes of the electronic and atomic subsystems. Finally, the relevance of SHI phenomena for effects induced by fission fragments in nuclear fuels and how this knowledge can be applied to better estimate damage risks in nuclear materials is discussed.

# 1. INTRODUCTION

## 1.1 Swift Heavy Ions: Definition and Fundamental Properties

Swift heavy ions (SHIs) are available at large accelerator facilities that are able to produce beams of high mass ions with kinetic energies in the MeV-GeV range and above. In many solids SHIs release sufficient energy to generate long, nanometer-sized damage trails often denoted as 'latent tracks' because they are not discernable by the naked eye or optical microscopy. The initial interest in ion tracks goes back to the late 1950s, when Young reported the etching of tracks from fission fragments in LiF [1], and shortly after, when Silk and Barnes published the first transmission electron microscopy images of fission tracks in mica [2]. At that time, these discoveries generated a boom in track research and motivated numerous applications in nuclear detector physics, geochronology, archaeology, and many other fields [3,4]. This interest was further stimulated with the advent of large heavy-ion accelerators in the 1980s and inspired intensive and systematic basic research on track formation as well as applied projects in material science, nanophysics, biophysics, and simulations of cosmic ray effects [5–8].

SHIs are usually characterized by their specific energy in units of MeV per nucleon (MeV/u). The energy per nucleon scales with the ion velocity which directly impacts the energy loss ($dE/dx$) along the ion path. Ions of the same specific energy have a similar range, typically on the order of 10 μm at ~1 MeV/u up to 1 mm at ~100 MeV/u [9]. The term SHI predominantly applies to ions with mass equal to and above that of carbon, with velocities comparable to and higher than the Bohr velocity, *i.e.*, ions faster than the velocity of the electrons in the Bohr orbit. In this velocity regime, the projectiles mainly interact with the target electrons, resulting in dense electronic excitations and ionizations of the target atoms (electronic stopping). Energy loss by elastic collisions with target atoms (nuclear stopping) is up to 2-3 orders of magnitude smaller and thus plays only a minor role for track formation. Since the majority of interactions occur with target electrons, no large-angle scattering of the ion projectiles occurs; this results in straight, highly parallel ion tracks (Fig. 1). Transmission electron microscopy (TEM) studies and small angle X-ray scattering (SAXS) experiments revealed that the track shape is nearly cylindrical with a constant diameter for the majority of the ion range, while over the last micrometer or so, before the ions come to rest, the track diameter narrows and the shape changes into a cigar-shaped and irregular form [8,10,11]. This behavior is related to elastic collisions with the target atoms which become increasingly important when the ion has slowed down and is no longer able to efficiently ionize the target material. At intermediate energies, where both nuclear and electronic stopping contribute, interesting phenomena occur that have recently been studied (see Section 2.4).

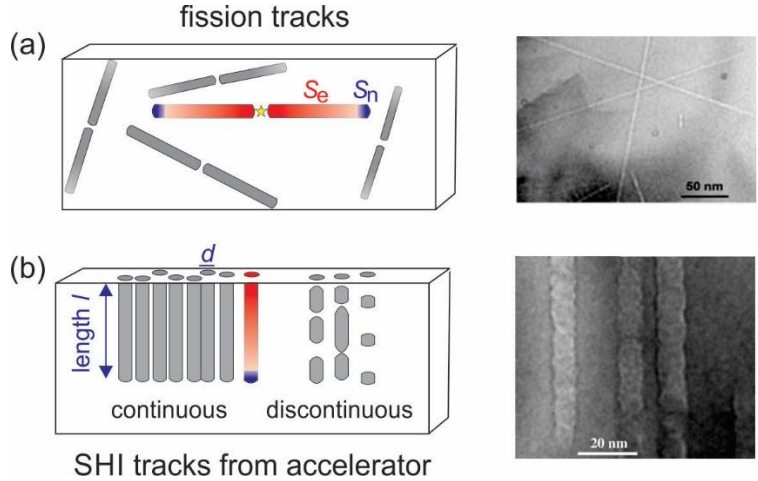

*Figure 1: Schematic representation and TEM image of tracks from (a) spontaneous and induced fission events and (b) from irradiations at large accelerator facilities. Fission tracks have random directions whereas SHI tracks are parallel oriented. Indicated in the schematics is the segment of a track at which the electronic dE/dx (red) and the nuclear energy loss (blue) dominates (see Fig. 2b). The TEM images are (a) fission tracks in apatite [12] and in (b) 2.2-GeV Au ion tracks in apatite [13].*

The electronic d$E$/d$x$ of SHIs follow a typical shape known as the *Bragg* curve with a peak at energies around 1-3 MeV/u. With increasing beam energy above this peak, the energy loss continuously decreases, as shown in Fig. 2a. At even higher energies, the curve rises slightly again (not shown) due to Cherenkov radiation and other relativistic effects, but this energy regime is beyond the scope of the present review. The *Bethe* equation [14] gives a good approximation of the electronic d$E$/d$x$ *versus* ion energy above the Bragg peak:

$$-\frac{dE}{dx} = \frac{4\pi k_0^2 Z_{eff}^2 e^4 n}{mc^2 \beta^2} \cdot \left[ ln\left(\frac{2mc^2\beta^2}{I(1-\beta^2)}\right) - \beta^2 \right] \qquad (1)$$

where $k_0$ is the Boltzmann constant, $Z_{eff}$ the effective charge state of the ion projectile, $e$ the elementary charge, $n$ the electron density of the target material, $mc^2$ the electron rest mass, $\beta$ the ion velocity, and *I* the mean ionization potential of the target material.

When a SHI penetrates into a solid, it slows down and simultaneously its energy loss increases towards the *Bragg* maximum. At even lower velocities (low-energy side of the *Bragg* maximum), the ion picks up more and more electrons reducing its effective charge state. This region of the d$E$/d$x$ curve (Fig. 2a) is no longer accurately described by the *Bethe* equation but by the LSS (Lindhard, Scharf, and Schiøtt) theory [15]. The strong dependence of the ion velocity and charge state (Eq. 1) creates the typical shape of the ion energy loss curve (Figure The long section of almost constant energy loss followed by a sudden and localized peak (Fig. 2b) can be taken advantage of, in particular, for application of SHIs in tumor therapy: the beam energy is selected such that ions have reduced d$E$/d$x$ in healthy tissue and the maximum d$E$/d$x$ in the tumor region [16,17].

There exist various codes to calculate the energy loss and penetration depth of SHIs as a function of their energy. A frequently used code is SRIM (The Stopping and Range of Ions in Matter) [9], but it should be mentioned that the calculations do not take into account structural properties of a certain target. This shortcoming has to be considered when the ions travel, *e.g.*, parallel to a particular interatomic plane in a single crystal. Under such channeling conditions, the energy loss is significantly reduced because the projectile encounters fewer electrons and the ions can travel much larger distances within the crystal [18].

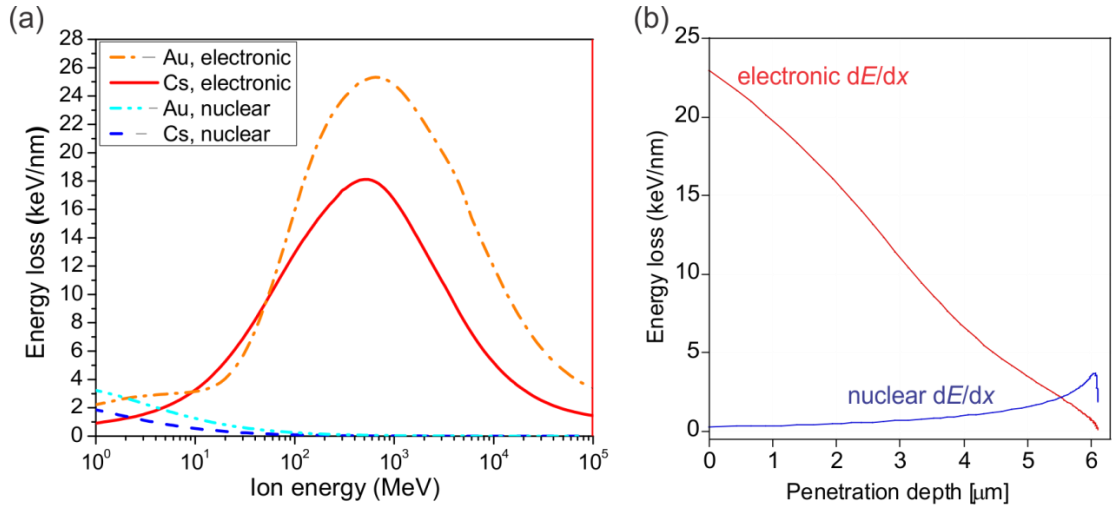

*Figure 2: (a) Electronic and nuclear energy loss of Cs and Au ions in graphite as a function of the ion kinetic energy illustrating the characteristics of the energy loss curves of SHIs and their dependence on the ion charge state, $Z_{eff}$. (b) Energy loss of a typical fission fragment in $UO_2$ (80-MeV Xe) as a function of penetration depth with a color scheme corresponding to the two dE/dx regions in Figure 1. Both data sets were obtained from SRIM calculations [9].*

The energy deposited by SHIs into the electronic subsystem of solids can be enormous, ranging from a few up to several tens of keV/nm. The dense electronic excitation nearly instantly produces a shower of energetic primary electrons ($10^{-17}$-$10^{-15}$ s) which initiate *via* secondary ionizations an extended electron cascade spreading radially ($10^{-15}$-$10^{-13}$ s) and leaving positive holes behind (see Section 2.1). In the following stage, the excited electrons lose energy to the lattice through electron–phonon interactions on a time scale of a few picoseconds ($10^{-12}$ s).

Energy transfer from the electrons and holes to the atoms drives the local atomic structure to a far from equilibrium state. Details of the following processes and the response of a given solid strongly depend on structural and thermal material properties [19]. The two-step process, *i.e.*, the energy transfer from the electronic to the atomic subsystem, and the interplay between multiple length and time scales is very complex and makes it difficult to model details of the mechanism by which the electronic excitation is converted into atomic motion and finally into stable tracks. The physical mechanisms involved include charge neutralization, local melting, shock waves, recrystallization, and the emission of secondary neutral or charged particles, and are still subject of intensive debates (see Section 3).

A characteristic feature of SHI irradiation is the formation of tracks requires a critical minimum electronic $dE/dx$. This threshold strongly depends on the material and slightly increases with ion velocity (see Section 2.3). It can be below 1 keV/nm for polymers and up to a few tens of keV/nm for metals. There exist a number of materials such as Cu, Ag, and Au, or crystalline Si and Ge, in which monoatomic ions of the highest energy losses are not able to produce tracks. In metals, the large heat conductivity of the electrons dissipates the deposited energy before the track has time to form [20]. From a structural point of view, track formation is in general more difficult in materials with a simple crystal structure due to a more efficient damage recovery and recrystallization process. In contrast, tracks readily form in complex systems, in materials with polymorphism and electronic defects, and if radiation can induce radiolysis in combination with volatile radiation products. Figure 3 presents examples for different material classes and their sensitivity scaling with the track-formation threshold. Ion tracks form readily in most insulators, in particular if they are amorphizable, including organic materials, phosphates, silicates, oxides, but also alkali and earth-alkali halides and other ceramics. The most sensitive materials are polymers where SHI irradiation leads to chain scissions and the formation of small volatile fragments that leave the sample through outgassing. In many oxides, the tracks consist of amorphous cylinders embedded in the crystalline matrix (SiO$_2$, apatite, mica, etc.) [21]. In other materials, the track structure can be more complex, *e.g.*, consisting of an amorphous core surrounded by a disordered crystalline shell (*e.g.*, Gd$_2$(Ti,Zr)$_2$O$_7$ and Gd$_2$TiO$_5$) [22]. Various solids remain crystalline but show beam-induced crystalline-to-crystalline phase transitions (*e.g.*, ZrO$_2$ and Ln$_2$O$_3$ with Ln = Sm, Gd, and Ho) [23–25]. Tracks even exist in amorphous materials such as in vitreous SiO$_2$ and metallic glasses, where the amorphous state within the track slightly differs from the disorder of the surrounding matrix. In semiconductors the effects of SHIs are not completely understood, but it seems that the band gap has some influence (see Section 2.3). Finally, it should be noted that some insulators (UO$_2$, ThO$_2$, CeO$_2$, MgO, etc.) are extremely resistant under SHI irradiation and their response includes mainly the creation of isolated defects and strain (see Section 2.3 and Section 4) [26].

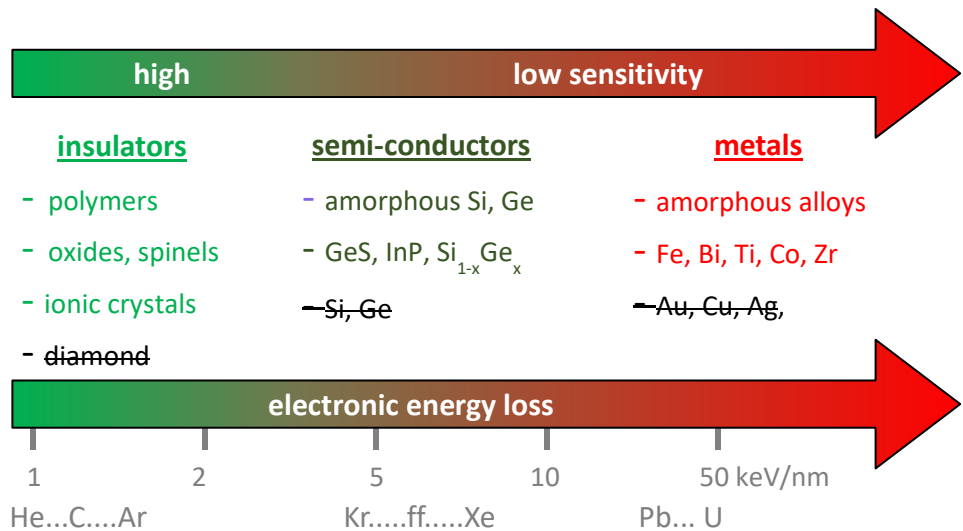

*Figure 3: Track registration sensitivity and range of electronic energy-loss threshold for different material classes illustrating high sensitivity and low thresholds for insulators and low sensitivity and high dE/dx thresholds for metallic systems. Track formation is facilitated in materials with a complex structure, for systems which easily form electronic defects or decompose by radiolysis. For some systems (strikethrough materials) no track formation was observed so far (e.g., diamond).*

The diameters of ion tracks are typically a few nanometers, with the exact value being material specific. Depending on the material and its radiation sensitivity, the track size becomes larger with increasing energy loss of the projectile. Above a critical threshold, each ion creates an individual, continuous, homogeneous cylindrical track, whereas close to the threshold, as well as in rather insensitive materials (metals and semiconductors), the damage morphology can be discontinuous and tracks may consist of a sequence of dotted damage fragments (Fig. 1).

Additionally, the track diameter depends on the ion velocity and is, for similar d$E$/d$x$ values, larger for low-velocity ions.

**1.2 Sources of Swift Heavy Ion Beams**

In nature, swift heavy ions appear in nuclear fission processes and in cosmic radiation. Cosmic particle radiation is primarily composed of energetic protons with a small (~1%) contribution of heavy projectiles, with C, O, Mg, Si, and Fe being prominent examples. Their energies cover an extremely broad regime with a maximum at about 500-1000 MeV/u. This flux is dominated by protons and the abundance of heavier particles decreases exponentially with mass. Despite their low abundance, heavier particles can play a critical role in radiation effects because they deposit their energy at much higher energy densities compared to that of protons and need to be included in radiation risk considerations for satellites, spacecrafts, and humans in space.

Energetic heavy particles are also produced whenever radionuclides decay *via* fission processes (spontaneous fission in nature and induced fission in nuclear reactors). In minerals containing elements such as uranium and thorium, fission-track damage accumulates over millions of years. Fission projectiles have a typical mass between 75 and 155 u and an energy between 70 and 120 MeV (~1 MeV/u) corresponding to a range in solids of ~10 μm. Irradiation experiments with fission-fragment sources are challenging due to their broad mass, energy, and emission angle distributions (Fig. 1).

Today, there are worldwide some 40,000 accelerator facilities that produce ion beams in the low to medium energy (keV-MeV) regime which are used for many different applications, including ion implantation/doping of materials for chip fabrication and ion-beam analysis (IBA). Swift heavy ion beams with MeV-GeV energies are only available at a limited number of accelerators due to high construction and operational costs. Large-scale ion facilities exist for instance at GSI (Darmstadt, Germany), GANIL (Caen, France), IMP/CAS (Lanzhou, China), and JINR (Dubna, Russia). Specially designed beamlines allow the irradiation of samples under well-controlled conditions with parameters such as ion species, beam flux, sample temperature, atmospheric conditions, etc. being adjusted and monitored. Some facilities provide beamlines with *in situ* characterization techniques such as X-ray diffraction, Raman, infrared, or UV–VIS spectroscopy, and scanning electron microscopy [27]. Motivated by fundamental nuclear and particle physics, several new, large facilities are currently under construction for relativistic beams of the highest intensities. Some of them will offer unique research opportunities for non-nuclear research communities including materials research, plasma physics, and biophysics [28]: FAIR facility in Darmstadt (Germany) [29], NICA (Russia) [30], RAON (Korea) [31], and HIAF (China) [32].

**1.3 Applications of Swift Heavy Ions**

Over the past several decades, ion-beam facilities that were developed for nuclear physics research have impacted many other fields, including atomic physics, plasma physics, material science, condensed matter physics, geosciences, environmental physics, and bio-medical sciences. The number of SHI-beam applications in basic research as well as for industrial applications is still growing. A unique feature of MeV-GeV heavy ions is that each individual projectile produces a cylindrical nanometer-sized track. Given the high kinetic energy, no straggling occurs and ion tracks can be regarded as well aligned high aspect ratio nanostructures (Fig. 1). Heavier projectiles are preferred as a structuring tool, because they produce tracks of continuous damage with the largest track diameters as a result of their high energy loss. The penetration depth of ions in a material can be adjusted by the beam energy. For instance, 10 MeV/u ions can be used to irradiate a 100-µm thick stack of sample with almost constant energy loss. Samples with a large area are irradiated with a collimated ion beam which is scanned over several cm$^2$, or alternatively, the beam is defocused to a large spot size. The track density is adjusted by the fluence which typically ranges from a single ion impact per sample up to the regime of multiple track overlap (typically above $\sim 10^{13}$ ions/cm$^2$). Tracks are usually stochastically distributed over the exposed target area, but can also be precisely placed on predefined positions using a heavy-ion microprobe. The application of targeted irradiations includes writing specific patterns [33], testing microelectronic circuits [34], and delivering a preset number of ions to the nuclei of individual living cells [35]. In the following subsections we present a few examples of exciting research topics based on SHIs.

*Ion Track Nanotechnology*

Soon after the discovery of ion tracks, it became clear that they can be selectively attacked by a suitable chemical etchant, converting the damage of each track into an individual nanopore. The etching time determines the pore size and specific etching conditions control the pore shape and geometry (cylindrical, conical, double conical, etc) [36]. The discovery of track-etching has triggered applications in a wide range of scientific and industrial areas. In geoscience, for instance, the evaluation of the number density of fission tracks accumulated over time in minerals (such as mica, apatite, and zircon) has become a standard dating method for geological samples [4].

During the last decade, the rapidly increasing activities and recent developments in nanoscience have boosted the interest in track-etched channels in polymers in particular (Fig. 4) [6,7,37–39]. Several small companies produce track-etched membranes commercially. They pay for beam time at accelerator facilities (*e.g.*, Caen, Louvain-la Neuve, Dubna, Lanzhou, Jyväskylä, and Brookhaven) and process large quantities of track-etched membranes with pores of extremely uniform diameter.

Within basic research, transport properties of membranes with a single track-etched nanopore became of great interest to mimic conditions of ion channels in biological cell membranes. On a nanoscale, charges on the pore wall can lead to ionic selectivity so that a nanopore with, e.g., negative surface charges will predominantly transport positively charged ions and vice versa [36]. The asymmetrical geometry of nanopores allows preferred transportation of chemical and biological species in one direction similar to an ionic diode [40,41]. Using different chemical and physical modification or decoration strategies, it is possible to tailor systems with sensor properties [42–45]. Ion track membranes are also of interest as nanofluidic devices for biosensing, separation of drug molecules, desalination, electro osmosis, electrochemical energy storage in batteries, and fuel cells and supercapacitors, just to name a few topics.

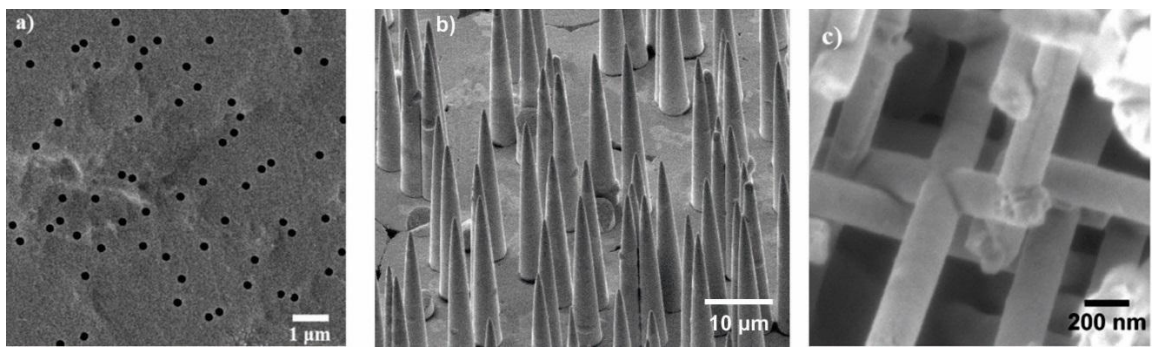

*Figure 4:* *Scanning electron microscopy images of (a) polycarbonate membrane with cylindrical nanopores produced by chemical track etching [46], (b) Au nanocones electrodeposited in conical nanopores [47], and (c) a mechanically stable Au nanowire network produced by ion irradiations under different angles of beam incidence [46].*

Another very active research field in nanoscience uses the ion-track technology for the fabrication of nanowires by filling the pores of ion-track membranes *via* electrodeposition. The combination of chemical etching, electrochemical deposition, and surface modification techniques led to the development of an enormous flexibility to synthesize tailored nanostructures of various metals and semiconducting compounds and exploit size dependent physical and chemical properties of materials at the nanoscale (Fig. 4). Compared to other template-based techniques, ion track membranes allow the control of the size, geometry, aspect ratio, and surface morphology of the nanowires. Recent investigations provide results on size effects on optical, electrical, and thermal properties, surface plasmon resonances, and thermal instabilities [48–51].

*Functional Bulk Material Testing*

Studies on the radiation hardening of bulk materials under extreme conditions including radiation dose, stress, temperature, and pressure, have received significant attention in the context of nuclear materials and of, more recently, next-generation accelerators. Functional materials under a constant flux of radiation that may modify their properties need to be tested and appropriate mitigation methods have to be developed. Irradiation experiments with SHIs are an efficient and controlled way to test bulk materials and identify specific physical and structural property changes. For reliability tests and lifetime estimates, it is important to understand (*i*) if and how a given material responds to extreme radiation fields, (*ii*) what is the nature of the specific damage, (*iii*) how does the track size depend on the energy loss of the ions, and (*iv*) how can thermal treatment mitigate the radiation damage? An important research area addresses the radiation hardening of materials for new high-power accelerator facilities (*e.g.*, FAIR, FRIB, ESS, etc.), operational limits of materials in high dose and high energy density environments, lifetime predictions, and the development of new material solutions for extreme cases. Unique material requirements also arise with the development of facilities with the highest pulse intensities. The dynamic response (pressure wave propagation and damping) under high-power beam impacts of beam intercepting devices (collimators, targets, and beam dump materials) are currently tested with SHIs and provide helpful experimental data for benchmarking respective finite element simulations [52–55].

Large-scale heavy ion facilities providing beams of several mm range even allow studies on the behavior of materials under multiple extreme conditions by irradiating samples pressurized and heated in diamond anvil cells (Fig. 5). The relativistic ions penetrate one of the two diamond anvils (thickness ~2 mm) until they reach the pressurized sample. Effects of radioactive decay events in compressed and heated minerals of Earth's interior can be correlated with material properties under geodynamic processes. It is also possible to synthesize new materials that are otherwise not accessible. As recently demonstrated, irradiations under pressure can give access to thermodynamic pathways in the phase diagram such that otherwise unstable, high-pressure phases can be recovered upon pressure release [56,57].

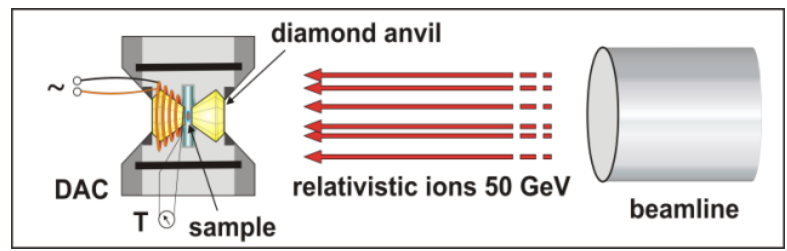

*Figure 5:* *Scheme of high-pressure irradiation experiments with relativistic heavy ions. In order to reach the sample (size ~100 μm) pressurized between two diamond anvils (2-3 mm thick), the initial beam energy must have an energy of at least 200 MeV/u. Temperature can be controlled by heating wires or intense laser heating (not shown). This set-up allows coupling of multiple extreme conditions: ion irradiation, pressure, and temperature. Figure has been modified from* [56].

*Cosmic Radiation Simulation and SHI-Induced Sputtering Processes*

Over the last few decades, the simulation of cosmic radiation by ground-based studies with SHI beams has become an important topic at large accelerator facilities. Electronic components in space missions are sensitive to radiation effects and require reliability tests before being installed. High-energy particles produce local charges and electron cascades that generate multiple types of errors, including the increase of leakage currents, local memory errors (*e.g.*, single event upsets), and single event transient errors at the system level. In some cases, more dramatic and destructive events like dielectric rupture, burn-out, or latch-up render the component completely unusable. For safe operation of satellites or devices for exploration of the solar system, the importance of accelerator-based studies and the need to develop countermeasures are acknowledged by all space agencies and give radiation tests at accelerator facilities high priority [58,59]. Nowadays, space qualified and radiation-hard components including commercial off the shelf (COTS) components undergo rigorous screening to ensure their survival in space. The advantage of SHIs are their large penetration depths that allows tests without disassembling the electronic components.

Another emerging and exciting research area is linked to cosmic radiation laboratory experiments on interstellar dust analogues. The challenge is to understand to what extent galactic radiation is responsible for the formation of large molecular compounds such as polycyclic aromatic hydrocarbon-like molecules and fullerenes observed in the interstellar medium. For example, when bombarding soot targets, hydrocarbons were detected containing up to several tens of carbon atoms and ionic fullerenes similar to large molecular compounds that were identified in space [60,61]. The release of large molecules from cryogenic films by sputtering and/or desorption is investigated by *in situ* mass spectrometry in combination with time-of-flight measurements [62,63]. This method is similar to conventional secondary-ion mass spectroscopy (SIMS) with keV ion beams but the release of particles and molecules is driven by electronic excitation leading to, in some cases, enormous yields. A variety of desorption and sputtering experiments under electronic energy loss conditions revealed many different new phenomena including extreme yields (thousands of sputtered particles per incoming projectile) and a strong dependence on the material class [64,65]. A more practical interest in better understanding these processes is linked to the dynamic vacuum conditions in accelerator sections, where beam losses stimulate extensive release of gas when SHIs hit the wall of the beam tube or the vacuum chamber [66].

## 2. FUNDAMENTALS OF SWIFT HEAVY ION MATTER INTERACTIONS
### 2.1 Electronic Energy Loss and Energy Dissipation

A swift heavy ion (SHI) passing through a solid induces a large variety of effects in the target material. During the slowing down process, the SHI travels a typical interatomic distance in the target within the characteristic time of ~$10^{-18}$ s [67]. Within this short interaction time, the atoms of the target as well as the electrons (except for deep-shell electrons of heavy elements) are essentially immobile and represent a group of fixed charges in space. The energy deposited in the target dissipates at a much later time, when the SHI has already departed from the point of interaction. The energy relaxation of the target includes the following processes: (*i*) Auger and radiative decay of holes in the electronic shells left after ionization (~$10^{-15}$ s) [68], (*ii*) transport of delta electrons away from the site of ionization (from $10^{-15}$ s to $10^{-12}$ s) [69], (*iii*) atomic motion due to energy transfer from the excited electrons (typically $10^{-13}$ s to $10^{-12}$ s) [70], (*iv*) in the case of sufficient energy transfer, formation of transient atomic disorder (~$10^{-12}$ s) [71], (*v*) partial or full recovery of the track (~$10^{-10}$ s) [71], and (*vi*) macroscopic relaxation of defects and strain in a track halo ($10^{-9}$ s to $10^{-6}$ s). The sequence from the original energy deposition to the final relaxation of defects and strain spans over ten orders of magnitude in time [72,73]. The faster processes trigger the slower ones, forming the initial conditions for their complex kinetics. We will consider these stages in more detail below.

When a SHI enters a target, the sudden change of the surrounding potential destabilizes the electronic system of the ion. One of the immediate effects is the loss (or capture) of electrons by the projectile, depending on its charge state and the expected equilibrium charge at its current velocity. If the initial SHI charge is higher than the equilibrium value, it will capture target electrons to reduce its charge and, in the opposite case, electrons will be stripped off. Electron loss typically occurs at much higher rates than electron capture, which means that a SHI with a lower charge than its equilibrium value will reach its equilibrium charge faster and within a shorter distance from the impact at the surface [74–76]. This disparity can be understood by the multitude of available states for the electrons in the target, which makes an electron loss process very probable; whereas, only a relatively small number of states are available in the shells of a SHI into which electrons can be captured. A SHI is often treated as a point-like charge travelling through a solid. Accounting for the finite size of the projectile usually leads to only minor modifications of its interaction with target atoms [77]. The SHI charge begins to oscillate around the equilibrium value once it is reached. Simultaneously, more complex processes take place, such as the creation of so-called convoy electrons that are not fully captured but follow the path of the projectile, carrying some energy out of the system (they are essentially analogous to electrons in the Rydberg states of the ion) [78]. Attracted electrons may also re-scatter off the ion and be re-emitted with different velocities – a process known as "Fermi shuttle" [79], which often can be disregarded as it has only little impact on the SHI penetration depth and target response.

The charge state of an ion directly affects the energy deposition, because the stopping power is proportional to $Z_{eff}^2$ (effective charge). For SHIs, the kinetic energy is deposited to the electronic system of the target material leading to ionizations and emission of electrons. The maximal kinetic energy of an emitted electron occurs in a head-on collision. For a non-relativistic ion, it can be estimated as follows:

$$E_e = E_i \frac{4M_i m_e}{(M_i+m_e)^2} - I_p, \qquad (2)$$

where $E_i$ is the initial ion energy, $M_i$ is its mass, $m_e$ is the electron rest mass, and $I_p$ is the ionization potential (energy of the atomic orbital) of the shell of an atom where the electron is being ionized from. For typical ion energies produced at large accelerators, the energy transfer to the electrons can reach up to a few tens of keV or higher. The electrons produced directly by an SHI impact are commonly referred to as delta electrons (formerly delta rays [80]). The energy distribution of the excited electrons follows approximately a ~$1/E^2$ law typical for Rutherford scattering, except for the low-energy region where other effects cause some deviations [69]. The angular distribution of the delta electrons is approximately proportional to $\cos^2(\theta)$, which implies that the minority of the electrons travel along the ion trajectory while the majority move perpendicular to it. This process transfers the energy from the initially excited region radially away from the ion path [69]. Typically, a SHI impact ionizes a few electrons from the same atom within a few angstrom distance from its trajectory, leaving a highly charged target ion behind [81]. The initial effect produced by the projectile can be regarded as a nanometer-sized non-equilibrium plasma column embedded in a solid target.

When passing through a material, a quickly moving charged projectile also excites collective electronic modes, called plasmons [82]. In the plasma physics community, such collective effects are known as wake-field potential generation. Plasmons can persist over longer time scales in metals with low damping rates, while they quickly decay in insulators into one or a few electron-hole pairs at femtosecond timescales. The presence of collective processes within the electronic system can significantly alter the energy dissipation [83] and is related to distant collisions beyond a few angstrom, but still within a nanometer from the initial ion path.

Electrons at the surface of the target can be emitted if they possess sufficient energy to overcome the work function of the material and the transient induced electromagnetic fields [84]. This effect is particularly important for grazing incident ions, but plays a minor role in near-perpendicular trajectories, for which only the first few nanometers below the sample surface are subject to electron emission. Auger decays of deep-shell holes in ionized target atoms produce secondary electrons at lower energies, which spread outward quickly with an almost isotropic geometry. Auger decays of holes in atoms surrounded by other atoms can be different from those in isolated atoms [85], including situations where a forbidden transition may become allowed [86]. Such processes, which involve electrons from neighboring atoms, are generally known as interatomic Auger decays (discussed in the solid state community as Knotek-Feibelman processes [87], or interatomic Coulombic decay (ICD), excitation–transfer–ionization (ETI), and a class of related processes in ab-initio chemistry [88]). Electrons emitted *via* Auger decay create two new holes in upper shells and have a kinetic energy that is defined by the energy level differences of the involved shells. Such cascades are completed when all holes are in the valence band of the insulator (or the conduction band of a metal), where they are mobile and can travel within the sample analogously to electrons within the conduction band. Radiative decays of deep-shell holes emit photons, which carry energy away from the ion trajectory much faster than electrons. Since the photon attenuation length is significantly larger than the delta-electron range, such radiative processes transport energy into a much larger volume [89,90]. The radiative decay dominates the Auger decay in very heavy elements, but the energy deposition of SHIs are in general not high enough to generate a significant amount of photons and the energy redistribution due to this channel is only of minor importance [91].

The electrons liberated by SHIs slow down and lose their energy through elastic and inelastic scattering with the target atoms. Elastic scattering transfers the electron energy into kinetic energy of atoms without electronic excitations, while inelastic scattering leads to impact ionization, releasing secondary electrons which can further ionize and thus multiply the number of electrons [92,93]. The electron cascades typically last about a hundred femtoseconds and their lifetime increases with the initial energy of the delta-electrons and thus, with the energy of the ion beam [94]. During this stage, the electronic system is in a highly non-equilibrium condition, not adhering to the Fermi-Dirac distribution. The thermalization of electrons towards the Fermi-Dirac distribution occurs over a minimum time frame of the cascade lifetime or even longer.

Partial thermalization among the slower secondary electrons proceeds faster than the total equilibration of the entire electronic ensemble, and the electron energy distribution function approximately follows the so-called 'bump-on-hot-tail' shape [70,95]. Photon emission due to electron Bremsstrahlung only plays a role at relativistic electron energies [96] and thus is often neglected in SHI research.

Elastic electron scattering transfers energy to the target atoms, which leads to atomic movement and heats the system. However, electrons with kinetic energies above a few tens or a hundred eV pass the atoms too fast to be impacted by this motion, so the scattering takes place on a system of essentially fixed atoms [97,98]. Only slower electrons will experience an effect of the correlated atomic motion and couple to phonons [99]. Valence holes in insulators and semiconductors can scatter analogously to electrons in the conduction band. Inelastic scattering by impact ionization will take place if the kinetic energy of a hole within the valence band is larger than the band gap of the material (a typical case for semiconductors, but rarely for insulators) [100]. Elastic scattering of holes provides atoms of the target with kinetic energy which adds to the energy obtained through elastic electron scattering, with both processes having a comparable contribution [100].

Following the completion of the ionization cascades, the kinetic energies of all electrons and holes fall below a threshold that depends on the band gap of the material, and can only scatter elastically at target atoms, phonons, pre-existing defects, grain boundaries, and among themselves. Electron-hole pairs will eventually recombine, releasing their potential energy *via* several channels: (*i*) the energy is transferred to another free electron nearby through a three-body recombination (sometimes also called Auger recombination), a process that is only important in the case of sufficiently high carrier densities as it scales with the hole density and the square of the electron density [101], (*ii*) the energy is released through photon emission, (*iii*) the energy transfers to the atomic system and either produces several phonons or knocks off an atom from its equilibrium position, forming a point defect [19], (*iv*) or recombination proceeds *via* an exciton, an intermediate bound state of an electron-hole pair. This exciton has its own complex kinetic pathways of relaxation, essentially leading to photon emission or point defect production, similar to (*ii*) and (*iii*) [102]. The characteristic time of electron-hole recombination is strongly material and condition dependent, ranging from a few hundred femtoseconds (*e.g.*, in SiO$_2$ [103]) up to microseconds in scintillators [104].

Atoms provided with sufficient energy may overcome their kinetic barriers and produce disorder in the bulk [105,106] or detach from the surface [107]. Destabilization of the atomic lattice is a result of electronic excitation and may take place around the SHI trajectory due to: (*i*) increase of the kinetic energy of atoms due to elastic scattering of atoms with electrons from ionization events – finally leading to thermal melting [108], (*ii*) severe modification of the interatomic potential due to electronic excitations and resulting in breaking of chemical bonds – inducing non-thermal melting [109,110], (*iii*) change of atomic charge states leading to Coulomb repulsion between neighboring ionized target atoms – called Coulomb explosion [111], and (*iv*) defect accumulation under prolonged beam exposure in the case when an individual ion impact is insufficient to cause damage [112]. The relative importance of these processes for track formation is still debated, with the currently prevalent opinion that the thermal melting (*i*) plays the dominant role [113]. It is based on the notion that Coulomb explosion (*iii*) is only important for finite size systems such as nanoclusters or molecules, or near the sample surface. In the bulk, this effect is not expected to play a significant role due to the presence of a large number of electrons in the surrounding material that can quickly neutralize the unbalanced charges [70,114,115]. The process of non-thermal melting (*ii*) requires high electronic excitation levels that are maintained for a sufficiently long time (over a few hundred femtoseconds [110,116,117]), which also limits its effect in the case of SHI tracks where the density of excited electrons drops rapidly due to electronic transport outside the nanometer-sized track region [118]. However, even if non-thermal processes are insufficient to damage the material directly, it may contribute alongside with thermal melting to the track-formation process. The modeling of track formation is discussed in more detail in Section 3.

**2.2 Characterization of Swift Heavy Ion Damage**

In principle, most characterization techniques available in materials science can be utilized to study radiation effects induced by SHIs, such as the nature of defects, the size of individual tracks, and beam-induced changes of specific material properties. For systematic investigations, one or several of the following irradiation parameters are varied: ion species, energy, energy loss, and applied fluence. The damage cross section or track size can be determined using two different approaches: (*i*) directly by measuring the diameters of individual tracks through microscopic techniques and (*ii*) indirectly by quantifying beam-induced material changes with increasing fluence. Transmission electron microscopy (TEM), the most common direct characterization technique, gives atomic-scale information on the nature of the damage as well as on the morphology and diameter of the tracks. When imaged in top-view mode (electron beam parallel to the ion track orientation), specific damage characteristics can be identified including single defects [119], new crystalline or amorphous phases [120,121], and complex core-shell morphologies [22]. The diameter can be accurately determined if the track is imaged with atomic-scale resolution. Imaging with the electron beam perpendicular to the ion track (cross-section view) reveals the damage morphology which strongly depends on $dE/dx$: a continuous damage trail with a constant diameter is formed at high $dE/dx$, whereas below or close to a material specific $dE/dx$ threshold, the track consists of discontinuous fragments with varying diameter [122]. TEM requires great effort in sample preparation and one has to ensure that tracks in the material under study are stable when exposed to the electron beam of the microscope. At the surface of many materials, nm-sized craters or hillocks appear at the ion impact site [123,124]. These features provide information on the track formation process and have been investigated in detail, particularly for $CaF_2$ and LiF by scanning force microscopy (SFM) [125] and for highly oriented pyrolytic graphite by scanning tunneling microscopy (STM) [126]. More recent efforts use TEM imaging to resolve the atomic-scale structure within such hillocks [127]. The mentioned microscope techniques (TEM, SFM, and STM) provide direct information on individual ion tracks, but to obtain a reliable mean value of the track diameter, it takes great effort to measure a statistically meaningful number of tracks. The analysis technique used to address this problem is small angle X-ray scattering (SAXS) because it is nondestructive and can measure millions of ion tracks simultaneously. Due to their cylindrical shape and parallel orientation, ion tracks are ideal scattering objects for SAXS. Synchrotron-based SAXS investigations have become increasingly important yielding mean track diameters with unprecedented accuracy and information on density changes in the tracks in crystalline and non-crystalline materials [21].

Information on the damage cross section (and thus indirectly on track diameter) is generated by monitoring damage as a function of fluence until severely overlapping tracks are obtained. The analysis can be made by techniques such as X-ray diffraction (XRD), Rutherford backscattering, Moessbauer spectrometry, profilometry, or optical absorption spectroscopy [128–134]. It is important to note, however, that different analytical techniques probe specific properties and may have different sensitivities [135]. An example of such an indirect approach is shown for an amorphizable material in Figure 6, where the irradiation-induced fraction of amorphized material is plotted as a function of fluence. The amorphous fraction within this irradiated sample was deduced from XRD measurements by quantifying the broad diffuse scattering background around the reflection ascribed to the crystalline part of the sample [136].

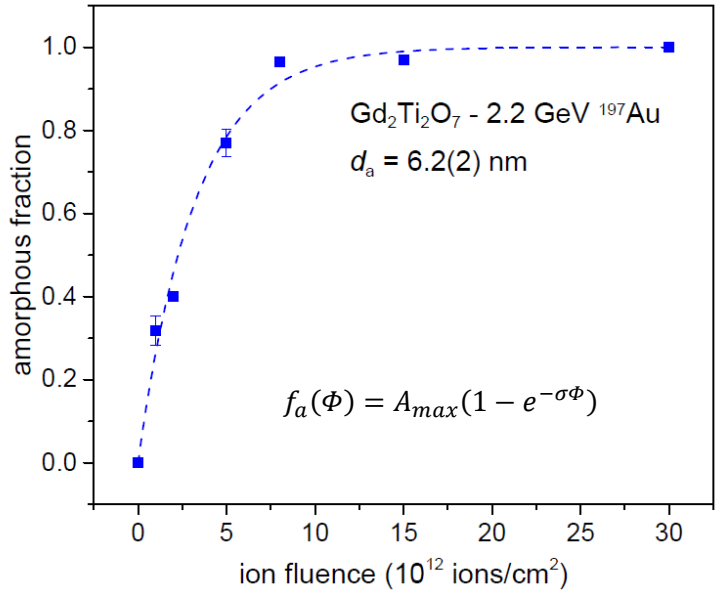

*Figure 6: Typical evolution of damage (here amorphous fraction $f_a$) as a function of fluence $\phi$. To analyze the damage cross section σ, the data are fit by a direct impact model (each ion produces a track) based on an exponential function which considers the effect of overlapping tracks. By assuming cylindrical symmetry one obtains the amorphous track diameter $d_a$. Figure has been modified from [137].*

For SHIs, the damage *versus* fluence curve has the characteristic saturation shape: in the low fluence regime the damage increases linearly because each ion contributes new damage to the damage fraction. As soon as tracks start to overlap, the increase becomes sublinear and at severely overlapping tracks the damage finally saturates when the entire sample is made amorphous. The damage increase with fluence considering track overlapping is modeled by an exponential equation (Fig. 6) where the product of the damage cross section and fluence appears in the exponent [138–140]. From a fit to the data, with the damage cross section, σ, being a free parameter, the track diameter is deduced assuming cylindrical track geometry. The direct impact model assumes that the impact of an individual ion leads to a localized material modification (*e.g.*, amorphization) as opposed to the defect-accumulation model for low-energy ions, where a phase change requires in most cases the accumulation of a sufficient number of individual defects until the damaged structure finally abruptly changes into the new state [77].

For complex track morphologies, specific modified accumulation models have been proposed [141,142]. Quantification of the track sizes by indirect techniques is complementary to direct track observations. Indirect methods are particularly useful for characterizing discontinuous tracks which are difficult to assess otherwise. If the track-formation efficiency (track per incident ion) is smaller than unity, direct track observation methods fail but the indirect technique can still provide a damage cross section and a mean effective track diameter. Moreover, comparing track diameters from direct and indirect methods can help to further assess the track damage morphology [143].

Radiation damage from ballistic collisions and swift heavy ions can be analyzed by the same methods and the approach is often very similar. However, the much larger penetration depth of energetic ion beams is advantageous for certain analytical methods and allows experiments with bulk samples that are not feasible with low-energy ions. A TEM specimen usually has a thickness of ~100 nm which is much thicker than the spatial extent of a damage cascade and orders of magnitude thinner than the typical range of SHIs. This has consequences for the imaging of the damage morphology in both cases. There is always undamaged material beneath and above a damage cascade which obscures the detailed analysis of structural defect features. Tracks created by SHIs fully penetrate the TEM specimen and can be inspected for top-view images that solely show the damage of an individual ion track with no overlap of the undamaged matrix (Fig. 7). Such measurements revealed a remarkably complex damage structure in some ceramics consisting of a distinct core-shell morphology [22].

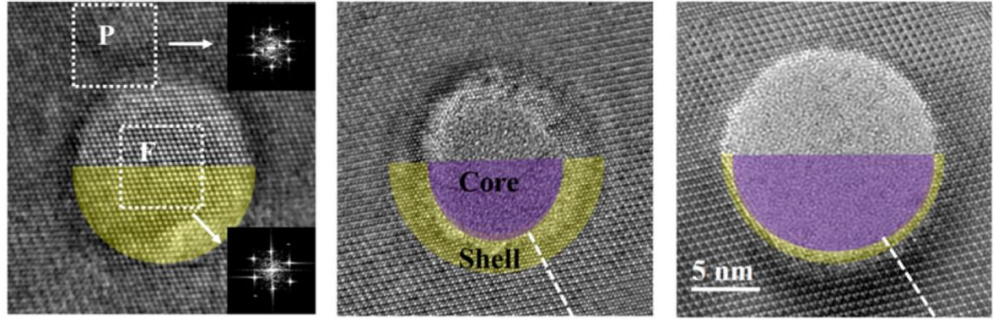

*Figure 7:* Top-view TEM images of (left) $Gd_2Zr_2O_7$, (middle) $Gd_2ZrTiO_7$, and (right) $Gd_2Ti_2O_7$ pyrochlore oxides irradiated with 12-MeV $C_{60}$ cluster ions. The track morphology is complex and depends strongly on the chemical composition of the target material. This Figure has been modified from [22].

The large penetration depth of SHIs is also beneficial for XRD characterization because thick material with fully penetrating ion tracks can be studied at a synchrotron source in transmission mode to obtain high-quality XRD and X-ray absorption spectroscopy (XAS) data [144]. Beyond well-established track characterization techniques, neutron scattering has recently been demonstrated to be an effective probe for studying track properties. Due to their small scattering cross sections, neutron scattering experiments require high neutron fluxes and much larger sample quantities than other techniques. Today, a few pulsed neutron sources, such as the Spallation Neutron Source at Oak Ridge National Laboratory, operate dedicated beamlines for producing high-quality neutron total scattering data, and therefore excellent pair distribution functions (PDFs). At such facilities the necessary sample mass can be reduced to just a few tens of milligrams. With some effort, this amount can be obtained in irradiation experiments with ~10 MeV/u ions of ~100 μm range. Neutron total scattering experiments are particularly useful for analyzing radiation effects in oxides. Information from the local atomic configuration and the long-range structure combined with high sensitivity of neutrons to both cations and anions provides unique insight into the damage structure of SHI irradiated nuclear ceramics and other materials [145,146].

**2.3 Materials Dependence on Swift Heavy Ion Induced Effects**

Among all materials studied under irradiation in the electronic energy loss regime, insulators have attracted the most attention due to their high sensitivity to damage in these conditions. This Section presents the relevant parameters and specific effects in detail for SHI damage in insulators, followed by a description of the damage response features of metallic and semiconducting materials. The material systems presented were selected to highlight specific SHI phenomena, which are described in the proceedings of various Swift Heavy Ion in Matter (SHIM) conferences [147–156] and in recent review articles [157–159].

*SHI-Induced Phase Changes in Crystalline Insulators*

In crystalline insulators two different types of SHI-induced phase changes are observed, the most common of which is the crystalline-to-amorphous transformation that occurs in many amorphizable materials, such as $SiO_2$ [160], $Gd_3Ga_5O_{12}$ [161], $Gd_2Ti_2O_7$ [22,141,143,162–165], whereas various non-amorphizable materials change from one crystalline phase into another, such as $ZrO_2$ (monoclinic-to-tetragonal) [24,166], $Y_2O_3$, $Sm_2O_3$, $Gd_2O_3$, $Ho_2O_3$ (cubic-to-monoclinic) [25,167], and $Gd_2Zr_2O_7$ (pyrochlore-to-fluorite) [22]. In general, amorphous tracks are easily observed using TEM and track diameters can be directly measured [22,168,169]. Imaging of crystalline tracks embedded in a matrix with a different crystalline structure is more challenging, [22] and the respective track diameters are typically deduced from XRD measurements as described in the previous section [24,136,166,167]. Track radii for various materials irradiated with ions of ~5 MeV/u (high-energy shoulder of the Bragg curve) and measured by different analytical techniques are summarized in Figure 8 [158]. For a given energy loss $dE/dx$, the track size in amorphizable materials tends to be larger than in non-amorphizable materials. The figure also illustrates that tracks are formed only above a critical electronic energy loss. The track formation threshold depends on the target material and is a specific characteristic of the SHI. Interestingly, the track size and thresholds for a number of amorphizable ceramics are rather similar, despite their different structures and chemical compositions [164].

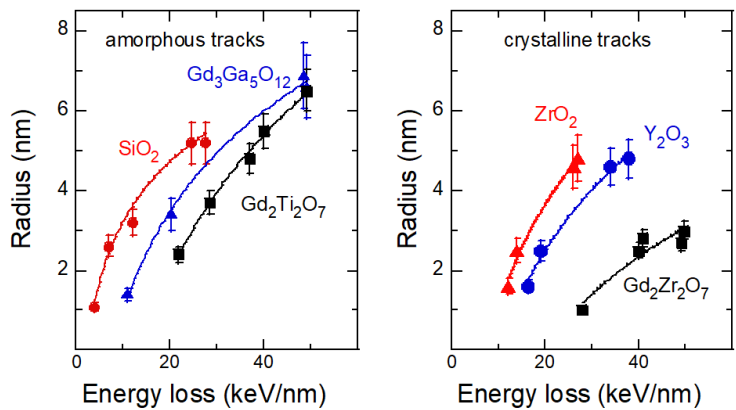

*Figure 8: Track radii as a function of energy loss for (a) amorphizable insulators $SiO_2$, $Gd_3Ga_5O_{12}$, $Gd_2Ti_2O_7$ [162], and (b) non-amorphizable insulators $ZrO_2$ [166], $Y_2O_3$ [167], and $Gd_2Zr_2O_7$ [162] for ion-beam energies between 5 and 10 MeV/u.*

*SHI-induced phase changes in insulators of high ionicity*

Tracks also exist in various non-amorphizable materials. The tendency to undergo amorphization is directly related to the strength of ionic bonding, *i.e.*, materials with a higher degree of ionicity are less easily amorphized. Non-amorphizable materials include some of the above mentioned oxides and ionic crystals. Many ionic crystals of alkaline and alkaline earth halides have been irradiated with different light and heavy ions and investigated in great detail with various techniques. In these crystals, radiation induced defect creation is based on the exciton process [170,171], and primary defects are transformed into color centers by trapping electrons (or holes). At high concentrations, single defects aggregate into smaller or larger defect clusters. Due to the extremely high energy density in SHI irradiations, the resulting damage morphology is quite complex. For example, in LiF which has the highest degree of ionicity and was most extensively studied as a model system, tracks have a shell-like structure. Simple color centers (known from irradiations with lowly ionizing radiation) are mainly formed in a large track halo (few tens of nm), while the track core (2-3 nm) consists of small defect aggregates (Li and F clusters) [172,173]. Other SHI-induced effects in ionic crystals are the appearance of surface hillocks, swelling, and fragmentation of the samples into smaller grains (Fig. 9) [172,174].

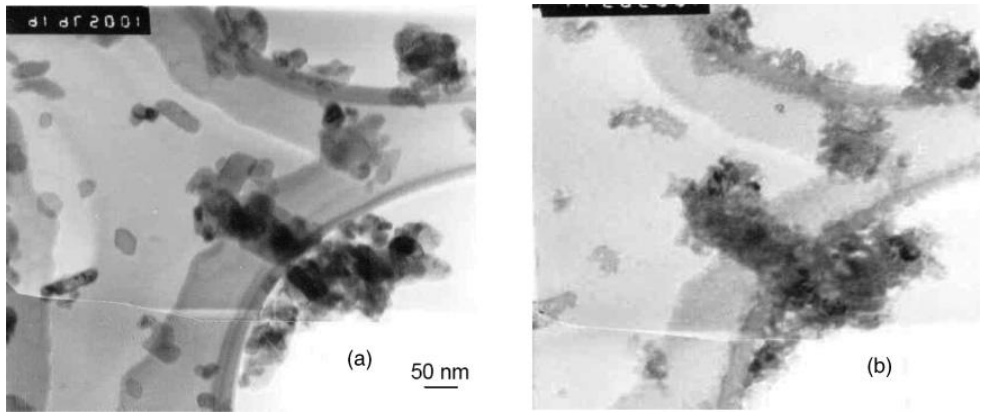

*Figure 9: TEM images of $CaF_2$ grains (a) before and (b) after irradiation with 4.1 MeV/u Pb ions of fluence $10^{12}$ ions/cm$^2$ [174] (copy right obtained from APS).*

The sensitivity of several ionic crystals in the electronic energy loss regime was also tested by measuring sputtering yields, which should in principle follow the same criteria as track formation in the bulk material. The experiments revealed that sputtering requires a higher critical energy loss than track formation. This is in clear contrast to the nuclear collision regime, where the energy necessary to create a defect is about five times larger than the energy required to sputter atoms from the surface. Furthermore, it was found that the sputtering rates of ionic crystals can be extremely high (thousands of sputtered particles per incoming ion) and have an unusual angular distribution consisting of a jet component perpendicular to the sample surface that is superimposed on an isotropic component [175,176].

*The Velocity Effect*

The previous section compares track sizes in different insulators irradiated with ions of about the same velocity (specific energy ~5 MeV/u). Over time it was discovered that the velocity of the ions has a direct impact on the range of the electron cascade developed in the initial stage of ion-target interaction. This aspect is not taken into account by the d$E$/d$x$ value which represents the linear energy transfer along the ion path but ignores into which radial distance the energy is deposited *via* the electron cascade. The range of the electron cascade is directly connected to the maximum energy transfer to the electrons as given by Eq. (2), defined by the kinetic energy of the projectile. At higher ion velocities, the energy is spread into a larger volume. Comparing ions of the same d$E$/d$x$ but different velocity, the resulting energy density becomes lower as ion velocity is increased. This so-called *velocity effect* was initially observed in $Y_3Fe_5O_{12}$ [129,177] and is illustrated exemplarily for $Gd_2Ti_2O_7$ pyrochlore [164,178] and vitreous $SiO_2$ [21,179,180] in Figure 10. Irradiation experiments with high-velocity ions were performed at large accelerator facilities (GANIL and GSI) and with small-velocity ions at Megavolt tandem accelerators (*e.g.*, Aramis at Orsay).

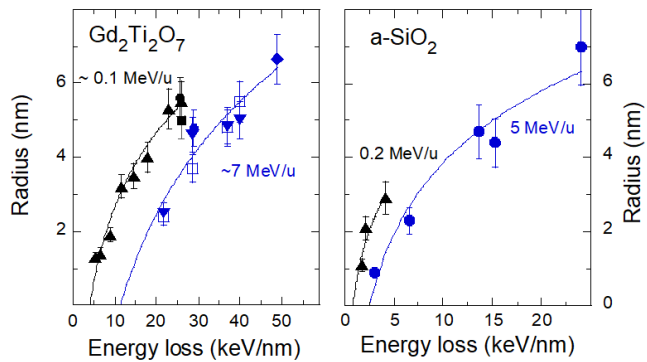

*Figure 10: Track radii as a function of energy loss in (left) crystalline $Gd_2Ti_2O_7$ and (right) vitreous $SiO_2$ irradiated with ions of low (black) [178,179] and high (blue) [164,180] velocities as indicated.*

The consequence is that for the same $dE/dx$, the large energy density deposited by low-velocity ions leads to larger track diameters as compared to high-velocity ions. Another important aspect of the velocity effect is that the $dE/dx$ threshold for track creation increases with the ion velocity. An example of low velocity ions are fission fragments (specific energy of 0.5-1 MeV/u). They induce larger tracks or more severe damage than ions of several MeV/u as typically used at large accelerators [25,181]. Irradiation with $C_{60}$ cluster ion beams of ~0.1 MeV/u is a unique case for damage formation. Due to the collective effect of the 60 constituents, the electronic energy loss of the cluster projectile is extremely high (50 keV/nm and higher). The low velocity in combination with the enormous $dE/dx$ leads to extreme energy densities which are sufficiently large to form tracks even in materials that are inert under monoatomic SHI irradiation (*e.g.*, Si) [182].

*Damage Morphology of Ion Tracks*

The damage morphology along the track may deviate significantly from a homogeneous cylinder [168]. Close to the track formation threshold, tracks in $Y_3Fe_5O_{12}$ consist of spherical defects with a radius of ~1.5 nm. With increasing $dE/dx$, they merge into a discontinuous trail of damage fragments with the same radius and finally become more and more continuous but with the radius fluctuating along the track [168,183]. Finally, at even higher $dE/dx$ values, the track has an almost cylindrical shape and homogeneous damage morphology of constant radius. The discontinuous track morphology is usually ascribed to statistical variations of the electronic energy loss fluctuating around the energy loss threshold [184,185]. Track etching is generally performed for continuous tracks with a radius above a critical value [186,187]. A more detailed insight into the longitudinal track morphology was obtained by a TEM study imaging details along the full track length [10,178]. For the track region at which the electronic and nuclear energy losses are comparable this information is helpful to better distinguish spherical damage regions from discontinuous tracks (electronic energy loss) and damage cascades that are induced by elastic collisions (nuclear energy loss).

It is only recently that the damage morphology across the ion track could be investigated using high-resolution TEM or SAXS. First evidence of a core-shell effect was found in amorphous SiO$_2$ using SAXS measurements which provide information on radial changes of the electron density within the tracks [21]. According to the small-angle scattering pattern, the tracks consist of a low-density amorphous core which is surrounded by an over-dense shell [21]. The measured track radii of the core and shell are presented in Figure 11. The velocity effect has not only an influence on the size of the core but also of the shell. Densification of amorphous SiO$_2$ under SHI was initially observed by Busch *et al*. [188] using infrared spectroscopy and later complemented by Rotaru *et al*. [180] who deduced a track size that is in agreement with the shell radius from the SAXS measurements.

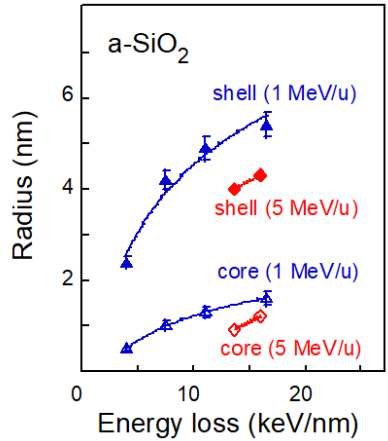

*Figure 11: Track radius versus electronic energy loss for vitreous SiO$_2$ at two different beam energies: the radii for the core and shell of the tracks was deduced from SAXS measurements [21]. Lines are guides to the eye.*

*SHI-Induced Radiation Effects in Metals*

The radiation response of metals to SHI irradiation has been studied by Dunlop *et al*. in the 90's [189–193]. Metals are non-amorphizable and most of them (*e.g*., Cu, Ag, Ni, Pt, and W) are insensitive to SHI-induced electronic excitation and ionization effects. Irradiation experiments were usually performed at temperatures below 90 K to avoid annealing of induced point defects. Examples of track radii deduced from fluence-dependent resistivity measurements are shown for Fe, Ti, and the semimetal Bi [189–191,194] in Figure 12. In contrast to pure metals, some metallic alloys are amorphizable (*e.g*., Ni$_3$B [5,195] and NiZr$_2$ [191]) and tracks were successfully imaged using TEM [196]. Other conducting materials show effects under SHI irradiation, such as the high-$T_c$ superconductors YBa$_2$Cu$_3$O$_7$ and Bi$_2$Sr$_2$Ca$_2$Cu$_2$O$_7$. Ion tracks have been used to induce pinning centers in such compounds [197–200].

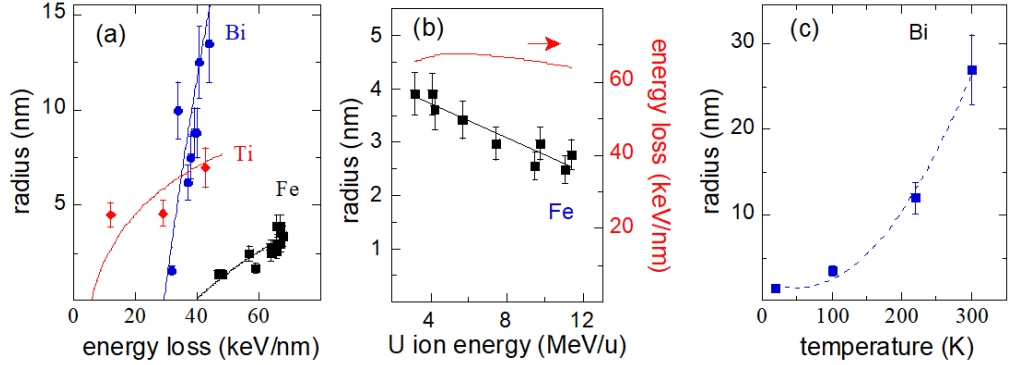

*Figure 12:* *(a) track radii versus electronic energy loss for Fe* [189], *Ti* [190], *and Bi* [194]. *(b) Track radii and electronic energy loss versus U beam energy in Fe* [189]. *(c) Track radii in Bi for an irradiation with ions of dE/dx = 37 keV/nm as a function of irradiation temperature* [194].

For Fe, the velocity effect was confirmed in an experiment with U ions where the track radii became steadily smaller with increasing beam energy at almost constant energy loss (Fig. 12b) [189]. Also the irradiation temperature plays a critical role as shown for Bi in Figure 12c [201], which exhibit a pronounced increase in track radius from ~5 to ~25 nm in the temperature range from 20 to 300 K.

*SHI-Induced Radiation Effects in Semiconductors*

Due to its importance for electronic applications, Si was one of the first semiconductors irradiated with monoatomic SHIs, but no track formation has been identified [202–204]. Tracks in the wide-bandgap semiconductor GeS were first revealed by Vetter *et al.* using high resolution TEM [205] followed by systematic irradiation studies on III-V semiconductors by Wesch *et al.* [206–210]. The evolution of the track radius of several semiconductors with increasing d*E/*d*x* is presented in Figure 13 for high-velocity monoatomic ions and low-velocity $C_{60}$ clusters. The examples highlight again the velocity effect which is also active in semiconductors. Interestingly, in contrast to the heaviest monoatomic ions, $C_{60}$ cluster ions of low velocity but very high electronic energy loss do produce amorphous tracks in Si [182,211] and Ge [212].

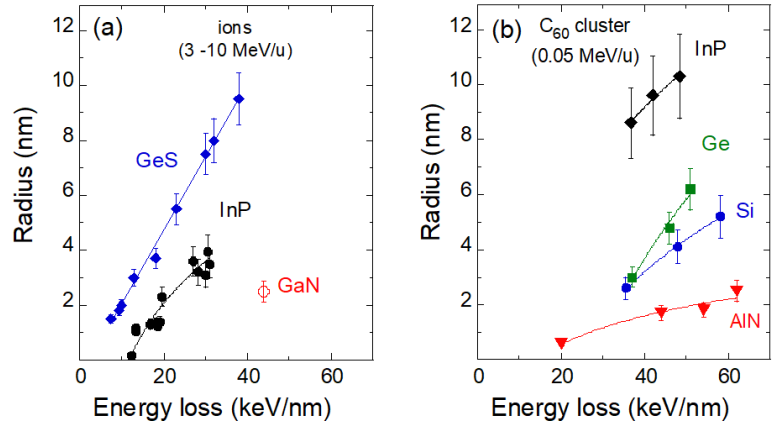

*Figure 13: Track radii as a function of electronic energy loss for various semiconductors: (a) GeS [205], InP [213] and GaN [214] irradiated with high velocity ions in the energy range from 3-10 MeV/u. (b) InP [213], Ge, Si [182,211], and AlN [214,215] irradiated with low velocity $C_{60}$ clusters of 0.05 MeV/u.*

*SHI-induced radiation effects in non-crystalline materials*

Ion tracks also form in non-crystalline materials. Based on existing data, it seems that in the electronic energy loss regime amorphous materials are more sensitive than the corresponding crystalline phase. This applies for track formation in insulators, semiconductors, and metals as well as for electronic sputtering. This phenomenon was first observed by Klaumünzer *et al*. [216,217] in amorphous metallic systems such as PdSi, showing a pronounced anisotropic growth perpendicular to the beam direction. A comparison of track sizes in crystalline ($SiO_2$ [160], Ge [212], and Fe [189]) and corresponding amorphous (a-$SiO_2$ [180], a-Ge [218], and a-$Fe_{85}B_{15}$ [195]) materials is shown in Figure 14. In amorphous materials, the track size is consistently larger than in the crystalline counterpart. Such a pronounced effect on the material structure is not known from the nuclear d$E$/d$x$ regime. Sputtering experiments in the electronic stopping regime confirm larger rates for vitreous $SiO_2$ (a-$SiO_2$) compared to crystalline quartz (c-$SiO_2$) [176,219] (Fig. 14c). The higher sensitivity of amorphous materials indicated by the lower track formation threshold is important when considering radiation effects under extreme SHI radiation conditions. For high-fluence irradiations, the sensitivity (track size and threshold) of the initially crystalline material may be quite different than the completely amorphized material.

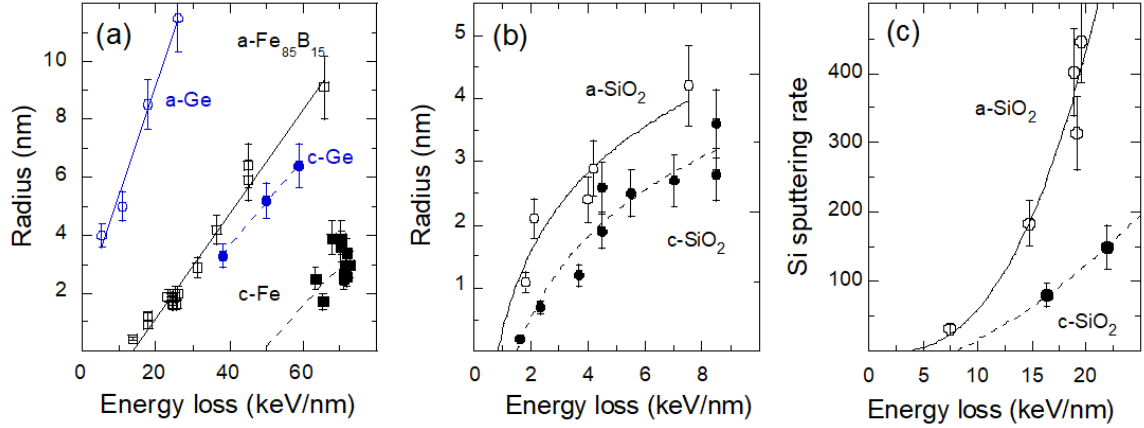

*Figure 14: Comparison of amorphous and crystalline materials: track size versus electronic energy loss of (a) amorphous $Fe_{85}B_{15}$ [5] and crystalline Fe [189] irradiated with 10 MeV/u ions and amorphous (a-Ge) [218] and crystalline (c-Ge) Ge [212] irradiated with 1 MeV/u and 0.05 MeV/u ions, respectively, (b) vitreous ($a$-$SiO_2$) [180] and crystalline ($c$-$SiO_2$) $SiO_2$ [160] irradiated with 1 MeV/u ions, and (c) sputtering rate for $a$-$SiO_2$ [219] and $c$-$SiO_2$ [176] for beam energy of 1 MeV/u.*

Various phenomena can be explained by this crystalline-amorphous structure sensitivity. Titanium irradiated with U ions, for instance, changes from the equilibrium hcp alpha-phase to the high-pressure omega-phase only when exposed to very high fluences ($>10^{13}$ $U/cm^2$). Another example is $Al_2O_3$ which undergoes a two-step process forming first a disordered phase at intermediate fluences and converting into an amorphous surface at high fluences (0.7 MeV/u Xe ions above $2.5\times10^{12}$ $cm^{-2}$) [220–222]. A similar process was also observed for $MgAl_2O_4$ spinel [223].

In summary, the response of materials to irradiation with SHIs shows a strong dependence on various beam and material parameters. Most sensitive are insulators, followed by semiconductors and metals. Amorphous materials are more sensitive than their crystalline counterparts. The track size becomes larger with increasing electronic energy loss, *i.e.*, heavy ions produce larger tracks than light ions. At a given $dE/dx$, sensitive materials have larger track diameters than less sensitive materials. The velocity of the ions also has some influence on the track-formation process because it determines the radial energy density deposited around the trajectory: a high (slow) ion velocity leads to large (small) radial energy spread and thus to a low (high) energy density. Characteristic for ion irradiations in the electronic energy loss regime is the existence of a $dE/dx$ threshold required for track formation. This threshold increases with decreasing sensitivity of the different material classes (low threshold for polymers and high threshold for metals). The irradiation temperature also has an influence on the size of ion tracks; a few examples show increasing track diameter with increasing irradiation temperatures [201,224].

**2.4 Interplay of Nuclear and Electronic Energy Loss and Multi-Beam Phenomena**

If a SHI slows down to energies below the Bragg-peak, the electronic d$E$/d$x$ decreases rapidly and the nuclear d$E$/d$x$ becomes increasingly important for defect production. There is an intermediate energy regime where the nuclear and electronic energy losses have a similar value and may influence each other. Experimental evidence of the following different scenarios have been reported [225]: (*i*) competitive, i.e. 1+1 < 1, (*ii*) cooperative, 1 < 1+1 < 2, (*iii*) additive, i.e. 1+1 = 2, and (*iv*) synergetic 1+1 > 2. Example for a competitive process are in Fe, Cu, and Ni where the electronic energy loss anneals defects that are induced by nuclear collisions [189,192,226–228]. A cooperative process was observed for the damage cross section of crystalline and vitreous $SiO_2$ [21,179,229–231], and a synergetic process seems to be active for surface sputtering of Ti [232].

Another interesting aspect concerns multi-beam experiments in which two (or multiple) different beams are applied simultaneously or in sequence. Several experiments investigated the question of whether and to what extent sequentially applied beams influence the damage effect [233–238]. In $SrTiO_3$ perovskite, for example, the damage induced by a first ion beam (nuclear energy loss) obviously reduces the track-formation threshold for the second ion beam (electronic energy loss), and the higher the pre-damage ratio, the larger the track diameter. Conversely, reduced SHI-track formation in $MgAl_2O_4$ was induced by pre-irradiation with ions of nuclear d$E$/d$x$ [239].

# 3. MODELLING OF SWIFT HEAVY ION TRACKS
## 3.1 Modeling of Electronic Processes

Simulating the interactions of ion projectiles with target atoms requires complex, multiscale modeling. It is currently impossible to fully treat such a system with first principles methods, but with a set of approximations many details of the process can be described. State-of-the-art simulation techniques that can be utilized for SHI-matter interactions include non-equilibrium Green's functions [240] and time-dependent density functional theory [241]. Based on the current computational capacities, these approaches are limited to small systems, such as thin-layer materials or molecules, and to relatively low ion energies. Treating bulk materials with the same degree of precision is not achievable and a more simplified approach is commonly used (with a few notable exceptions, that still required some simplifications, *e.g.*, [242,243]).

The first approximation considers the SHI dynamics to be uncorrelated with the target dynamics, which allows them to be treated separately. The SHI is usually assumed to be a charged (semi-)classical particle. Several interaction channels with the target are considered to account for the electronic energy loss. There are various semi-empirical models available to calculate the equilibrium charge state of a SHI within a medium. The original model by Bohr, later adjusted by Barkas (see *e.g.*, review [244]), is currently one of the most common approximations:

$$Z_{eff}(v) = Z_{ion} \left[ 1 - \exp\left(-\frac{v}{v_0} Z_{ion}^{-2/3}\right) \right] \quad (3)$$

where $Z_{ion}$ is the atomic number of the projectile and $v$ is its velocity with $v_0 = Ac$ and $c$ being the speed of light in vacuum. In Bohr's original model, $A$ denotes the fine structure constant $\alpha = 1/137$, whereas the Barkas model assumes $A = 1/125$. The equilibrium charge expression of Barkas is tuned to reproduce the ion energy loss, but the charge itself is underestimated. A more complex model was proposed by Schiwietz and Grande [245], which was implemented in the CasP code [246] to reproduce the equilibrium charge state in agreement with experimental values over a wide range of targets and SHI velocities. Similar models are also used in other standard codes which simulate ion ranges in solids, such as GEANT4 [247] and FLUKA [248]. The charge state equilibration of ions along their path and the associated charge oscillations can be simulated, for example, with the ETACHA code [74] and its recent extension [249], or by means of a matrix formalism [76].

The ion energy loss can be calculated with different approaches, most commonly based on the first Born approximation. Assuming the wave function of an incident particle to be a plane wave, the Born approximation allows splitting the cross section of scattering on a multicomponent target into the scattering cross section associated with an individual atom and the dynamic structure factor (DSF) [97]. This significantly simplifies the calculation of the stopping power. Assuming the target is in thermodynamic equilibrium, the DSF can be recast in terms of the complex dielectric function (CDF, $\varepsilon(\omega,q)$) of the target employing the fluctuation dissipation theorem [250]. The differential cross section of scattering $\sigma$ in the non-relativistic case is then written as [91]:

$$\frac{d^2\sigma}{d(\hbar\omega)d(\hbar q)} = \frac{2(Z_{eff}(v)e)^2}{n_{at}\pi\hbar^2 v^2} \frac{1}{\hbar q} \left[1 - \exp\left(-\frac{\hbar\omega}{k_B T}\right)\right]^{-1} Im\left(\frac{-1}{\varepsilon(\omega,q)}\right) \quad (4)$$

where the imaginary part of the inverse CDF $Im\left(\frac{-1}{\varepsilon(\omega,q)}\right)$ is also called the loss function; $\hbar\omega$ is the energy transferred into the system and $\hbar q$ is the transferred momentum ($\hbar$ is the Planck constant); $Z_{eff}$ is the effective charge of the incident particle defined by Eq. (3); $e$ is the electron charge; $v$ is the velocity of the particle; $n_{at}$ is the atomic or molecular density (depending on the normalization of the loss function); $T$ is the temperature of the sample and $k_B$ is the Boltzmann constant. Moments of the cross section produce the mean free path (MFP, $\lambda$) and electronic energy loss, $S_e$, of the incident particle:

$$\lambda^{-1} = n_{at} \int_{E_{min}}^{E_{max}} \int_{q_-}^{q_+} \frac{d^2\sigma}{d(\hbar\omega)d(\hbar q)} d(\hbar\omega) d(\hbar q) \quad (5)$$

$$S_e = -\frac{dE}{dx} = n_{at} \int_{E_{min}}^{E_{max}} \int_{q_-}^{q_+} \hbar\omega \frac{d^2\sigma}{d(\hbar\omega)d(\hbar q)} d(\hbar\omega) d(\hbar q) \quad (6)$$

where $E_{min} = E_{gap}$, the band gap of the target material; $E_{max}$ is the maximal possible energy transferred from the projectile to an electron, as defined in Section 2.1, or set equal to $E_{max} = E_e/2$ ($E_e$ is the kinetic energy of an electron) in case of an incident electron accounting for the identity of electrons, or $E_{max} = E_h$ in case of valence holes ($E_h$ is the kinetic energy of a hole). The momentum integration limits are defined as $q_\pm = \sqrt{2m^2/(m_e\hbar^2)}\left(\sqrt{E} \pm \sqrt{E - \hbar\omega}\right)$, with $m$ being the mass of the incident particle (SHI, electron, or the effective mass of a valence hole) and $m_e$ is the electron rest mass [69,251].

To calculate the stopping power of an incident SHI, the loss function of the target must be defined, which can be obtained from (*i*) *ab-initio* calculations such as the density functional theory (see e.g., [252]), (*ii*) experimental data (e.g., [253,254]), or (*iii*) from calculations within some model approximations. The input data from experiments such as optical coefficients (CDF for q=0) are often extended into the q>0 region with an appropriate model. The most commonly applied models of this type are the Ritchie and Howie's model [255], Ashley's model [256], and Penn's algorithm [257]. The first one can be most straightforwardly extended from incident electrons to ions, and thus is widely utilized for calculations of the energy loss of ions [258]. Also the Mermin model with analogous experimental input is quite popular in the ion beam community and even more so in the plasma physics community [259,260].

Once the loss function is defined, cross sections and their moments can be obtained for ions, electrons, and holes in the target. They are known to reproduce the experimental and tabulated data well at energies above 20-50 eV for electrons (see *e.g.*, [69]). Ion energy loss calculations are usually cross-checked against SRIM tables (which in turn were compared to experimental data) [261], the NIST database [262], or IAEA databases [263]. Electron mean free paths and ranges can be compared with values in the NIST database [264], the X-ray data booklet [89], or to previously published data.

For modeling track formation and the overall response of a specific material, the dynamics must be taken into consideration. The ion penetration depth and its energy transfer to the electronic system of the target, as well as the early response of that system, can be appropriately modeled with the help of Monte Carlo (MC) simulations [17,265]. Using an event-by-event MC method, one can trace the SHI penetration and electron kinetics in the target. The charged particle is assumed to travel according to its classical trajectory between two scattering events. The energy loss occurs only at the end of the current particle path. The distance traveled is sampled with the help of a random number ($\gamma$ within the range from 0 to 1) assuming the Poisson law:

$$l = -\lambda \ln(\gamma) \qquad (7)$$

This law relies on the assumption of a homogeneous distribution of the scattering centers in the media and does not capture the structure of the target and corresponding geometric effects. In the simplest case, the trajectory traveled is along a straight line, neglecting long range interactions between particles in the absence of an external field. Otherwise, the trajectory must be simulated according to Newton's law, similarly to the molecular dynamics (MD) simulation technique [17] outlined further below.

Each scattering event is also sampled according to the relative probabilities of various scattering channels. Elastic scattering of a projectile, photon emission, inelastic scattering on the different shells of each element of the target (and other channels if included) create the pool of possibilities from which one is selected for a given scattering event. The relative probabilities are defined from the specific cross sections normalized to the total scattering cross section [265,266]. After selecting a specific scattering channel, the energy transfer to the target particle is sampled based on a random number and the integral of the cross section:

$$\gamma\sigma = \int_{E_{min}}^{\delta E} \int_{q-}^{q+} \frac{d^2\sigma}{d(\hbar\omega)d(\hbar q)} d(\hbar\omega)d(\hbar q) \qquad (8)$$

where Eq.(8) must be solved for the transferred energy δE. The momentum transfer can be either sampled from the double differential cross section [265] or estimated with the help of models for the momentum-energy dispersion curve [69]. The transferred energy is then given to the corresponding particle – a photon, an electron, a target atom, etc. This modeling approach creates the initial conditions for the electronic system of the target during the inelastic scattering of a SHI. An electron that received the energy δE has a kinetic energy $E_{kin} = \delta E - I_p$, with $I_p$ being the ionization potential of the shell the electron is being ionized from. The released delta electrons then start to propagate in the target, which are modeled in an analogous manner. Valence holes can also be modeled in the same way with their own effective mass. A model for an effective hole mass in dielectrics was proposed based on the density of states in a material [267] and successfully tested for ion beam simulations [91]. Further electron scattering *via* inelastic channels leads to excited secondary electrons that are described in the same manner as discussed above.

A simplified approach of so-called condensed collisions is often used for relativistic ion impacts. This procedure does not track each individual energy loss event, but rather splits the inelastic energy channels into two contributions: (*i*) close collisions which form the next generations of fast electrons, a process which is modeled with the same approach as described above and (*ii*) distant collisions, which occur along the trajectory between close collisions and act as fictitious friction force to decrease the particle energy following the energy loss function.

This shortens the simulation times, but information is lost on how the energy is deposited in detail and further redistributed in the target through distant collisions. Such an algorithm is used in most standard codes [247,248].

Elastic scattering channels of delta electrons – scattering on target atoms without excitation of electrons – provide target atoms with kinetic energy. The cross sections of such processes can be extracted from experimental data on optical phonon CDF [69,268]. If such data is not available, the elastic scattering cross section can be calculated from the lattice DSF through classical molecular dynamics simulations [269]. However, often even simpler models are used which treat scattering on target atoms the same as scattering on individual atoms (e.g., Mott's cross section [91,270]). Auger decays of deep shell holes can be sampled with the Poisson law of the characteristic decay times which can be found for most elements in the EPICS database [271]. Since exact data are not always available, it is often assumed that the decay times in solids are the same as in individual atoms. An electron released in an Auger decay will join the ensemble of other primary and secondary electrons and is included in the algorithm of the MC simulation. A photon released in a radiative decay of a deep shell hole, or in a Bremsstrahlung or Cherenkov emission, can be modeled as a particle traveling with the speed of light until a scattering event occurs. The most essential scattering channel for nonrelativistic photons is the photo-absorption process and the characteristic mean free paths can be found in the EPICS databases [271] and in Henke's tables [90] (also available online [272]).

An example of the electron kinetics during a cascade as modeled with the Monte Carlo code TREKIS [69,91] is shown in Figure 1. A typical electron distribution is created *via* various processes as described above: (*i*) the first front of the radial distribution is formed by the photon transport, exciting new electrons *via* photo-absorption, (*ii*) the second wave front is formed by delta electrons for which the fastest are from a dissipative wave, leaving behind a trace of secondary electrons *via* impact ionization, and (*iii*) the third front of electrons is created by plasmon decays which trigger another dissipating wave with typical energies corresponding to the plasmon energy in the solid (typically around 20-30 eV in dielectrics). The majority of the slow electrons produced in the close proximity of the projectile are due to Auger-decays of holes and impact ionizations [91]. It should be emphasized that the wave-like behavior of electrons during the cascades (<100 fs) is of non-equilibrium nature, which cannot be described in general with equilibrium thermodynamics [70].

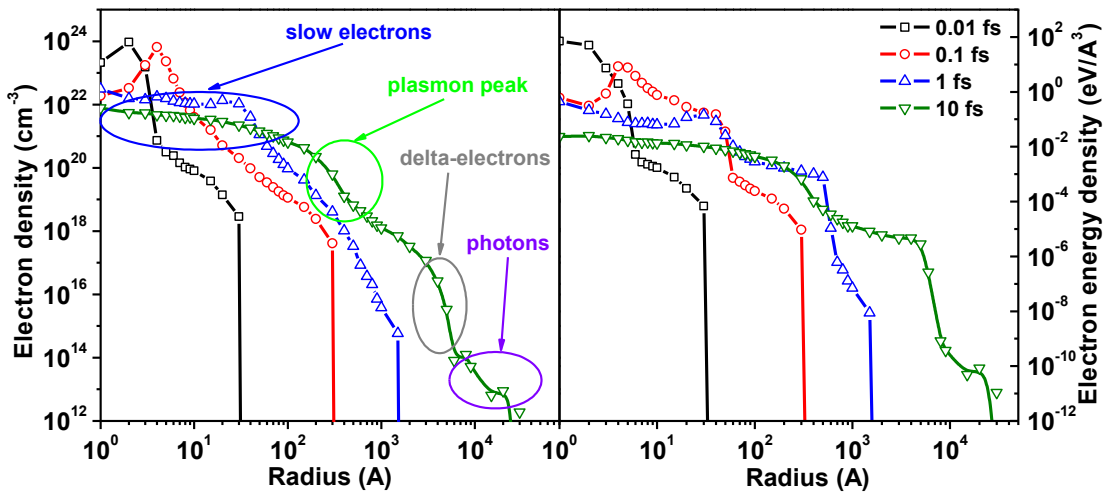

*Figure 15: Radial electron distributions (left) and their energy density (right) at different times after passage of a 2187 MeV Au ion in LiF. Contributions from different processes are highlighted at the 10 fs curve. Figure adapted from [91].*

An interesting consequence of the electron transport and redistribution of deposited energy is the *velocity effect* mentioned above [273]. For the same energy loss, the velocity of the ions has a direct impact on the size of the resulting ion tracks. This arises from the radial distribution of the electron spectra: for higher ion velocities, more energetic delta electrons are produced, which travel faster and transport more energy farther from the ion path. This reduces the energy density. The evolution of the delta-electron spectra as a function of the ion energy (velocity) and its consequence on the produced tracks were recently studied theoretically in detail [105]. With knowledge on the dynamics, the electronic excitation of the target, and the transport of electrons, holes, and photons, one can address track formation. This requires information on how the energy is transferred to the atomic system of the target. As discussed in Section 2.1, there are multiple channels through which the atomic system is affected by the excited electrons. It is beyond current approaches to rigorously account for all of them. Earlier track-formation modeling efforts that focused only on one of these channels are, for example:

(*i*) The Coulomb-explosion model which is based on the assumption that the charge non-neutrality is the sole driving force for track formation. The model was proposed in the 1960s [274] and was further revised, in particular in attempts to combine it with other modelling approaches [111]. The charge distribution from MC is used as initial condition to further evaluate the atomic (ionic) response.

(*ii*) The two-temperature model (TTM) considers only the exchange of the kinetic energy between electrons and atoms of the target. The model was initially proposed in the 1950s [108,275] and further developed within the laser physics community, where its applicability is on stronger grounds [276]. It then returned to the ion-beam community when the importance of electron degeneracy was realized, which however proved insufficient to describe the physics of track formation [277]. The simplicity of the model led to its widespread use in the past three decades. It utilizes MC or similar data sets as initial conditions for the electrons and follows the evolution with coupled thermo-diffusion equations for the electronic and atomic (phononic) system. Adjustable parameters are used empirically to improve the agreement with experimental results on track diameters [278]. This model will be described in detail in section 3.2.

(*iii*) Molecular dynamics (MD) simulations focus on the atomic response to the energy transfer from the electron system [241]. The energy transfer from electrons to atoms can be described in MD simulations in different ways. The dose distributions from MC simulations [279] or empirical analytical formulas [280] were used as initial conditions for MD simulations [111] (as well as TTM models). This approach did not produce sufficient agreement with experimental data and systematically underestimated the track size. As for the TTM model, adjustable parameters were introduced in the MD models, such as the electron-phonon coupling parameter [278] or the deposited dose profile [106] with the aim to increase the energy deposition into the atomic system within the track center. This technique will be further described in section 3.3.

A simplified and efficient model with no need of fitting parameters was recently proposed [281,282] and proved to be sufficiently accurate. In this approach, the atoms are modeled with MD simulations once they receive kinetic energy from electrons and holes which are described within the MC scheme as described above. It has been shown that the hole contribution is an essential parameter that was neglected in previous attempts to model SHI tracks [100]. The model assumes further that the potential energy of electron-hole pairs, which is responsible for non-thermal effects, is also transferred to the atoms within a short time frame.

This energy is particularly important at the periphery of the track as it creates slower decaying radial energy density profiles. Figure 16 shows a typical radial distribution of energy transferred to the atoms as just described using the TREKIS code [69,91]. The conversion of the potential energy of electron-hole pairs into the kinetic energy of atoms avoids complex manipulations of the interatomic potentials. The initial condition for atoms within a MD model is then the kinetic energy transfer from electrons and holes and the potential energy of holes at the end of the electronic cascades (~100 fs) [71,281,282]. Based on this approach ion tracks have been simulated in a wide range of insulators and SHI energies. The obtained track dimensions are in very good agreement with experimental data. Since no fitting parameters are used, this model has full predictive power.

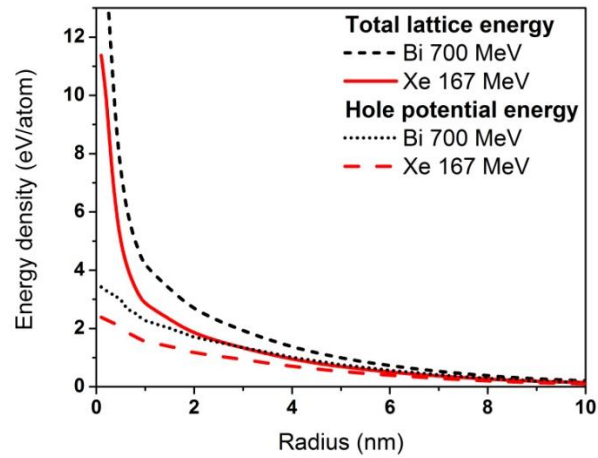

*Figure 16. Example of radial energy density of total lattice energy and potential energy of valence holes in tracks of 700 MeV Bi ions (black) and 167 MeV Xe ions (red) in $Al_2O_3$ at 100 fs. Reproduced from* [281].

**3.2 Two-Temperature Model: A Macroscopic Description of Track Formation**

The electronic kinetics after the passage of a SHI and the subsequent energy transfer to the atoms are complex processes. A theoretical description is still under active development. For practical purposes, simpler semi-empirical models are often used. By introducing (as few as possible) adjustable parameters, such models provide quick and simple estimations of the required parameters, such as the track radius. The inelastic thermal spike (i-TS) model – a particular realization of the two-temperature model – describes track formation as a transient thermal process that allows under certain assumptions the calculation of the track size. The mathematical approach is based on two equations for thermal diffusion in the electron and atom subsystems coupled *via* the electron-phonon coupling parameter. The numerical solutions have been established for ion tracks in different metallic materials [20,194,227], in a large variety of insulators [161,283], and, more recently, in semiconductors [215,284]. The model has also been modified to account for various ion-beam conditions and sample geometries, including multilayer systems [285–289], a full 3D simulation of gold particles embedded in amorphous $SiO_2$ [290–293], and surface processes by low energy highly charged ions [207]. Many details of the thermal spike model are summarized by C. *Dufour et al*. [294], and critically reviewed by S. Klaumunzer [295].

The key elements of the i-TS model are the two heat diffusion equations that link the electron and atom subsystems and depend on time *t* and radial distance *r*. In cylindrical geometry, the two coupled differential equations are:

$$C_e(T_e)\frac{\partial T_e}{\partial t} = \frac{1}{r}\frac{\partial}{\partial r}\left[rK_e(T_e)\frac{\partial T_e}{\partial r}\right] - g(T_e)(T_e - T_a) + A(r,t) \qquad (9)$$

$$C_a(T_a)\frac{\partial T_a}{\partial t} = \frac{1}{r}\frac{\partial}{\partial r}\left[rK_a(T_a)\frac{\partial T_a}{\partial r}\right] + g(T_e)(T_e - T_a) + B(r,t) \qquad (10)$$

where $T_{e,a}(r,t)$, $C_{e,a}(r,t)$ and $K_{e,a}(r,t)$ are the temperature, the specific heat, and the thermal conductivity for the respective electronic (index e) and atom (index a) subsystems. $A(r,t)$ is the initial ion energy deposited to the electrons [279] and $B(r,t)$ is the energy directly deposited to the atoms by ballistic collisions (nuclear energy loss) if applicable [232]. The only free parameter in this model is the electron-phonon coupling parameter $g(T_e)$ [108]. The need to have at least one adjustable parameter stems from the fact that other important effects in electron (as well as hole and photon) transport and energy exchange mechanisms in SHI tracks, described in Section 3.1, are not captured by a thermodynamic approximation.

The thermal equilibrium is assumed to be reached within $2\times10^{-15}$ s for the slow electrons near the SHI trajectory, and within $2\times10^{-13}$ s for the atoms [296]. Fast electrons are not thermalized at such times (Section 3.1). They are thus excluded from the low-energy electron ensemble and only contribute to the source term $A(r,t)$. The following description outlines the approach for how the i-TS model is utilized to obtain information on ion track sizes:

(*i*) The energy of the ion initially deposited to the electronic subsystem is taken from energy loss codes [297–299]. The radial distribution of this electron energy $A(r,t)$ around the ion path is provided by Monte Carlo simulations formalized by Waligorski *et al.* [279]. Based on the initial energy distribution, a cylinder radius $r_e$ is defined, which contains 66% of this energy loss. The value of $r_e$ is material specific and depends on the velocity of the ions. By considering that the initial radial energy distribution of the electrons gets larger with increasing ion velocity, the thermal spike model has well reproduced experimentally determined track radii for different ion velocities in a variety of materials [21,179,180].

(*ii*) The energy in the electron subsystem relaxes first *via* electron-electron collisions (Eq. 9) and eventually *via* electron-atom collisions. This latter process is characterized by the electron-phonon coupling strength $g$ which is connected to the electron-phonon mean free path $\lambda$ by $\lambda^2 = K_e(T_e)/g$. Although $g(T_e)$ can be calculated for various materials, especially metals [99], it proved necessary to replace it with an adjustable parameter instead. For metallic materials, it can then be obtained from Kaganov *et al.* [108] as being proportional to $T_d^2/K_e(T_a)$, where $T_d$ is the Debye temperature and $K_e(T_a)$ the electronic thermal conductivity [194]. These values are tabulated in the papers of Wang *et al.* [20,227]. The situation is more complicated for insulators because (in contrast to metals) there exist no free electrons and the thermal parameters of the electronic subsystem are not available. $\lambda$ has thus been deduced by determining the best fit value from i-TS calculations compared to the evolution of the experimentally determined track radii with increasing electronic energy loss. Figure 17 presents $\lambda$ for various insulators indicating a correlation of $\lambda$ with the band gap energy $E_g$ [161,283].

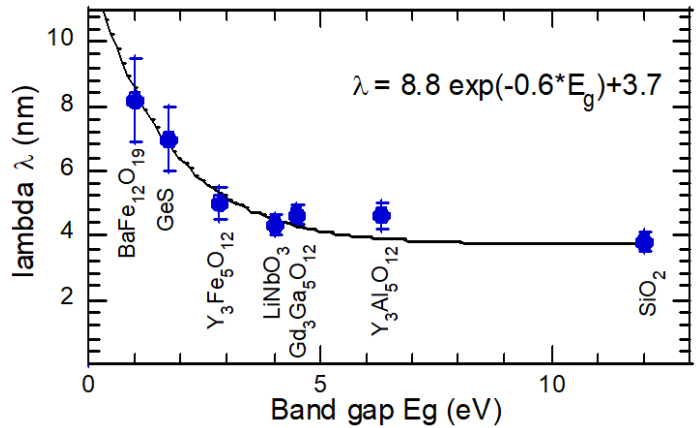

*Figure 17: Electron-phonon mean free path $\lambda$ from i-TS calculations versus band gap energy $E_g$ for a wide range of insulators* [283].

(*iii*) The calculation of the atomic temperature (Eq. 10) is usually based on a superheating scenario. This means that with increasing energy transfer to the atomic system, the temperature increases at the melting, as well as vaporization, temperatures. This process has been experimentally observed in fs-laser experiments [300].

(*iv*) In the thermal-spike model, track formation is directly linked to quenching of a molten phase. The formation of a molten track thus requires that the deposited energy is high enough to reach the melting temperature and to provide the latent heat for the solid–liquid phase change. The track radius is defined by the largest radial zone which contains sufficient energy to reach the molten state. The model is adjusted such that this criterion fits the experimental data, without considering relaxation processes that may lead to recrystallization and recovery of the transient damage [71].

In the past decades, thermal spike calculations for a large number of materials have been made. The interested reader is thus referred to existing publications and reviews on this topic [20,157,158,161,194,289,301]. The application of the thermal spike model for semiconductors has raised some critical questions. Mixing of Ni and Si has been observed under 3 MeV/u Au ion irradiation [302], despite the fact that both Ni [303] and Si [182,211] are insensitive to electronic excitation if irradiated individually. The $\lambda$ values from thermal spike modeling for a number of semiconductors are summarized in Figure 18. Some values follow the same trend as observed for insulators (InN, InP and GaN) while others deviate significantly (Ge, Si, AsGa, AlN). It remains an open question whether the i-TS model fails to predict track sizes in these materials.

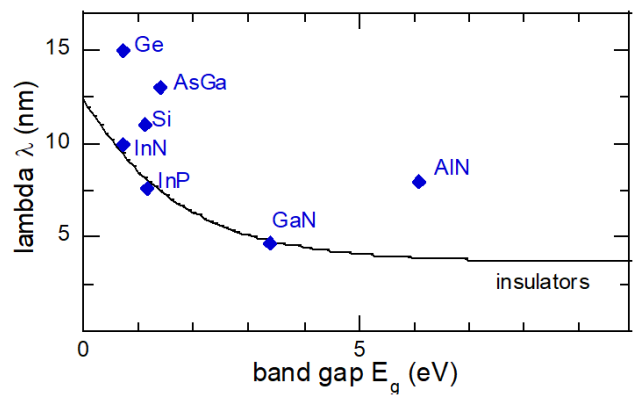

*Figure 18: Electron-phonon mean free path λ from i-TS calculations versus bad gap energy deduced for various semiconductors [218]. The solid line presents the curve (λ = 8.8 exp(-0.6 $E_g$)+3.7) for insulators presented in Fig. 17.*

### 3.3 Molecular Dynamics Simulations

Molecular dynamics (MD) simulations have proven to be suitable for modelling radiation effects in various materials [304]. Unlike the binary collision approximation [9], which describes reasonably well only the initial high-energy (short-distance interaction) phase of collision cascades, the MD method enables the full dynamics of all atoms that either took part in the collision cascade or were in its vicinity until the full dissipation of energy deposited by the ion in the material *via* elastic scattering. Newton's equations, solved iteratively for all atoms in the system at every MD step, can describe accurately not only energetic interactions between colliding atoms (recoils), but also thermal lattice vibrations of all atoms in the system, the equilibrium case being in the form of phonons. The spectrum of resulting phonons depends on the size of the simulation cell and the number of atoms. In adequately large systems, the spectrum of generated phonons is sufficiently close to that observed in experiments, resulting in the accurate simulated value of the lattice heat conduction, especially in crystal structures. The lattice heat conduction provides a channel for energy dissipation, delivering the excess energy to the borders of the simulation cell where it is gradually removed from the system by applying specific thermostats. The latter trick is used to imitate the bulk lattice heat conduction [305].

It is important to note that MD simulations, by definition, do not include electronic effects in solids. The electronic stopping power can, for instance, be introduced as a friction force, as in the PARCAS MD code or *via* the electron-phonon coupling as suggested by Q. Hou *et al*. [306]. Another way to take into account the electronic effects in MD simulations is to solve the heat diffusion equation for the electronic subsystem and couple the electronic energy as heat to the lattice dynamics by applying the Langevin thermostat [307]. The input of the energy from the electrons provides kinetic energy to the atoms. To cool the system, energy excess is gradually removed, and the motion of the atoms is monitored long after the collision phase is completed. This is beneficial for studies of radiation defects in materials as the thermal vibrations of atoms allow the system to relax, eliminating all metastable states and recovering shallow defects. After the energy is fully dissipated from the system, the atoms may assume the positions, comprising the final structure with a stable and, hence, surviving set of defects. Single defects can eventually develop into more complex defect structures such as voids, dislocation loops, and precipitates at longer time scales accessible by other simulation methods such as kinetic Monte Carlo [1,308] or phase-field simulations [309].

The non-thermal melting discussed above (also formerly known as *cold melting* or *lattice destabilization* model) induces significant modifications of interatomic interactions due to electronic excitations. Since the energetics of bond formation change when the electrons are in excited states, the equilibrium states of bonds will also change. This mechanism does not require large kinetic energies of atoms as the change in potential energy will promote the displacement of atoms. The mechanism assumes that the excitation lasts a sufficiently long time to cause fast melting at temperatures that do not exceed the melting point and often can be as low as room temperature (this gave rise to the name of the mechanism). But this mechanism has an inherent problem of implementation in classical atomistic simulations as it requires a modification of the interatomic potentials depending on the excited state of the atoms. A few potentials were developed to take into account the bond softening process [310,311]. However, in metals, the modification of the interatomic potential did not introduce significant difference in the results obtained with the unmodified potential [310]. There are also some attempts to simulate the process of non-thermal melting in SHI tracks using *ab-initio* MD simulations [118]. While the simulations were performed for a very small system, which is not comparable to the experiment, the authors concluded that stable disorder in the $Al_2O_3$ lattice can be expected at electronic temperatures above 10 eV (~100,000 K).

*Molecular Dynamics Simulations of SHI Tracks*

Processes induced by a SHI in the electron subsystem occur on very different time scales than atomic processes. Electron dynamics is in general not modeled within the MD algorithm. This significantly limits the application of MD simulations to describe the track-formation process. A few attempts used MD simulations with interatomic potentials replaced by pure Coulomb repulsion to mimic the stripping of electrons from atoms by the wake of the passing SHI. Such simulations were initially performed by Bringa *et al.* [312,313] for polymers, for which the Coulomb explosion is considered to be a plausible mechanism. In this approach, the damage formation is not linked to the electronic energy loss, but to the ionization per unit length which is used as a fit parameter to define the number of ionized atoms within a track. A more advanced MD model to simulate Coulomb explosion in ionic materials (e.g., LiF) was used by Cherednikov *et al.* [314,315]. They applied a hybrid model of particle-in-cell (PIC) and MD simulations. The ion dynamics were calculated with a classical MD algorithm using a Buckingham-like potential to account for dynamic changes of the ion charge. This charge $q_i$ was estimated by taking into account the local electron density $n_e$ within the track ($q_i = eZ_i - \Omega n_e$, where $q_i$ and $eZ_i$ are the current and total charge state of the ion and $\Omega$ is the atomic volume of the same ion). Due to the complexity of the problem, the electron temperature was not taken into account and the ionization was assumed to proceed through the formation of two $F^+$ ions and four electrons per LiF monolayer. The ion energy was distributed between the potential energy of the $F^+$ ions and the kinetic energy of the electrons. The de-excitation time of the excited electrons was used as a free parameter. No details of the electron dynamics were considered, and the electrons were treated as a classical fluid. Within this simplified model, the authors found indications of a Coulomb explosion. As a result of the high electric field built up within the track, the positive $Li^+$ ions are ejected from the ion path with relatively high kinetic energies. The ion motion eventually results in strong thermo-elastic waves, which leads to a disordered track with 60% of the initial material density. Despite this insight into track formation from MD simulations based on the Coulomb explosion process, more work is needed to enable quantitative predictions which can be compared with experimental results. It should be reiterated that the Coulomb explosion scenario is considered unrealistic for many solids, because the charge neutralization time is too short to cause significant displacements of atoms [316].

In alternative MD approaches by Urbassek *et al.* [317] and others [312,313], the atoms of the irradiated material were energized within a cylindrical volume by instantaneously giving them a fixed energy corresponding to the electronic stopping power deposited within the length of the simulation cell. The simplicity of the model and reasonable results helped to evaluate the radiation response [318,319], particularly in cases when electronic parameters of a material were not accessible [164,313,318,320,321]. In some of the simulations, a fixed energy per atom across the track length was introduced and only the velocity directions were randomized [319,322]. In other cases, the transferred energy followed a Gaussian profile (with a standard deviation of about 1–2 nm) [323] to account for experimental results of a complex track morphology [143]. In different oxides, this simplified process of energy transfer to the lattice provided valuable insight into track formation, which is characterized by a competition of heat and mass transfer, fast quenching of the molten phase, and final recrystallization. However, this simplified MD approach does not allow accurate prediction of the track size for a particular energy deposition in a given material but still requires adjustment to experimental results [312,317].

More recent approaches combined the MD simulation with input from the energy distribution of the electron system from the thermal spike model [160]. The equation for heat diffusion of the electron system (Eq. 9) is solved numerically. At the time of maximum electron-phonon coupling ($t_{max}$ ~100 fs), the temperature profile $T_a(r,t_{max})$ from the thermal spike calculation is translated into kinetic energies of the atoms within a simulation cell [21,324,325]. The thermal spike model uses equilibrium parameters of the system and some potentials do not accurately reproduce the melting point. It is thus required to adjust $T_a(r,t_{max})$ to the melting point based on the potential in use.

*Concurrent Two-Temperature MD Model*

In the two-temperature MD model (2T-MD) the continuum calculation of heat transport in the lattice before time $t_{max}$ needs to be combined with the discrete calculation of the same process after the energy profile is instantaneously deposited to the MD cell. In the approach developed by Duffy and Rutherford [307], the entire dynamics between the atoms and electrons is simulated concurrently by applying the inhomogeneous Langevin thermostat (Fig. 19). Two limits – continuum (electron dynamics) and discrete (atomic dynamics) – are superimposed by using a grid with points enclosing a sufficient number of atoms to represent a thermodynamic ensemble. The sizes of the grid points have to be selected as a compromise between the accuracy of electronic temperature calculations and statistical fluctuations inevitably appearing if the grid point size is chosen too small. The electronic temperature in such an approach is found in the same manner as in the inelastic thermal spike model (Eq. 9), but on a rigid grid and the temperature of the electronic subsystem $T_e^j$ is found in every grid point *j*.

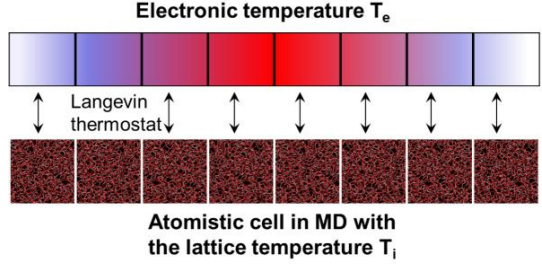

*Figure 19: Illustration of concurrent two-temperature MD model to introduce the effect of the inelastic thermal spike. The electronic heat diffusion is calculated in continuous limit on a grid, while the atomic dynamics is simulated directly by MD.*

This inhomogeneous Langevin thermostat provides the mechanism for the energy transfer between the electronic and the lattice subsystems in MD simulations [326]. This thermostat is applied to the electronic subsystem (the target temperature is $T_e^j$) with energy transferred to the atomistic lattice by a stochastic force term:

$$m_i \frac{d^2 r_i}{dt^2} = \mathbf{F}_i - \gamma_i v_i + \widetilde{\mathbf{F}}_i \qquad (11)$$

In this equation, the force $\mathbf{F}_i$ is acting on an atom $i$ as a result of atomic interactions and the term $-\gamma_i v_i + \widetilde{\mathbf{F}}_i$ describes the friction force that removes energy from atom $i$ due to the electronic energy loss $\gamma_i v_i$ and the stochastic energy gain from electron-phonon coupling $\widetilde{\mathbf{F}}_i$. The force describing the energy gain is $\widetilde{\mathbf{F}}_i = \sqrt{\Gamma_i}\, \widetilde{A}_i$ with $\widetilde{A}_i$ being a vector providing a random direction to $\widetilde{\mathbf{F}}_i$. The parameter $\Gamma_i$ is given by the electron-phonon coupling as $\Gamma_i = 6\gamma_i k_B T_e^j$. Here, $k_B$ and $T_e^j$ are the Boltzmann constant and the electronic temperature in a given grid point $j$, respectively. The coefficient $\gamma_i$ is implemented to account for friction, if the velocity of the energetic target atoms exceeds a cut-off value $v_{cut}$. Hence, $\gamma_i = \gamma_{e-ph} + \gamma_{Se}$, if $v > v_{cut}$, and otherwise, $\gamma_i = \gamma_{e-ph} = \frac{mV}{3Nk_B} G$ with $V$ being the volume and $N$ the number of atoms in the grid point. The implementation of this algorithm is not straightforward as the electron-phonon coupling is only applicable to the thermal motion of the atoms and the collective motion of the atoms (*i.e.*, all atoms in the grid point move together) is excluded. A simpler and more straightforward approach to introduce electronic interactions within MD simulations was suggested by Ivanov and Zhigilei [327] based on a model proposed by Caro and Victoria [328,329]. The stochastic nature of the directions of atoms resulting from the electron-phonon scattering is not taken into account, but the energy is coupled within the non-adiabatic MD simulations as:

$$m_i \frac{d^2 r_i}{dt^2} = \mathbf{F}_i + \xi m_i v_i \qquad (12)$$

with $\xi$ being the electron-phonon coupling term described by $\xi = \frac{VG(T_e - T_a)}{\sum_k (m_k v_k^2)}$ summed over all atoms $k$ within volume V. Since the energy is added in the direction of the stochastic atomic motion, the resulting distribution will not differ significantly from the one obtained using the Langevin thermostat approach.

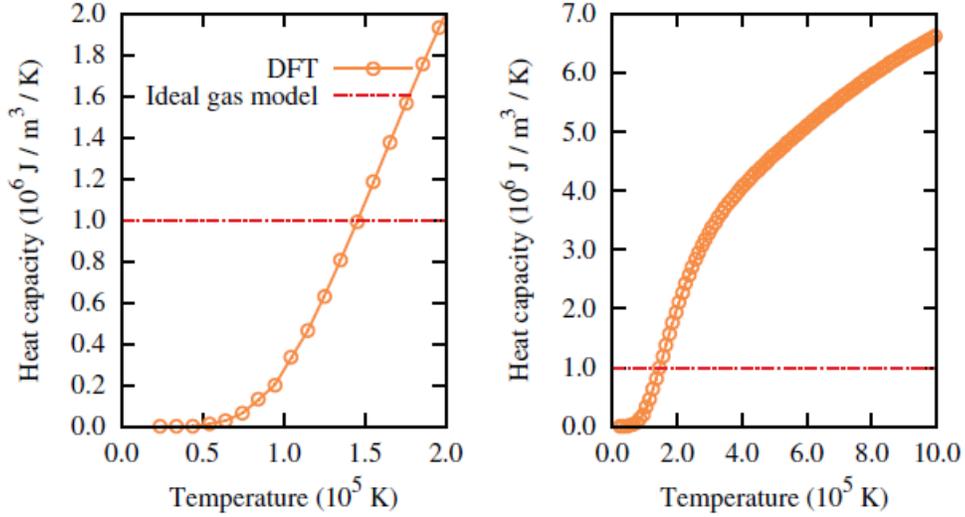

***Figure 20*** *(left) electronic temperature dependence of the specific heat capacity $C_e$ as calculated with DFT; (right) same data in the high temperature range. The dash-dotted lines show the commonly used $C_e$ from the free-electron gas model assuming two excited electrons. Figure adopted from* [330].

The two-temperature MD model for track formation was originally developed for metals and was based on the free electron gas model with a linear dependence of the specific heat capacity on electron temperature. A comparison of the $C_e(T_e)$ function from finite-temperature DFT calculations based on the ideal gas model shows that the latter does not capture the heat storage dynamics in non-metallic systems, which may affect the dynamics of the heat exchange. Although the electron-phonon coupling and specific heat capacity for metals can theoretically be obtained from finite-temperature DFT calculations [99], there appears a large deviation from the linear dependence of $C_e(T_e)$ and the constant value of electron-phonon coupling [99]. Accurate calculations of the electronic parameters in the 2T-MD model can reveal a significantly different behavior of metals under SHI irradiation as compared to the simplified free electron gas model [331].

In non-metallic materials with a band gap, the above approximations are not valid over the entire range of electronic temperatures. Figure 20 compares the specific heat capacity from finite-temperature *ab-initio* calculations for quartz with values obtained from the ideal free electron gas model. It is obvious that this parameter is not constant but has a pronounced dependence on the electron temperature. The situation is further complicated in band gap materials, since the number of charge carriers may vary in space and time, which is not the case for metals. Daraszewicz and Duffy [326] suggested to incorporate the carrier conservation equation into the inelastic thermal spike model to enable the simulation of track formation in Ge irradiated with $C_{60}$ clusters by implementing:

$$\frac{dN}{dt} + \nabla J = G_e - R_e \qquad (13)$$

where $N$ is the concentration of electron-hole pairs, $J$ denotes the charge carrier density, and $G_e = A(r(v_{ion}, t))$ and $R_e = \frac{C_{e-h}}{\tau_{eph}}(T_e - T_a)$ are the source and sink terms of electrons and holes. The carrier density is related to the concentration of electron-hole pairs, electronic temperature, and the band gap of the material. The complexity of this model and the limited available potentials prevent, at the present stage, a detailed comparison of modeling results to experimental data and further work is still required.

*MD Simulations Compared with Experimental Track Data*

Structural disorder induced by the high energy deposition of SHI was already evidenced in early MD simulations [312,313,317]. These simulations aimed to explain the high sputtering yields observed under electronic excitations which greatly exceeded those measured in the nuclear energy loss regime. The track radius was a fitting parameter and the track structure was not further analyzed.

The application of MD simulations led to an improved understanding of the atomic dynamics. It became clear that efficient heat conduction of the lattice with a strong recrystallization ability plays a significant role in the final size and morphology of the track. This explains why specific materials such as diamond, ZnO [332], SiC [333,334], and some pyrochlore oxides [323] are fairly resistant under SHI irradiation. MD simulations also revealed that the mobility of certain species plays a crucial role, such as O atoms migrating less in $CeO_2$ than in $UO_2$ [335,336], and explains the tendency of materials such as $Al_2O_3$, MgO, and YAG to recrystallization [71]. Recent MD simulations combined with the i-TS approach helped to understand the track size in complex pyrochlore oxides and details in the observed damage morphology that strongly depend on the chemical composition (Fig. 21) [183].

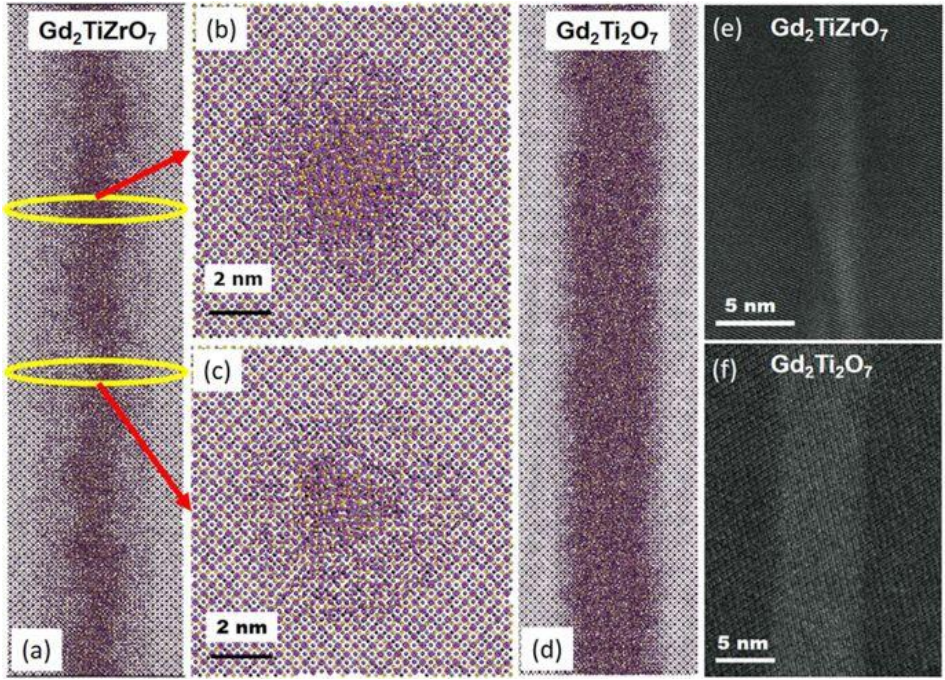

*Figure 21: MD simulations (a-d) and TEM images (e,f) of tracks in $Gd_2TiZrO_7$ and $Gd_2Ti_2O_7$ irradiated with 2.3 GeV Pb ions. Details of the damage morphology change with slight variation of the chemical composition. The figure is reprinted from* [183].

Another excellent agreement between MD simulation and experimental data has been achieved for tracks in vitreous $SiO_2$. SAXS measurements revealed a well-pronounced core-shell track structure [21]. As in many other materials, the mass density in amorphous tracks was known to be smaller compared to the surrounding matrix [160,337–339]. However, the new SAXS measurements revealed a subtle but significant density difference within the track region itself, and with complementing MD simulations, structural details and the core-shell formation process are now better understood. Figure 22 shows the radial density profiles after thermal quenching for various energy loss values together with the profile deduced from SAXS data [21]. The simulations reveal that the sudden thermal expansion of energized atoms within the narrow region along the ion path creates a pressure wave towards the cooler surrounding. This leads to a densification of the outer shell region at the expense of the track interior resulting in an under-dense core. Due to the high viscosity of vitreous $SiO_2$, the density gradient freezes during the quenching phase to room temperature [160].

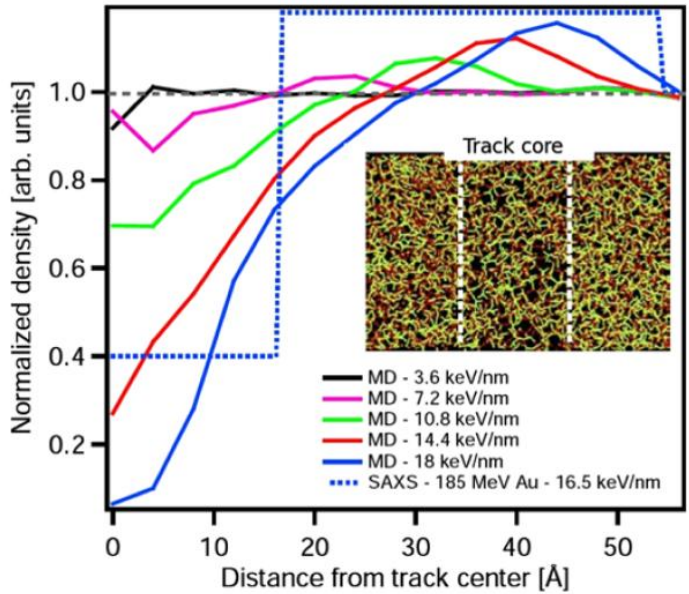

*Figure 22: MD simulations for tracks in vitreous $SiO_2$ showing the radial density profiles of tracks for various energy loss values. Due to the pressure wave induced by the ion, the track consists of an underdense core surrounded by an overdense shell. The simulations are in excellent agreement with the profile used to explain SAXS measurements (dotted line)* [21].

A similar trend was recently observed in amorphous $Si_3N_4$ [340]. Tracks in amorphous germanium show also an under-dense core and over-dense shell which leads to an interesting bow-tie shaped structure (Fig. 23) [112]. Here again, MD simulations provide insight into the underlying processes: the solid-to-liquid phase transformation is followed by a volume contraction because the molten phase has a higher density than solid Ge. This opens up free volume, which organizes itself into voids. Gradually the high-density liquid solidifies, but the temperature within the track core remains sufficiently high to trigger a transition from a high-density amorphous state to a more relaxed low-density amorphous state. The bow-tie structure results from radially inward re-solidification and expansion of material into the void with a shape well reproduced by simulations (Fig. 23) [106,341]. The produced voids are potential precursors of the formation of a macroscopic sponge-like porosity observed under high fluence irradiations [342].

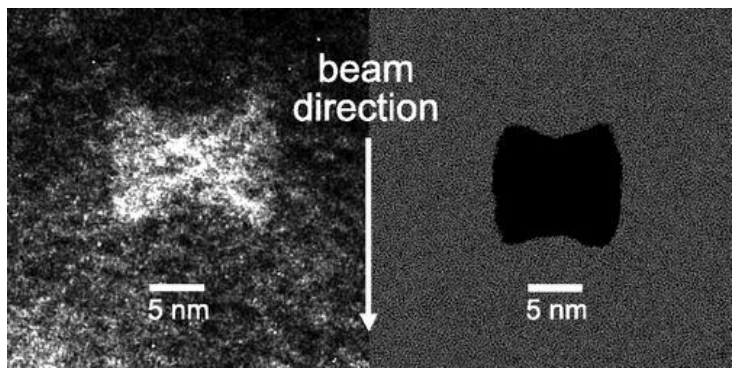

*Figure 23: TEM image of a bow-tie-shaped void observed in amorphous germanium after irradiation with 185 MeV Au ions (left) compared with MD simulations (right). The figure is a reprint from* [106]*.*

In diamond-like carbon, SHIs trigger atomic rearrangements that lead to graphitization, increased electrical conductivity, and a significant reduction in $sp^3$ bonds within the track [53,325,343,344]. This disordered phase along the track has a lower density which gradually increases from the track core outward until it reaches the value of the undamaged surrounding bulk material (**Error! Reference source not found.**).

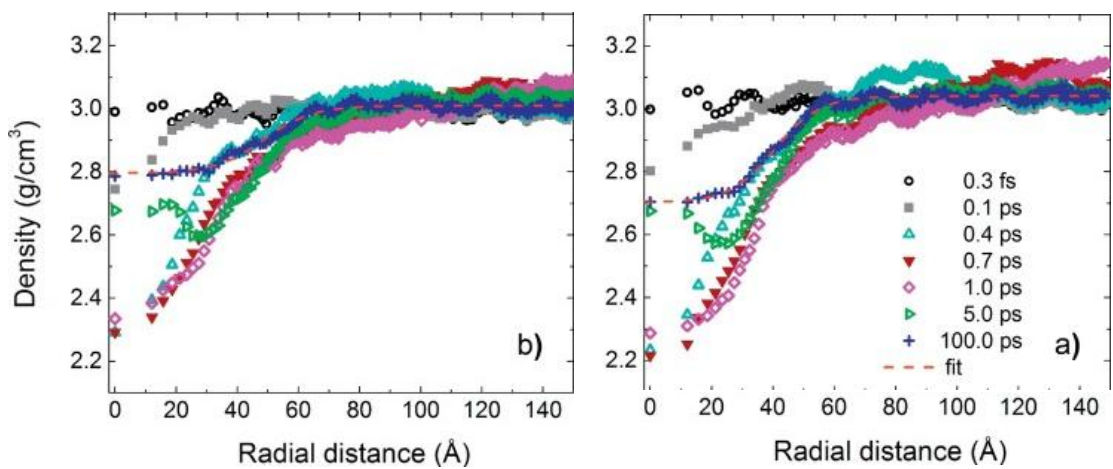

*Figure 24: Evolution of radial density during track formation in diamond-like carbon. The fit uses a Fermi-like function to obtain the final track radius. The figure is adopted from* [53]*.*

The hybrid i-TS+MD model approach describes various other trends such as the evolution of the track size in $SiO_2$ with increasing electronic energy loss [325] and the temperature dependence of track radii measured by SAXS in crystalline quartz and apatite [224]. Experimentally measured track effects in $SrTiO_3$ pre-damaged with low-energy ions [233] were modeled and attributed to increased electron-phonon coupling due to a defect-induced reduction of thermal conductivity in the electronic and atomic subsystems. It has also been shown that SHIs can act as a heat source and induce annealing of pre-existing damage and epitaxial crystallization. This so-called SHIBIEC effect was initially observed in SiC [321], studied in detail with MD simulations [333,345], and later confirmed in a number of other materials [236].

In summary, MD simulations of SHI-matter interactions have provided valuable insight into atomic-level mechanisms that are induced in many materials as a result of the large energy transfer from electronic excitations to the atomic system. Several methods have been developed to handle this energy-transfer mechanism that is not intrinsic to the MD system. Despite the discrepancies between the different approaches, the significantly improved insight into the overall complex situation by means of MD simulations has to be emphasized.

## 4. SWIFT HEAVY ION RELATED RESEARCH IN NUCLEAR MATERIALS

Irradiation experiments with SHIs have relevance for nuclear materials research, particularly to simulate the effects of fission fragments in nuclear fuels. The primary sources of radiation damage in nuclear fuels are fissile actinides, such as uranium, which fission into two fragments of ~100 MeV kinetic energy. It is interesting to note that end-of-range ballistic collisions by fission fragments are almost entirely responsible for the displacement damage near the end of the $UO_2$ fuel life (4 to 6 years) in current nuclear power plants [346]. Only about ~20 displacements per atom (dpa) are from collisions with fission neutrons, while fission fragments produce ~1200 to ~1800 dpa between the center and outer periphery of the fuel pellet. This damage process, which is based on elastic collisions, has been studied in great detail and is well understood. However, many open questions remain on the effects of the electronic energy loss and the extremely high energy densities that are induced along the major part of the trajectory of fission fragments. As shown in the schematic of Figure 1a and the corresponding energy-loss curve in Figure 2b, a typical fission fragment with a mass of 132 atomic mass units and an energy of 80 MeV has an electronic energy loss that dominates for about 5.5 μm of the total 6 μm range in $UO_2$. No tracks have been observed for fission fragments with an electronic energy loss of 18-22 keV/nm in $UO_2$ and tracks have been only confirmed at the sample surface [347–349]. However, MD simulations have shown that the electronic energy deposition of fission fragments produces $10^4$ Frenkel pairs per projectile on the uranium sublattice and $1.4\times10^4$ Frenkel pairs on the oxygen sublattice [350]. The intense ionization processes during the slowing down of fission fragments can also influence the formation and stability of fission-gas bubbles and alter the chemical composition of the $UO_2$ fuel *via* radiation enhanced diffusion. Some studies suggest that fission fragments can result in the stabilization of low-swelling metastable phases [351,352] and fine precipitate structures that act as noble gas nucleation sites (thereby suppressing fuel swelling) [353], as well as the formation of favorable gas bubble architectures [193,354–356].

Experimentally, large accelerator facilities for SHIs provide the most suitable conditions for simulating fission-fragment type damage in nuclear fuel because the irradiation conditions are well-controlled and the ion species as well as their energies can be adjusted. This allows irradiation experiments at various electronic stopping powers, including tests below and above the threshold for track formation. Ion-beam experiments of $UO_2$ with SHIs of an electronic d*E/*d*x* that is slightly higher than that of fission fragments (22-29 keV/nm) evidenced track formation in the bulk [357]. The tracks are not amorphous but contain defects, and the threshold for track formation is relatively high in $UO_2$ as compared to other insulators. This was explained by diffusion processes of overlapping ion tracks, which leads to enhanced defect recovery and a radiation resistance that is similar to metallic materials rather than insulators [358,359]. The enhanced diffusion under irradiation was attributed to the strong electron-phonon coupling in $UO_2$ [360]. Recent irradiation experiments with SHIs have shown that the electronic energy loss induces redox effects in uranium oxide systems that directly dictate the material's response [26,181]. For example, irradiation of $UO_3$ with 167 MeV Xe ions causes the reduction of the uranium cation charge state ($U^{6+} \rightarrow U^{5+}$ and $U^{4+}$) and of the oxygen stoichiometry ($UO_3 \rightarrow UO_{2+x}$) [181]. Alternatively, irradiation of microcrystalline $UO_2$ induces oxidation, while irradiation of nanocrystalline $UO_{2+x}$ induces reduction. In each of these cases, the structure reflects these changes by displaying increased disorder and heterogeneous microstrain [26]. In the case of nanocrystalline $UO_{2+x}$, which is produced in nuclear reactor fuel at the outer periphery of fuel pellets [361], the microstructure is affected by these redox effects, showing grain growth with increasing ion fluence, potentially limiting further grain subdivision in the high burnup structure of nuclear fuels. These results suggest that the enhanced resistance of $UO_2$ to irradiation in the electronic loss regime may be attributed to the flexibility of its electronic structure and the fact that the fluorite structure can accommodate defects without amorphization.

Further valuable insight into the complex radiation response of $UO_2$ was obtained by comparing the SHI radiation response to isostructural analogues such as $CeO_2$ and $ThO_2$. For example, bulk and nanocrystalline samples of $CeO_2$, $ThO_2$, and $UO_2$ retain their crystallinity under irradiation and show only defect accumulation to various degrees [26]. However, the process of defect accumulation and its effect on the microstructure depend strongly on the chemical composition and the grain size of the material (Fig. 25).

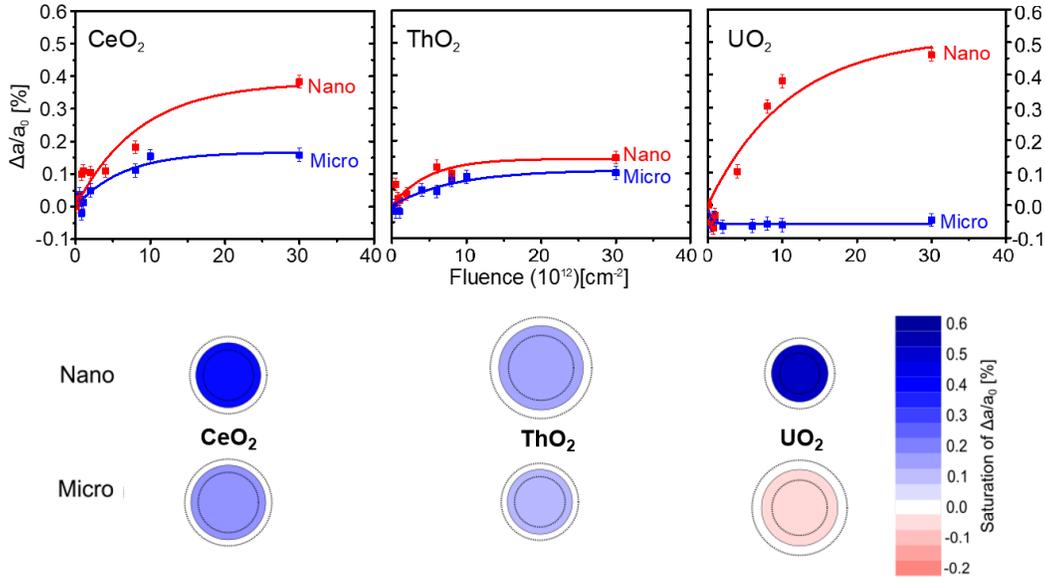

*Figure 25: (Top) Radiation-induced changes in unit-cell parameter as a function of ion fluence for microcrystalline and nanocrystalline $CeO_2$, $ThO_2$, and $UO_2$, based on refinement of corresponding XRD patterns. The solid lines represent fits to the data points based on a single-impact model. (Bottom) Relative track cross section displaying the degree of swelling (blue) and contraction (red) of the unit-cell parameter at the highest fluence. The intensity of color in the schematic scales with the degree of swelling or contraction. This Figure has been modified from [26].*

$ThO_2$ shows the smallest difference in defect-induced swelling between nano- and micro-sized samples, while $UO_2$ displays the most dramatic difference. Accumulation of simple defects in $ThO_2$, regardless of grain size, is attributed to its lack of redox response, due to the thorium cations' propensity to not easily deviate from their tetravalent state [362]. Bulk $UO_2$ shows only small changes in the unit cell volume during irradiation, yet nanocrystalline $UO_2$ exhibits significant swelling. In $CeO_2$, the structural response is driven by reduction mechanisms because cerium cations easily change their valence state ($Ce^{4+} \rightarrow Ce^{3+}$) [363]. Irradiation-induced reduction in $CeO_2$ can even cause the formation of a secondary $Ce_{11}O_{20}$ phase [26,181].

In nanocrystalline materials the production of this phase is accelerated as a function of ion fluence ascribed to the direct expulsion of oxygen from the grains. This hypothesis is corroborated by microscopy images of SHI-irradiated $CeO_2$, where the morphology of individual ion tracks is shown to consist of an oxygen-vacancy rich core and a periphery of an increased number of oxygen interstitials [119]. The affected area of damage is highly dependent on the energy and velocity of the incident ions [364] as well as on the irradiation temperature [365–367]. According to a recent neutron total scattering study, which probed the local ordering in swift heavy ion-irradiated $CeO_2$, the oxygen interstitials may cluster to form small peroxide-like defects [145]. Isochronal annealing between 300 and 1100 K revealed two-stage and one-stage defect recovery processes for irradiated $CeO_2$ and $ThO_2$, respectively, indicating that the morphology of the defects produced with SHI irradiation of these two materials differs significantly. These results suggest that the cation electronic configuration plays a significant role not only in the defect production behavior (Fig. 25) but also in the defect recovery mechanisms of the fluorite-structure oxides [368,369].

In summary, the radiation tolerance of fuel-type materials depends on the presence and efficiency of redox reactions, such that damage can be inhibited (microcrystalline $UO_2$) or exacerbated ($CeO_2$ and nanocrystalline $UO_2$) by altering the cation valence which drives microstructural modifications. Irradiations with SHI are essential for the design and performance prediction of actinide-bearing nuclear fuels because they allow systematic investigations on material-specific beam-induced redox behavior and microstructural changes. Future research must address the complex and evolving chemical and structural behavior of fuel materials. This should include the effect of stress, temperature, thermal gradients, and the interplay of radiation effects due to electronic and nuclear energy losses. This also requires a variation of the experimental conditions (ion species, energy, fluence, etc.) including coupled extreme environments (temperature and pressure). There are many different potential fuel forms that are currently being considered, for example, UC and UN compounds. As compared with traditional oxide fuel, these materials contain much greater uranium densities and will experience, therefore, more extreme ionizations from exposure to intense radiation fields of fission fragments. Investigations on the behavior of such advanced reactor fuels under well-controlled SHI irradiation conditions are important to affirm their performance and safe use in next-generation reactor systems.

Apart from nuclear fuels, irradiation with fission-fragment species is of limited relevance for other nuclear materials, such as structural components in a reactor or waste forms, because in those applications the radiation damage is predominantly caused by ballistic collisions of either neutron irradiation [370] or low-energy recoil nuclei originating from alpha-decay events of incorporated radionuclides [371]. However, SHIs can be extremely useful to produce defects in sufficiently large nuclear sample volumes for advanced bulk characterization techniques. For example, neutron total scattering experiments are very suitable to elucidate damage details in irradiated waste form materials, but the technique requires ~100 mg of irradiated sample mass. This is achievable with SHIs due to their large range [372]. As potential candidate materials, pyrochlore oxides ($A_2B_2O_7$) were intensively investigated. Depending on their chemical composition, they disorder or amorphize under intense irradiation. Interestingly, for many pyrochlore compositions, the final material modifications due to SHI irradiation (electronic stopping) is consistent with radiation effects induced by keV – MeV ions (nuclear stopping). Recent neutron pair distribution function analysis revealed that for amorphized and disordered pyrochlore compositions, the local pyrochlore structure is transformed into similar short range atomic configurations [146]. They both are locally best fit by an orthorhombic weberite-type structure, even though at longer length scales the two compositions have distinctly different structures – aperiodic *versus* disordered-crystalline. Thus, the resistance of a material to amorphization may not depend primarily on defect formation energies, but rather on the compatibility of its structure with mesoscale modulations of the local order such that the long-range periodicity is maintained. This is a significant insight into the damage behavior in an important group of nuclear materials and emphasizes how experiments using SHIs can contribute to fundamental energy and materials sciences. In principle, in-reactor neutron irradiations can also produce large quantities of damaged materials, but a major drawback is the prolonged irradiation time (months to years) and the very high activation levels which make post-irradiation analysis very challenging. In addition to understanding irradiation effects, exploring the behavior of nuclear materials under coupled extremes has become increasingly important and SHIs are an essential tool for such endeavors. For example, molten salt reactors are reemerging as potential candidates for advanced nuclear reactors due to the various economic and operational benefits associated with this technology [373]. Realization of such reactor designs are limited in part by knowledge gaps in material degradation pathways within the coupled, harsh environments of high temperature, chemistry, stress, and intense irradiation. SHIs are able to penetrate into a dedicated heated sample chamber that contains the molten salt together with the structural material and allow *in situ* tests of radiation-induced effect (Fig. 5).

There is no doubt that there has been an increased interest of the use of SHI beams for nuclear materials research. This is reflected by the recent proposed efforts to connect a new SHI accelerator (MeV/u) to the existing Advanced Photon Source (APS) and utilize the existing hard X-ray analysis techniques to monitor SHI-induced damage processes (XMAT). Similar plans exist at the National Synchrotron Light Source II (NSLS-II). This review article intended to provide a comprehensive description of fundamental phenomena and applications of SHI irradiations. We hope that this contribution is a useful complement to the large body of literature on low-energy ion irradiation effects in nuclear materials and will help to further implement SHI studies in the nuclear materials research community.

## AUTHOR CONTRIBUTIONS

## ACKNOWLEDGEMENTS

M.L. acknowledges supported by the U.S. Department of Energy, Office of Science, Basic Energy Sciences, under Award DE-SC0020321. Partial financial support from the Czech Ministry of Education, Youth and Sports, Czech Republic [grants numbers LTT17015, LM2015083] is acknowledged by N.M.

## REFERENCES

[1]    W. M. Young and E. W. Elcock, Proc. Phys. Soc. **89**, 735 (1966).
[2]    E. C. H. Silk and R. S. Barnes, Philos. Mag. A J. Theor. Exp. Appl. Phys. **4**, 970 (1959).
[3]    P. Henderson, Mineral. Mag. **42**, 306 (1978).
[4]    G. A. Wagner and P. den Haute, in *Fission-Track Dating* (Springer Netherlands,

Dordrecht, 1992), pp. 59–94.

[5] A. Audouard, E. Balanzat, S. Bouffard, J. C. Jousset, A. Chamberod, A. Dunlop, D. Lesueur, G. Fuchs, R. Spohr, J. Vetter, and L. Thomé, Phys. Rev. Lett. **65**, 875 (1990).

[6] M. Toulemonde, C. Trautmann, E. Balanzat, K. Hjort, and A. Weidinger, Nucl. Instruments Methods Phys. Res. Sect. B Beam Interact. with Mater. Atoms **216**, 1 (2004).

[7] C. Trautmann, in *Ion Beams Nanosci. Technol.*, edited by R. Hellborg, H. J. Whitlow, and Y. Zhang (Springer Berlin Heidelberg, Berlin, Heidelberg, 2010), pp. 369–387.

[8] P. Apel, Nucl. Instruments Methods Phys. Res. Sect. B Beam Interact. with Mater. Atoms **208**, 11 (2003).

[9] J. Ziegler, M. D. Ziegler, and J. Biersack, Nucl. Instruments Methods Phys. Res. Sect. B Beam Interact. with Mater. Atoms **268**, 1818 (2010).

[10] W. Li, M. Lang, A. J. W. Gleadow, M. V Zdorovets, and R. C. Ewing, Earth Planet. Sci. Lett. **321**–**322**, 121 (2012).

[11] D. Schauries, B. Afra, T. Bierschenk, M. Lang, M. D. Rodriguez, C. Trautmann, W. Li, R. C. Ewing, and P. Kluth, Nucl. Instruments Methods Phys. Res. Sect. B Beam Interact. with Mater. Atoms **326**, 117 (2014).

[12] W. Li, L. Wang, K. Sun, M. Lang, C. Trautmann, and R. C. Ewing, Phys. Rev. B - Condens. Matter Mater. Phys. **82**, (2010).

[13] W. Li, L. Wang, M. Lang, C. Trautmann, and R. C. Ewing, Earth Planet. Sci. Lett. **302**, 227 (2011).

[14] E. Segrè, H. Staub, H. A. Bethe, and J. Ashkin, *Experimental Nuclear Physics. Volume I Volume I* (John Wiley & Sons ; Chapman & Hall, New York; London, 1953).

[15] J. Lindhard, M. Scharff, and H. E. Schioett, (1963).

[16] L. L. Gunderson, J. E. Tepper, and J. A. Bogart, *Clinical Radiation Oncology* (Saunders/Elsevier, 2012).

[17] A. V. Solov'yov, *Nanoscale Insights into Ion-Beam Cancer Therapy*, 1st ed. (Springer International Publishing, 2017).

[18] D. Schmaus, S. Andriamonje, M. Chevallier, C. Cohen, N. Cue, D. Dauvergne, R. Dural, R. Genre, Y. Girard, K. O. Groeneveld, J. Kemmler, R. Kirsch, A. L'hoir, J. Moulin, J. C. Poizat, Y. Quere, J. Remillieux, and M. Toulemonde, Radiat. Eff. Defects Solids **126**, 313 (1993).

[19] N. Itoh, D. M. Duffy, S. Khakshouri, and a M. Stoneham, J. Phys. Condens. Matter **21**, 474205 (2009).

[20] Z. G. Wang, C. Dufour, E. Paumier, and M. Toulemonde, J. Phys. Condens. Matter **6**, 6733 (1994).

[21] P. Kluth, C. Schnohr, O. Pakarinen, F. Djurabekova, D. Sprouster, R. Giulian, M. C. Ridgway, A. Byrne, C. Trautmann, D. Cookson, K. Nordlund, and M. Toulemonde, Phys. Rev. Lett. **101**, 175503 (2008).

[22] J. Zhang, M. Toulemonde, M. Lang, J. Costantini, S. della-negra, and R. Ewing, J. Mater. Res. **30**, 2456 (2015).

[23] B. Schuster, M. Lang, R. Klein, C. Trautmann, R. Neumann, and A. Benyagoub, Nucl. Instruments Methods Phys. Res. Sect. B Beam Interact. with Mater. Atoms **267**, 964 (2009).

[24] A. Benyagoub, Phys. Rev. B **72**, 94114 (2005).

[25] C. L. Tracy, M. Lang, F. Zhang, C. Trautmann, and R. C. Ewing, Phys. Rev. B **92**, 174101 (2015).

[26] W. F. Cureton, R. I. Palomares, J. Walters, C. L. Tracy, C.-H. Chen, R. C. Ewing, G. Baldinozzi, J. Lian, C. Trautmann, and M. Lang, Acta Mater. **160**, 47 (2018).
[27] O. Baake, T. Seidl, U. H. Hossain, A. O. Delgado, M. Bender, D. Severin, and W. Ensinger, Rev. Sci. Instrum. **82**, 45103 (2011).
[28] M. Durante, A. Golubev, W.-Y. Park, and C. Trautmann, Phys. Rept. **800**, 1 (2019).
[29] T. Stöhlker, V. Bagnoud, K. Blaum, A. Blazevic, A. Bräuning-Demian, M. Durante, F. Herfurth, M. Lestinsky, Y. Litvinov, S. Neff, R. Pleskac, R. Schuch, S. Schippers, D. Severin, A. Tauschwitz, C. Trautmann, D. Varentsov, and E. Widmann, Nucl. Instruments Methods Phys. Res. Sect. B Beam Interact. with Mater. Atoms **365**, 680 (2015).
[30] V. Kekelidze, A. Kovalenko, R. Lednicky, V. Matveev, I. Meshkov, A. Sorin, and G. Trubnikov, Nucl. Part. Phys. Proc. **273–275**, 170 (2016).
[31] D. Jeon, I. Hong, J. W. Kim, R. Bodenstein, H. J. Cha, S. J. Choi, S. Choi, O. R. Choi, H. Do, B. H. Choi, C. Choi, J. Han, W. K. Han, M. O. Hyun, H. Jang, J. D. Joo, J. Joung, H. Jung, and E. S. Kim, J. Korean Phys. Soc. **65**, 1010 (2014).
[32] J. C. Yang, J. W. Xia, G. Q. Xiao, H. S. Xu, H. W. Zhao, X. H. Zhou, X. W. Ma, Y. He, L. Z. Ma, D. Q. Gao, J. Meng, Z. Xu, R. S. Mao, W. Zhang, Y. Y. Wang, L. T. Sun, Y. J. Yuan, P. Yuan, W. L. Zhan, J. Shi, W. P. Chai, D. Y. Yin, P. Li, J. Li, L. J. Mao, J. Q. Zhang, and L. N. Sheng, Nucl. Instruments Methods Phys. Res. Sect. B Beam Interact. with Mater. Atoms **317**, 263 (2013).
[33] P. R. Rajasekaran, C. Zhou, M. Dasari, K. O. Voss, C. Trautmann, and P. Kohli, Sci Adv **3**, e1602071 (2017).
[34] A. Evans, D. Alexandrescu, V. Ferlet-Cavrois, and K. Voss, in *2015 IEEE Int. Reliab. Phys. Symp.* (2015), p. SE.6.1-SE.6.6.
[35] B. E. Fischer, K.-O. Voss, and G. Du, Nucl. Instruments Methods Phys. Res. Sect. B Beam Interact. with Mater. Atoms **267**, 2122 (2009).
[36] P. Y. Apel, I. V Blonskaya, O. L. Orelovitch, P. Ramirez, and B. A. Sartowska, Nanotechnology **22**, 175302 (2011).
[37] P. Y. Apel, Radiat. Phys. Chem. **159**, 25 (2019).
[38] P. Wang, M. Wang, F. Liu, S. Ding, X. Wang, G. Du, J. Liu, P. Apel, P. Kluth, C. Trautmann, and Y. Wang, Nat. Commun. **9**, 569 (2018).
[39] P. Y. Apel, I. V Blonskaya, N. E. Lizunov, K. Olejniczak, O. L. Orelovitch, M. E. Toimil-Molares, and C. Trautmann, Small **14**, e1703327 (2018).
[40] Z. Siwy, P. Apel, D. Dobrev, R. Neumann, R. Spohr, C. Trautmann, and K. Voss, Nucl. Instruments Methods Phys. Res. Sect. B Beam Interact. with Mater. Atoms **208**, 143 (2003).
[41] I. Vlassiouk, T. R. Kozel, and Z. S. Siwy, J. Am. Chem. Soc. **131**, 8211 (2009).
[42] T. Ma, E. Balanzat, J. M. Janot, and S. Balme, Biosens Bioelectron **137**, 207 (2019).
[43] G. Perez-Mitta, J. S. Tuninetti, W. Knoll, C. Trautmann, M. E. Toimil-Molares, and O. Azzaroni, J Am Chem Soc **137**, 6011 (2015).
[44] G. Pérez-Mitta, A. G. Albesa, W. Knoll, C. Trautmann, M. E. Toimil-Molares, and O. Azzaroni, Nanoscale **7**, 15594 (2015).
[45] A. Spende, N. Sobel, M. Lukas, R. Zierold, J. C. Riedl, L. Gura, I. Schubert, J. M. Moreno, K. Nielsch, B. Stuhn, C. Hess, C. Trautmann, and M. E. Toimil-Molares, Nanotechnology **26**, 335301 (2015).
[46] L. Movsesyan, I. Schubert, L. Yeranyan, C. Trautmann, and M. E. Toimil-Molares, Semicond. Sci. Technol. **31**, 14006 (2015).

[47] L. Burr, Ion-Track Technology Based Synthesis and Characterization of Gold and Gold Alloys Nanowires and Nanocones, Technische Universitaet Darmstadt, 2016.
[48] M. E. Toimil-Molares, Beilstein J Nanotechnol **3**, 860 (2012).
[49] R. Delalande, L. Burr, E. Charron, M. Jouini, M. E. Toimil-Molares, and L. Belliard, Appl. Phys. Lett. **115**, 83103 (2019).
[50] L. Movsesyan, A. W. Maijenburg, N. Goethals, W. Sigle, A. Spende, F. Yang, B. Kaiser, W. Jaegermann, S. Y. Park, G. Mul, C. Trautmann, and M. E. Toimil-Molares, Nanomater. **8**, (2018).
[51] I. Schubert, C. Huck, P. Kröber, F. Neubrech, A. Pucci, M. E. Toimil-Molares, C. Trautmann, and J. Vogt, Adv. Opt. Mater. **4**, (2016).
[52] F. Pellemoine, M. Avilov, M. Bender, R. C. Ewing, S. Fernandes, M. Lang, W. X. Li, W. Mittig, M. Schein, D. Severin, M. Tomut, C. Trautmann, and F. X. Zhang, Nucl. Instruments Methods Phys. Res. Sect. B Beam Interact. with Mater. Atoms **365**, (2015).
[53] K. Kupka, A. A. Leino, W. Ren, H. Vázquez, E. H. Åhlgren, K. Nordlund, M. Tomut, C. Trautmann, P. Kluth, M. Toulemonde, and F. Djurabekova, Diam. Relat. Mater. **83**, 134 (2018).
[54] J. Habainy, Y. Lee, K. B. Surreddi, A. Prosvetov, P. Simon, S. Iyengar, Y. Dai, and M. Tomut, Nucl. Instruments Methods Phys. Res. Sect. B Beam Interact. with Mater. Atoms **439**, 7 (2019).
[55] C. Hubert, K. O. Voss, M. Bender, K. Kupka, A. Romanenko, D. Severin, C. Trautmann, and M. Tomut, Nucl. Instruments Methods Phys. Res. Sect. B Beam Interact. with Mater. Atoms **365**, 509 (2015).
[56] M. Lang, J. Lian, F. Zhang, B. W. H. Hendriks, C. Trautmann, R. Neumann, and R. C. Ewing, Earth Planet. Sci. Lett. **274**, 355 (2008).
[57] M. Lang, F. Zhang, J. Zhang, J. Wang, B. Schuster, C. Trautmann, R. Neumann, U. Becker, and R. C. Ewing, Nat. Mater. **8**, 793 (2009).
[58] M. R. Shaneyfelt, J. R. Schwank, P. E. Dodd, and J. A. Felix, IEEE Trans. Nucl. Sci. **55**, 1926 (2008).
[59] S. K. Hoeffgen, M. Durante, V. Ferlet-Cavrois, R. Harboe-Sorensen, W. Lennartz, T. Kuendgen, J. Kuhnhenn, C. LaTessa, M. Mathes, A. Menicucci, S. Metzger, P. Nieminen, R. Pleskac, C. Poivey, D. Schardt, and U. Weinand, IEEE Trans. Nucl. Sci. **59**, 1161 (2012).
[60] Dartois, E., Chabot, M., Pino, T., Béroff, K., Godard, M., Severin, D., Bender, M., and Trautmann, C., A&A **599**, A130 (2017).
[61] C. Mejía, M. Bender, D. Severin, C. Trautmann, P. Boduch, V. Bordalo, A. Domaracka, X. Y. Lv, R. Martinez, and H. Rothard, Nucl. Instruments Methods Phys. Res. Sect. B Beam Interact. with Mater. Atoms **365**, 477 (2015).
[62] F. Meinerzhagen, L. Breuer, H. Bukowska, M. Bender, D. Severin, M. Herder, H. Lebius, M. Schleberger, and A. Wucher, Rev Sci Instrum **87**, 13903 (2016).
[63] L. Breuer, P. Ernst, M. Herder, F. Meinerzhagen, M. Bender, D. Severin, and A. Wucher, Nucl. Instruments Methods Phys. Res. Sect. B Beam Interact. with Mater. Atoms **435**, 101 (2018).
[64] M. Toulemonde, W. Assmann, C. Trautmann, F. Grüner, H. D. Mieskes, H. Kucal, and Z. G. Wang, Nucl. Instruments Methods Phys. Res. Sect. B Beam Interact. with Mater. Atoms **212**, 346 (2003).
[65] M. Toulemonde, W. Assmann, D. Muller, and C. Trautmann, Nucl. Instruments Methods

Phys. Res. Sect. B Beam Interact. with Mater. Atoms **406**, 501 (2017).

[66] E. Mahner, L. Evans, D. Küchler, R. Scrivens, M. Bender, H. Kollmus, D. Severin, and M. Wengenroth, Phys. Rev. Spec. Top. - Accel. Beams **14**, 50102 (2011).

[67] N. A. Medvedev, A. E. Volkov, N. S. Shcheblanov, and B. Rethfeld, Phys. Rev. B **82**, 125425 (2010).

[68] O. Keski-Rahkonen and M. O. Krause, At. Data Nucl. Data Tables **14**, 139 (1974).

[69] N. A. Medvedev, R. A. Rymzhanov, and A. E. Volkov, J. Phys. D. Appl. Phys. **48**, 355303 (2015).

[70] N. Medvedev, A. E. Volkov, and B. Ziaja, Nucl. Instruments Methods Phys. Res. Sect. B Beam Interact. with Mater. Atoms **365**, 437 (2015).

[71] R. A. A. Rymzhanov, N. A. Medvedev, J. H. H. O'Connell, A. Janse van Vuuren, V. A. A. Skuratov, and A. E. E. Volkov, Sci. Rep. **9**, 3837 (2019).

[72] K. Nordlund, C. Björkas, T. Ahlgren, A. Lasa, and A. E. Sand, J. Phys. D. Appl. Phys. **47**, 224018 (2014).

[73] A. V. Solov'yov, E. Surdutovich, E. Scifoni, I. Mishustin, and W. Greiner, Phys. Rev. E **79**, 011909 (2009).

[74] J. P. Rozet, C. Stéphan, and D. Vernhet, Nucl. Instruments Methods Phys. Res. Sect. B Beam Interact. with Mater. Atoms **107**, 67 (1996).

[75] M. Imai, M. Sataka, K. Kawatsura, K. Takahiro, K. Komaki, H. Shibata, H. Sugai, and K. Nishio, Nucl. Instruments Methods Phys. Res. Sect. B Beam Interact. with Mater. Atoms **267**, 2675 (2009).

[76] O. Osmani and P. Sigmund, Nucl. Instruments Methods Phys. Res. Sect. B Beam Interact. with Mater. Atoms **269**, 813 (2011).

[77] N. A. Medvedev, R. A. Rymzhanov, and A. E. Volkov, Nucl. Instruments Methods Phys. Res. Sect. B Beam Interact. with Mater. Atoms **315**, 85 (2013).

[78] L. Sarkadi, G. Lanzanò, E. De Filippo, H. Rothard, C. Volant, A. Anzalone, N. Arena, M. Geraci, F. Giustolisi, and A. Pagano, Nucl. Instruments Methods Phys. Res. Sect. B Beam Interact. with Mater. Atoms **233**, 31 (2005).

[79] M. Canepa, G. Lulli, L. Mattera, F. Priolo, H. Rothard, G. Lanzanò, E. De Filippo, and C. Volant, Nucl. Instruments Methods Phys. Res. Sect. B Beam Interact. with Mater. Atoms **230**, 419 (2005).

[80] R. H. Ritchie and C. Claussen, Nucl. Instruments Methods Phys. Res. **198**, 133 (1982).

[81] J. Rzadkiewicz, O. Rosmej, A. Blazevic, V. P. Efremov, A. Gójska, D. H. H. Hoffmann, S. Korostiy, M. Polasik, K. Słabkowska, and A. E. Volkov, High Energy Density Phys. **3**, 233 (2007).

[82] A. Akkerman, M. Murat, and J. Barak, Nucl. Instruments Methods Phys. Res. Sect. B Beam Interact. with Mater. Atoms **321**, 1 (2014).

[83] L. G. Gerchikov, A. N. Ipatov, R. G. Polozkov, and A. V. Solov'yov, Phys. Rev. A **62**, 043201 (2000).

[84] R. A. Rymzhanov, N. A. Medvedev, and A. E. Volkov, Phys. Status Solidi **252**, 159 (2015).

[85] N. A. Medvedev, A. E. Volkov, B. Rethfeld, and N. S. Shcheblanov, Nucl. Instruments Methods Phys. Res. Sect. B Beam Interact. with Mater. Atoms **268**, 2870 (2010).

[86] L. S. Cederbaum, J. Zobeley, and F. Tarantelli, Phys. Rev. Lett. **79**, 4778 (1997).

[87] M. L. Knotek and P. J. Feibelman, Surf. Sci. **90**, 78 (1979).

[88] K. Gokhberg, A. B. Trofimov, T. Sommerfeld, and L. S. Cederbaum, Europhys. Lett. **72**,

228 (2005).

[89] A. Thompson, D. Vaughan, J. Kirz, D. Attwood, E. Gullikson, M. Howells, K.-J. Kim, J. Kortright, I. Lindau, P. Pianetta, A. Robinson, J. Underwood, G. Williams, and H. Winick, *X-Ray Data Booklet*, 2009th ed. (Center for X-ray Optics and Advanced Light Source, Lawrence Berkeley National Laboratory, Berkeley, CA, USA, 2009).

[90] B. L. Henke, E. M. Gullikson, and J. C. Davis, At. Data Nucl. Data Tables **54**, 181 (1993).

[91] R. A. Rymzhanov, N. A. Medvedev, and A. E. Volkov, Nucl. Instruments Methods Phys. Res. Sect. B Beam Interact. with Mater. Atoms **388**, 41 (2016).

[92] B. Gervais and S. Bouffard, Nucl. Instruments Methods Phys. Res. Sect. B Beam Interact. with Mater. Atoms **88**, 355 (1994).

[93] D. Emfietzoglou, A. Akkerman, and J. Barak, IEEE Trans. Nucl. Sci. **51**, 2872 (2004).

[94] N. Medvedev, Appl. Phys. B **118**, 417 (2015).

[95] D. A. Chapman and D. O. Gericke, Phys. Rev. Lett. **107**, 165004 (2011).

[96] J. Sempau, E. Acosta, J. Baro, J. M. Fernández-Varea, and F. Salvat, Nucl. Instruments Methods Phys. Res. Sect. B Beam Interact. with Mater. Atoms **132**, 377 (1997).

[97] L. Van Hove, Phys. Rev. **95**, 249 (1954).

[98] A. E. Volkov and V. A. Borodin, Nucl. Instruments Methods Phys. Res. Sect. B Beam Interact. with Mater. Atoms **146**, 137 (1998).

[99] Z. Lin, L. Zhigilei, and V. Celli, Phys. Rev. B **77**, 075133 (2008).

[100] R. A. Rymzhanov, N. A. Medvedev, and A. E. Volkov, Nucl. Instruments Methods Phys. Res. Sect. B Beam Interact. with Mater. Atoms **365**, 462 (2015).

[101] J. K. Chen, D. Y. Tzou, and J. E. Beraun, Int. J. Heat Mass Transf. **48**, 501 (2005).

[102] A. Lushchik, C. Lushchik, E. Vasil'chenko, and A. I. Popov, Low Temp. Phys. **44**, 269 (2018).

[103] P. Martin, S. Guizard, P. Daguzan, G. Petite, P. D'Oliveira, P. Meynadier, and M. Perdrix, Phys. Rev. B **55**, 5799 (1997).

[104] C. Cuesta, M. A. Oliván, J. Amaré, S. Cebrián, E. García, C. Ginestra, M. Martínez, Y. Ortigoza, A. Ortiz de Solórzano, C. Pobes, J. Puimedón, M. L. Sarsa, J. A. Villar, and P. Villar, Opt. Mater. (Amst). **36**, 316 (2013).

[105] R. A. Rymzhanov, S. A. Gorbunov, N. Medvedev, and A. E. Volkov, Nucl. Instruments Methods Phys. Res. Sect. B Beam Interact. with Mater. Atoms **440**, 25 (2019).

[106] M. C. Ridgway, T. Bierschenk, R. Giulian, B. Afra, M. D. Rodriguez, L. L. Araujo, A. P. Byrne, N. Kirby, O. H. Pakarinen, F. Djurabekova, K. Nordlund, M. Schleberger, O. Osmani, N. Medvedev, B. Rethfeld, and P. Kluth, Phys. Rev. Lett. **110**, 22 (2013).

[107] L. T. Chadderton, Radiat. Meas. **36**, 13 (2003).

[108] M. I. Kaganov, I. M. Lifshitz, and L. V. Tanatarov, Sov. Phys. JETP **4**, 173 (1957).

[109] C. W. Siders, Science (80-. ). **286**, 1340 (1999).

[110] S. K. Sundaram and E. Mazur, Nat. Mater. **1**, 217 (2002).

[111] E. M. Bringa and R. E. Johnson, Phys. Rev. Lett. **88**, 165501 (2002).

[112] A. M. Stoneham and N. Itoh, Appl. Surf. Sci. **168**, 186 (2000).

[113] F. F. Komarov, Physics-Uspekhi **60**, 435 (2017).

[114] J. Rossbach, J. R. Schneider, and W. Wurth, Phys. Rep. (2019).

[115] S. Li, S. Li, F. Zhang, D. Tian, H. Li, D. Liu, Y. Jiang, A. Chen, and M. Jin, Appl. Surf. Sci. **355**, 681 (2015).

[116] F. Tavella, H. Höppner, V. Tkachenko, N. Medvedev, F. Capotondi, T. Golz, Y. Kai, M. Manfredda, E. Pedersoli, M. J. Prandolini, N. Stojanovic, T. Tanikawa, U. Teubner, S.

Toleikis, and B. Ziaja, High Energy Density Phys. **24**, 22 (2017).

[117] G. Sciaini, M. Harb, S. G. Kruglik, T. Payer, C. T. Hebeisen, F.-J. M. zu Heringdorf, M. Yamaguchi, M. H. Hoegen, R. Ernstorfer, and R. J. D. Miller, Nature **458**, 56 (2009).

[118] R. A. Voronkov, N. Medvedev, R. A. Rymzhanov, and A. E. Volkov, Nucl. Instruments Methods Phys. Res. Sect. B Beam Interact. with Mater. Atoms **435**, 87 (2018).

[119] S. Takaki, K. Yasuda, T. Yamamoto, S. Matsumura, and N. Ishikawa, Nucl. Instruments Methods Phys. Res. Sect. B Beam Interact. with Mater. Atoms **326**, (2014).

[120] J. Zhang, M. Lang, R. Ewing, R. Devanathan, W. Weber, and M. Toulemonde, J. Mater. Res. - J MATER RES **25**, 1344 (2010).

[121] L. A. Bursill and G. Braunshausen, Philos. Mag. A **62**, 395 (1990).

[122] C. Houpert, M. Hervieu, D. Groult, F. Studer, and M. Toulemonde, Nucl. Instruments Methods Phys. Res. Sect. B Beam Interact. with Mater. Atoms **32**, 393 (1988).

[123] R. M. Papaleo, M. R. Silva, R. Leal, P. L. Grande, M. Roth, B. Schattat, and G. Schiwietz, Phys. Rev. Lett. **101**, 1 (2008).

[124] R. Neumann, J. Ackermann, N. Angert, C. Trautmann, M. Dischner, T. Hagen, and M. Sedlacek, Nucl. Instruments Methods Phys. Res. Sect. B Beam Interact. with Mater. Atoms **116**, 492 (1996).

[125] C. Müller, A. Benyagoub, M. Lang, R. Neumann, K. Schwartz, M. Toulemonde, and C. Trautmann, Nucl. Instruments Methods Phys. Res. Sect. B Beam Interact. with Mater. Atoms **209**, (2003).

[126] J. Liu, R. Neumann, C. Trautmann, and C. Müller, Phys. Rev. B **64**, (2001).

[127] N. Ishikawa, T. Taguchi, and N. Okubo, Nanotechnology **28**, 445708 (2017).

[128] C. Grygiel, H. Lebius, S. Bouffard, A. Quentin, J.-M. Ramillon, T. Madi, S. Guillous, T. Been, P. Guinement, D. Lelièvre, and I. Monnet, Rev. Sci. Instrum. **83**, 13902 (2012).

[129] A. Meftah, F. Brisard, J. M. Costantini, M. Hage-Ali, J. P. Stoquert, F. Studer, and M. Toulemonde, Phys. Rev. B **48**, 920 (1993).

[130] K. Schwartz, C. Trautmann, A. El-Said, R. Neumann, M. Toulemonde, and W. Knolle, Phys. Rev. B **70**, (2004).

[131] B. Schuster, F. Fujara, B. Merk, R. Neumann, T. Seidl, and C. Trautmann, Nucl. Instruments Methods Phys. Res. Sect. B Beam Interact. with Mater. Atoms **277**, 45 (2012).

[132] F. Studer and M. Toulemonde, Nucl. Instruments Methods Phys. Res. Sect. B Beam Interact. with Mater. Atoms **65**, 560 (1992).

[133] A. Egaña, V. Tormo-Márquez, A. Torrente, J. E. Muñoz-Santiuste, J. Olivares, and M. Tardío, Nucl. Instruments Methods Phys. Res. Sect. B Beam Interact. with Mater. Atoms **435**, 152 (2018).

[134] C. Trautmann, Phys. Rev. B **62**, (2000).

[135] M. Toulemonde and F. Studer, Philos. Mag. A **58**, 799 (1988).

[136] M. Lang, F. Zhang, R. Ewing, J. Lian, C. Trautmann, and Z. Wang, J. Mater. Res. **24**, 1322 (2009).

[137] J. Shamblin, C. L. Tracy, R. C. Ewing, F. Zhang, W. Li, C. Trautmann, and M. Lang, Acta Mater. **117**, 207 (2016).

[138] J. F. Gibbons, Proc. IEEE **60**, 1062 (1972).

[139] W. J. Weber, Nucl. Instruments Methods Phys. Res. Sect. B Beam Interact. with Mater. Atoms **166**–**167**, 98 (2000).

[140] K. Sickafus, (2007).

[141] G. Sattonnay, C. Grygiel, I. Monnet, C. Legros, M. Herbst-Ghysel, and L. Thomé, Acta Mater. **60**, 22 (2012).
[142] C. Tracy, J. Shamblin, S. Park, F. Zhang, C. Trautmann, M. Lang, and R. Ewing, Phys. Rev. B **94**, (2016).
[143] M. Lang, J. Lian, J. Zhang, F. Zhang, W. Weber, C. Trautmann, and R. Ewing, Phys. Rev. B **79**, (2009).
[144] M. Lang, C. Tracy, R. Palomares, F. Zhang, D. Severin, M. Bender, C. Trautmann, C. Park, V. Prakapenka, V. A. Skuratov, and R. Ewing, J. Mater. Res. **30**, 1366 (2015).
[145] R. Palomares, J. Shamblin, C. Tracy, J. Neuefeind, R. Ewing, C. Trautmann, and M. Lang, J. Mater. Chem. A **5**, (2017).
[146] J. Shamblin, C. L. Tracy, R. I. Palomares, E. C. O'Quinn, R. C. Ewing, J. Neuefeind, M. Feygenson, J. Behrens, C. Trautmann, and M. Lang, Acta Mater. **144**, 60 (2018).
[147] In (Radiation Effects and Defects in Solids 110, Caen, France, 1989).
[148] In (Radiation Effects and Defects in Solids 126, Bensheim, Germany, 1993).
[149] In *Nucl. Instr. Meth. Phys. Res. B 107* (Caen, France, 1996).
[150] In (Nucl. Instr. Meth. Phys. Res. B 146, Berlin, Germany, 1998).
[151] In (Nucl. Instr. Meth. Phys. Res. B 209, Catania, Italy, 2003).
[152] In (Nucl. Instr. Meth. Phys. Res. B 245, Aschaffenburg, Germany, 2006).
[153] In (Nucl. Instr. Meth. Phys. Res. B 267, Lyon, France, 2009).
[154] In (Nucl. Instr. Meth. Phys. Res. B 314, Kyoto, Japan, 2013).
[155] In (Nucl. Instr. Meth. Phys. Res. B 365, Darmstadt, Germany, 2015).
[156] In (Nucl. Instr. Meth. Phys. Res. B, Caen, France, 2019).
[157] M. Toulemonde, W. Assmann, C. Dufour, A. Meftah, and C. Trautmann, Nucl. Instruments Methods Phys. Res. Sect. B Beam Interact. with Mater. Atoms **277**, 28 (2012).
[158] P. Sigmund and R. Jeraj, *Ion Beam Science: Solved and Unsolved Problems (Part I, II)* (2008).
[159] A. Benyagoub and M. Toulemonde, J. Mater. Res. **30**, 1529 (2015).
[160] A. Meftah, F. Brisard, J. M. Costantini, E. Dooryhee, M. Hage-Ali, M. Hervieu, J. P. Stoquert, F. Studer, and M. Toulemonde, Phys. Rev. B **49**, 12457 (1994).
[161] A. Meftah, J. M. Costantini, N. Khalfaoui, S. Boudjadar, J. P. Stoquert, F. Studer, and M. Toulemonde, Nucl. Instruments Methods Phys. Res. Sect. B Beam Interact. with Mater. Atoms **237**, 563 (2005).
[162] M. Lang, M. Toulemonde, J. Zhang, F. Zhang, C. L. Tracy, J. Lian, Z. Wang, W. J. Weber, D. Severin, M. Bender, C. Trautmann, and R. C. Ewing, Nucl. Instruments Methods Phys. Res. Sect. B Beam Interact. with Mater. Atoms **336**, 102 (2014).
[163] M. Lang, F. Zhang, J. Zhang, J. Wang, J. Lian, W. J. Weber, B. Schuster, C. Trautmann, R. Neumann, and R. C. Ewing, Nucl. Instruments Methods Phys. Res. Sect. B Beam Interact. with Mater. Atoms **268**, 2951 (2010).
[164] M. Lang, R. Devanathan, M. Toulemonde, and C. Trautmann, Curr. Opin. Solid State Mater. Sci. **19**, 39 (2015).
[165] I. Jozwik-Biala, J. Jagielski, B. Arey, L. Kovarik, G. Sattonnay, A. Debelle, S. Mylonas, I. Monnet, and L. Thomé, Acta Mater. **61**, 4669 (2013).
[166] A. Benyagoub, Nucl. Instruments Methods Phys. Res. Sect. B Beam Interact. with Mater. Atoms **245**, 225 (2006).
[167] S. Hémon, V. Chailley, E. Dooryhée, C. Dufour, F. Gourbilleau, F. Levesque, and E.

Paumier, Nucl. Instruments Methods Phys. Res. Sect. B Beam Interact. with Mater. Atoms **122**, 563 (1997).

[168] C. Houpert, F. Studer, D. Groult, and M. Toulemonde, Nucl. Inst. Methods Phys. Res. B **39**, 720 (1989).

[169] F. Studer, M. Hervieu, J.-M. Costantini, and M. Toulemonde, Nucl. Instruments Methods Phys. Res. Sect. B Beam Interact. with Mater. Atoms **122**, 449 (1997).

[170] N. Itoh, Adv. Phys. **31**, 491 (1982).

[171] N. Medvedev and B. Ziaja, Sci. Rep. **8**, 5284 (2018).

[172] C. Trautmann, M. Toulemonde, K. Schwartz, J. M. Costantini, and A. Müller, Nucl. Instruments Methods Phys. Res. Sect. B Beam Interact. with Mater. Atoms **164**–**165**, 365 (2000).

[173] K. Schwartz, Phys. Rev. B **58**, 11232 (1998).

[174] M. Toulemonde, A. Benyagoub, C. Trautmann, N. Khalfaoui, M. Boccanfuso, C. Dufour, F. Gourbilleau, J. J. Grob, J. P. Stoquert, J. M. Costantini, F. Haas, E. Jacquet, K. O. Voss, and A. Meftah, Phys. Rev. B **85**, 54112 (2012).

[175] W. Assmann, M. Toulemonde, and C. Trautmann, in (Springer Berlin Heidelberg, Berlin, Heidelberg, 2007), pp. 401–450.

[176] M. Toulemonde, W. Assmann, C. Trautmann, and F. Grüner, Phys. Rev. Lett. **88**, 57602 (2002).

[177] J. M. Costantini, F. Brisard, J. L. Flament, A. Meftah, M. Toulemonde, and M. Hage-Ali, Nucl. Instruments Methods Phys. Res. Sect. B Beam Interact. with Mater. Atoms **65**, 568 (1992).

[178] I. Jozwik-Biala, J. Jagielski, L. Thomé, B. Arey, L. Kovarik, G. Sattonnay, A. Debelle, and I. Monnet, Nucl. Instruments Methods Phys. Res. Sect. B Beam Interact. with Mater. Atoms **286**, 258 (2012).

[179] M. Toulemonde, W. J. Weber, G. Li, V. Shutthanandan, P. Kluth, T. Yang, Y. Wang, and Y. Zhang, Phys. Rev. B **83**, 54106 (2011).

[180] C. Rotaru, F. Pawlak, N. Khalfaoui, C. Dufour, J. Perrière, A. Laurent, J. P. Stoquert, H. Lebius, and M. Toulemonde, Nucl. Instruments Methods Phys. Res. Sect. B Beam Interact. with Mater. Atoms **272**, 9 (2012).

[181] C. L. Tracy, M. Lang, J. M. Pray, F. X. Zhang, D. Popov, C. Y. Park, C. Trautmann, M. Bender, D. Severin, V. A. Skuratov, and R. C. Ewing, Nat. Commun. **6**, 9 (2015).

[182] A. Dunlop, G. Jaskierowicz, and S. Della-Negra, Nucl. Instruments Methods Phys. Res. Sect. B Beam Interact. with Mater. Atoms **146**, 302 (1998).

[183] R. Sachan, E. Zarkadoula, M. Lang, C. Trautmann, Y. Zhang, M. F. Chisholm, and W. J. Weber, Sci. Rep. **6**, 27196 (2016).

[184] M. Toulemonde, G. Fuchs, N. Nguyen, F. Studer, and D. Groult, Phys. Rev. B **35**, 6560 (1987).

[185] C. Trautmann, S. Bouffard, and R. Spohr, Nucl. Instruments Methods Phys. Res. Sect. B Beam Interact. with Mater. Atoms **116**, 429 (1996).

[186] M. Toulemonde, N. Enault, J. Y. Fan, and F. Studer, J. Appl. Phys. **68**, 1545 (1990).

[187] J. Jensen, A. Razpet, M. Skupiński, and G. Possnert, Nucl. Instruments Methods Phys. Res. Sect. B Beam Interact. with Mater. Atoms **245**, 269 (2006).

[188] M. C. Busch, A. Slaoui, P. Siffert, E. Dooryhee, and M. Toulemonde, J. Appl. Phys. **71**, 2596 (1992).

[189] A. Dunlop, D. Lesueur, P. Legrand, H. Dammak, and J. Dural, Nucl. Instruments Methods

Phys. Res. Sect. B Beam Interact. with Mater. Atoms **90**, 330 (1994).

[190] H. Dammak, D. Lesueur, A. Dunlop, P. Legrand, and J. Morillo, Radiat. Eff. Defects Solids **126**, 111 (1993).

[191] A. Barbu, A. Dunlop, D. Lesueur, and R. S. Averback, Europhys. Lett. **15**, 37 (1991).

[192] T. Iwata and A. Iwase, Nucl. Instruments Methods Phys. Res. Sect. B Beam Interact. with Mater. Atoms **61**, 436 (1991).

[193] B. Ye, L. Jamison, Y. Miao, S. Bhattacharya, G. L. Hofman, and A. M. Yacout, J. Nucl. Mater. **488**, 134 (2017).

[194] C. Dufour, A. Audouard, F. Beuneu, J. Dural, J. P. Girard, A. Hairie, M. Levalois, E. Paumier, and M. Toulemonde, J. Phys. Condens. Matter **5**, 4573 (1993).

[195] A. Audouard, A. Dunlop, D. Lesueur, N. Lorenzelli, and L. Thomé, Eur. Phys. J. AP **3**, 149 (1998).

[196] A. Barbu, A. Dunlop, D. Lesueur, R. S. Averback, R. Spohr, and J. Vetter, Int. J. Radiat. Appl. Instrumentation. Part D. Nucl. Tracks Radiat. Meas. **19**, 35 (1991).

[197] H. Watanabe, B. Kabius, K. Urban, B. Roas, S. Klaumünzer, and G. Saemann-Ischenko, Phys. C Supercond. **179**, 75 (1991).

[198] V. Hardy, D. Groult, M. Hervieu, J. Provost, B. Raveau, and S. Bouffard, Nucl. Instruments Methods Phys. Res. Sect. B Beam Interact. with Mater. Atoms **54**, 472 (1991).

[199] M. Toulemonde, S. Bouffard, and F. Studer, Nucl. Instruments Methods Phys. Res. Sect. B Beam Interact. with Mater. Atoms **91**, 108 (1994).

[200] R. Weinstein, R.-P. Sawh, D. Parks, and B. Mayes, Nucl. Instruments Methods Phys. Res. Sect. B Beam Interact. with Mater. Atoms **272**, 284 (2012).

[201] C. Dufour, F. Beuneu, E. Paumier, and M. Toulemonde, Europhys. Lett. **45**, 585 (1999).

[202] J. Auleytner, J. Bak-Misiuk, Z. Furmanik, M. Toulemonde, and J. Vetter, Radiat. Eff. Defects Solids **115**, 335 (1991).

[203] M. Toulemonde, J. Dural, G. Nouet, P. Mary, J. F. Hamet, M. F. Beaufort, J. C. Desoyer, C. Blanchard, and J. Auleytner, Phys. Status Solidi **114**, 467 (1989).

[204] J. Krynicki, M. Toulemonde, J. C. Muller, and P. Siffert, Mater. Sci. Eng. B **2**, 105 (1989).

[205] J. Vetter, R. Scholz, D. Dobrev, and L. Nistor, Nucl. Instruments Methods Phys. Res. Sect. B Beam Interact. with Mater. Atoms **141**, 747 (1998).

[206] W. Wesch, A. Kamarou, E. Wendler, A. Undisz, and M. Rettenmayr, Nucl. Instruments Methods Phys. Res. Sect. B Beam Interact. with Mater. Atoms **257**, 283 (2007).

[207] W. Wesch, O. Herre, P. I. Gaiduk, E. Wendler, S. Klaumünzer, and P. Meier, Nucl. Instruments Methods Phys. Res. Sect. B Beam Interact. with Mater. Atoms **146**, 341 (1998).

[208] A. Kamarou, W. Wesch, E. Wendler, A. Undisz, and M. Rettenmayr, Phys. Rev. B **73**, 184107 (2006).

[209] G. Szenes, Z. E. Horváth, B. Pécz, F. Pászti, and L. Tóth, Phys. Rev. B **65**, 45206 (2002).

[210] S. J. Zinkle, V. A. Skuratov, and D. T. Hoelzer, Nucl. Instruments Methods Phys. Res. Sect. B Beam Interact. with Mater. Atoms **191**, 758 (2002).

[211] B. Canut, N. Bonardi, S. M. M. Ramos, and S. Della-Negra, Nucl. Instruments Methods Phys. Res. Sect. B Beam Interact. with Mater. Atoms **146**, 296 (1998).

[212] A. Colder, O. Marty, B. Canut, M. Levalois, P. Marie, X. Portier, S. M. M. Ramos, and M. Toulemonde, Nucl. Instruments Methods Phys. Res. Sect. B Beam Interact. with Mater. Atoms **174**, 491 (2001).

[213] S. Dhamodaran, A. P. Pathak, A. Dunlop, G. Jaskierowicz, and S. Della Negra, Nucl. Instruments Methods Phys. Res. Sect. B Beam Interact. with Mater. Atoms **256**, 229 (2007).
[214] M. Sall, F. Moisy, J. G. Mattei, C. Grygiel, E. Balanzat, and I. Monnet, Nucl. Instruments Methods Phys. Res. Sect. B Beam Interact. with Mater. Atoms **435**, 116 (2018).
[215] M. Sall, I. Monnet, F. Moisy, C. Grygiel, S. Jublot-Leclerc, S. Della–Negra, M. Toulemonde, and E. Balanzat, J. Mater. Sci. **50**, 5214 (2015).
[216] S. Klaumünzer and G. Schumacher, Phys. Rev. Lett. **51**, 1987 (1983).
[217] M. Hou, S. Klaumünzer, and G. Schumacher, Phys. Rev. B **41**, 1144 (1990).
[218] S. Furuno, H. Otsu, K. Hojou, and K. Izui, Nucl. Instruments Methods Phys. Res. Sect. B Beam Interact. with Mater. Atoms **107**, 223 (1996).
[219] M. Toulemonde, W. Assmann, and C. Trautmann, Nucl. Instruments Methods Phys. Res. Sect. B Beam Interact. with Mater. Atoms **379**, 2 (2016).
[220] A. Ribet, J. G. Mattei, I. Monnet, and C. Grygiel, Nucl. Instruments Methods Phys. Res. Sect. B Beam Interact. with Mater. Atoms **445**, 41 (2019).
[221] A. Kabir, A. Meftah, J. P. Stoquert, M. Toulemonde, and I. Monnet, Nucl. Instruments Methods Phys. Res. Sect. B Beam Interact. with Mater. Atoms **266**, 2976 (2008).
[222] C. Grygiel, F. Moisy, M. Sall, H. Lebius, E. Balanzat, T. Madi, T. Been, D. Marie, and I. Monnet, Acta Mater. **140**, 157 (2017).
[223] V. A. S. S.J. Zinkle, H. Matzke, in *Microstruct. Process. Dur. Irradiat.*, edited by J. S. W. S.J. Zinkle, G.E. Lucas, R.C. Ewing (Warrendale, PA, 1999), pp. 299–304.
[224] D. Schauries, M. Lang, O. H. Pakarinen, S. Botis, B. Afra, M. D. Rodriguez, F. Djurabekova, K. Nordlund, D. Severin, M. Bender, W. X. Li, C. Trautmann, R. C. Ewing, N. Kirby, and P. Kluth, J. Appl. Crystallogr. **46**, 1558 (2013).
[225] M. Toulemonde, W. Assmann, Y. Zhang, M. Backman, W. J. Weber, C. Dufour, and Z. G. Wang, Procedia Mater. Sci. **7**, 272 (2014).
[226] Z. G. Wang, C. Dufour, E. Paumier, and M. Toulemonde, Nucl. Instruments Methods Phys. Res. Sect. B Beam Interact. with Mater. Atoms **115**, 577 (1996).
[227] Z. G. Wang, C. Dufour, E. Paumier, and M. Toulemonde, J. Phys. Condens. Matter **7**, 2525 (1995).
[228] A. Iwase and S. Ishino, J. Nucl. Mater. **276**, 178 (2000).
[229] M. Toulemonde, S. M. M. Ramos, H. Bernas, C. Clerc, B. Canut, J. Chaumont, and C. Trautmann, Nucl. Instruments Methods Phys. Res. Sect. B Beam Interact. with Mater. Atoms **178**, 331 (2001).
[230] M. Backman, F. Djurabekova, O. H. Pakarinen, K. Nordlund, Y. Zhang, M. Toulemonde, and W. J. Weber, Nucl. Instruments Methods Phys. Res. Sect. B Beam Interact. with Mater. Atoms **303**, 129 (2013).
[231] M. Backman, F. Djurabekova, O. Pakarinen, K. Nordlund, and W. Weber, Nucl. Instr. Meth. Phys. Res. B **303**, (2012).
[232] H. D. Mieskes, W. Assmann, F. Grüner, H. Kucal, Z. G. Wang, and M. Toulemonde, Phys. Rev. B **67**, 155414 (2003).
[233] W. J. Weber, E. Zarkadoula, O. H. Pakarinen, R. Sachan, M. F. Chisholm, P. Liu, H. Xue, K. Jin, and Y. Zhang, Sci. Rep. **5**, 7726 (2015).
[234] H. Xue, E. Zarkadoula, P. Liu, K. Jin, Y. Zhang, and W. J. Weber, Acta Mater. **127**, 400 (2017).
[235] H. Xue, E. Zarkadoula, R. Sachan, Y. Zhang, C. Trautmann, and W. J. Weber, Acta

Mater. **150**, 351 (2018).
[236] L. Thomé, A. Debelle, F. Garrido, S. Mylonas, B. Décamps, C. Bachelet, G. Sattonnay, S. Moll, S. Pellegrino, S. Miro, P. Trocellier, Y. Serruys, G. Velisa, C. Grygiel, I. Monnet, M. Toulemonde, P. Simon, J. Jagielski, I. Jozwik-Biala, L. Nowicki, M. Behar, W. J. Weber, Y. Zhang, M. Backman, K. Nordlund, and F. Djurabekova, Nucl. Instruments Methods Phys. Res. Sect. B Beam Interact. with Mater. Atoms **307**, 43 (2013).
[237] A. H. Mir, M. Toulemonde, C. Jegou, S. Miro, Y. Serruys, S. Bouffard, and S. Peuget, Sci. Rep. **6**, 30191 (2016).
[238] Y. Liu, X. Han, M. L. Crespillo, Q. Huang, P. Liu, and X. Wang, Materialia **7**, 100402 (2019).
[239] S. J. Zinkle and V. A. Skuratov, Nucl. Instruments Methods Phys. Res. Sect. B Beam Interact. with Mater. Atoms **141**, 737 (1998).
[240] K. Balzer, N. Schlünzen, and M. Bonitz, Phys. Rev. B **94**, 245118 (2016).
[241] A. V. Krasheninnikov and K. Nordlund, J. Appl. Phys. **107**, (2010).
[242] A. A. Correa, J. Kohanoff, E. Artacho, D. Sánchez-Portal, and A. Caro, Phys. Rev. Lett. **108**, 213201 (2012).
[243] A. P. Horsfield, A. Lim, W. M. C. Foulkes, and A. A. Correa, Phys. Rev. B **93**, 245106 (2016).
[244] H.-D. Betz, Rev. Mod. Phys. **44**, 465 (1972).
[245] G. Schiwietz and P. L. Grande, Nucl. Instruments Methods Phys. Res. Sect. B Beam Interact. with Mater. Atoms **90**, 10 (1994).
[246] P. L. Grande and G. Schiwietz, Nucl. Instruments Methods Phys. Res. Sect. B Beam Interact. with Mater. Atoms **267**, 859 (2009).
[247] Geant4 Collaboration, *Geant4 A Simulation Toolkit: Physics Reference Manual* (2017).
[248] A. Ferrari, P. R. Sala, A. Fassò, and J. Ranft, *Fluka: A Multi-Particle Transport Code* (Geneva, 2005).
[249] E. Lamour, P. D. Fainstein, M. Galassi, C. Prigent, C. A. Ramirez, R. D. Rivarola, J.-P. Rozet, M. Trassinelli, and D. Vernhet, Phys. Rev. A **92**, 042703 (2015).
[250] R. Kubo, Reports Prog. Phys. **29**, 255 (1966).
[251] A. Akkerman, T. Boutboul, A. Breskin, R. Chechik, A. Gibrekhterman, and Y. Lifshitz, Phys. Status Solidi B **198**, 769 (1996).
[252] C. Ambrosch-Draxl and J. O. Sofo, Comput. Phys. Commun. **175**, 1 (2006).
[253] E. D. Palik, *Handbook of Optical Constants of Solids* (Academic Press, San Diego, 1985).
[254] S. Adachi, *The Handbook on Optical Constants of Semiconductors: In Tables and Figures* (World Scientific Publishing Company, New Jersey, London, Singapore, 2012).
[255] R. H. Ritchie and A. Howie, Philos. Mag. **36**, 463 (1977).
[256] J. C. Ashley, J. Electron Spectros. Relat. Phenomena **46**, 199 (1988).
[257] H. Shinotsuka, S. Tanuma, C. J. Powell, and D. R. Penn, Surf. Interface Anal. **47**, 871 (2015).
[258] A. Akkerman, J. Barak, and D. Emfietzoglou, Nucl. Instruments Methods Phys. Res. Sect. B Beam Interact. with Mater. Atoms **227**, 319 (2005).
[259] M. Dapor, I. Abril, P. de Vera, and R. Garcia-Molina, Phys. Rev. B **96**, 064113 (2017).
[260] D. Emfietzoglou, I. Kyriakou, I. Abril, R. Garcia-Molina, I. D. Petsalakis, H. Nikjoo, and A. Pathak, Nucl. Instruments Methods Phys. Res. Sect. B Beam Interact. with Mater. Atoms **267**, 45 (2009).
[261] J. P. Ziegler, U. Biersack, and J. F. Littmark, *The Stopping and Range of Ions in Solids*

(Pergamon Press, New York, 1985).
[262] M. J. Berger, J. S. Coursey, M. A. Zucker, and J. Chang, *Stopping-Power and Range Tables for Electrons, Protons, and Helium Ions | NIST* (1998).
[263] C. Montanari, Int. At. Energy Agency / Nucl. Data Sect. (n.d.).
[264] C. J. Powell and A. Jablonsky, http://www.nist.gov/srd/nist71.cfm (2014).
[265] D. Bank, *PENELOPE – A Code System for Monte Carlo Simulation of Electron and Photon Transport A Code System for Monte Carlo* (2001).
[266] N. A. Medvedev, *Excitation and Relaxation of the Electronic Subsystem in Solids after High Energy Deposition Dissertation* (TU Kaiserslautern, Kaiserslautern, 2011).
[267] B. Y. Mueller and B. Rethfeld, Phys. Rev. B **87**, 035139 (2013).
[268] J.-C. Kuhr and H.-J. Fitting, J. Electron Spectros. Relat. Phenomena **105**, 257 (1999).
[269] S. A. Gorbunov, P. N. Terekhin, N. A. Medvedev, and A. E. Volkov, Nucl. Instruments Methods Phys. Res. Sect. B Beam Interact. with Mater. Atoms **315**, 173 (2013).
[270] T. M. Jenkins, W. R. Nelson, and A. Rindi, editors, *Monte Carlo Transport of Electrons and Photons* (Springer US, Boston, MA, 1988).
[271] D. E. Cullen, *A Survey of Atomic Binding Energies for Use in EPICS2017* (Vienna, 2018).
[272] (n.d.).
[273] K. Wittmaack, Nucl. Instruments Methods Phys. Res. Sect. B Beam Interact. with Mater. Atoms **380**, 57 (2016).
[274] R. L. Fleischer, P. B. Price, R. M. Walker, and E. L. Hubbard, Phys. Rev. **156**, 353 (1967).
[275] V. L. Shabansky and V. P. Ginzburg, Reports Acad. Sci. USSR **100**, 445 (1955).
[276] S. I. Anisimov, B. L. Kapeliovich, and T. L. Perel-man, J. Exp. Theor. Phys. (1974).
[277] I. A. Baranov, Y. V. Martynenko, S. O. Tsepelevich, and Y. N. Yavlinskii, Physics-Uspekhi **31**, 1015 (1988).
[278] M. Toulemonde, C. Dufour, and E. Paumier, Phys. Rev. B **46**, 14362 (1992).
[279] M. P. R. Waligórski, R. N. Hamm, R. Katz, and O. Ridge, Int. J. Radiat. Appl. Instrumentation. Part D. Nucl. Tracks Radiat. Meas. **11**, 309 (1986).
[280] E. Kobetich and R. Katz, Phys. Rev. **170**, 391 (1968).
[281] R. Rymzhanov, N. A. Medvedev, and A. E. Volkov, J. Phys. D. Appl. Phys. **50**, 475301 (2017).
[282] R. A. Rymzhanov, N. Medvedev, A. E. Volkov, J. H. O'Connell, and V. A. Skuratov, Nucl. Instruments Methods Phys. Res. Sect. B Beam Interact. with Mater. Atoms **435**, 121 (2018).
[283] M. Toulemonde, C. Dufour, A. Meftah, and E. Paumier, Nucl. Instruments Methods Phys. Res. Sect. B Beam Interact. with Mater. Atoms **166–167**, 903 (2000).
[284] A. Kamarou, W. Wesch, E. Wendler, A. Undisz, and M. Rettenmayr, Phys. Rev. B - Condens. Matter Mater. Phys. **78**, 1 (2008).
[285] Z. G. Wang, C. Dufour, S. Euphrasie, and M. Toulemonde, Nucl. Instruments Methods Phys. Res. Sect. B Beam Interact. with Mater. Atoms **209**, 194 (2003).
[286] P. Patra, S. A. Khan, M. Bala, D. K. Avasthi, and S. K. Srivastava, Phys. Chem. Chem. Phys. **21**, 16634 (2019).
[287] A. Chettah, Z. G. Wang, M. Kac, H. Kucal, A. Meftah, and M. Toulemonde, Nucl. Instruments Methods Phys. Res. Sect. B Beam Interact. with Mater. Atoms **245**, 150 (2006).
[288] A. Chettah, H. Kucal, Z. G. Wang, M. Kac, A. Meftah, and M. Toulemonde, Nucl.

Instruments Methods Phys. Res. Sect. B Beam Interact. with Mater. Atoms **267**, 2719 (2009).

[289] C. Dufour, V. Khomrenkov, Y. Y. Wang, Z. G. Wang, F. Aumayr, and M. Toulemonde, J. Phys. Condens. Matter **29**, 095001 (2017).

[290] G. Rizza, P. E. Coulon, V. Khomenkov, C. Dufour, I. Monnet, M. Toulemonde, S. Perruchas, T. Gacoin, D. Mailly, X. Lafosse, C. Ulysse, and E. A. Dawi, Phys. Rev. B **86**, 35450 (2012).

[291] C. Dufour, V. Khomenkov, G. Rizza, and M. Toulemonde, J. Phys. D. Appl. Phys. **45**, 065302 (2012).

[292] M. C. Ridgway, R. Giulian, D. J. Sprouster, P. Kluth, L. L. Araujo, D. J. Llewellyn, A. P. Byrne, F. Kremer, P. F. P. Fichtner, G. Rizza, H. Amekura, and M. Toulemonde, Phys. Rev. Lett. **106**, 95505 (2011).

[293] C. D'Orléans, J. P. Stoquert, C. Estournès, C. Cerruti, J. J. Grob, J. L. Guille, F. Haas, D. Muller, and M. Richard-Plouet, Phys. Rev. B **67**, 220101 (2003).

[294] C. Dufour and M. Toulemonde, in edited by W. Wesch and E. Wendler (Springer International Publishing, Cham, 2016), pp. 63–104.

[295] S. Klaumunzer, *Thermal-Spike Models for Ion Track Physics: A Critical Examination* (2006).

[296] T. D. de la Rubia, R. S. Averback, R. Benedek, and W. E. King, Phys. Rev. Lett. **59**, 1930 (1987).

[297] P. Sigmund, Eur. Phys. J. D **47**, 45 (2008).

[298] P. L. Grande and G. Schiwietz, Phys. Rev. A **58**, 3796 (1998).

[299] M. Breese, L. Rehn, C. Trautmann, and I. Vickridge, Nucl. Instruments Methods Phys. Res. Sect. B Beam Interact. with Mater. Atoms **267**, 1 (2009).

[300] P. Hermes, B. Danielzik, N. Fabricius, D. von der Linde, J. Kuhl, J. Heppner, B. Stritzker, and A. Pospieszczyk, Appl. Phys. A **39**, 9 (1986).

[301] C. Dufour, Z. G. Wang, E. Paumier, and M. Toulemonde, Bull. Mater. Sci. **22**, 671 (1999).

[302] K. P. Lieb, K. Zhang, V. Milinovic, P. K. Sahoo, and S. Klaumünzer, Nucl. Instruments Methods Phys. Res. Sect. B Beam Interact. with Mater. Atoms **245**, 121 (2006).

[303] H. Dammak and A. Dunlop, Nucl. Instruments Methods Phys. Res. Sect. B Beam Interact. with Mater. Atoms **146**, 285 (1998).

[304] K. Nordlund, J. Nucl. Mater. **520**, 273 (2019).

[305] M. P. Allen and D. J. Tildesley, *Computer Simulation of Liquids* (Clarendon Press ; Oxford University Press, Oxford [England]; New York, 1989).

[306] Q. Hou, M. Hou, L. Bardotti, B. Prevel, P. Mélinon, and A. Perez, Phys. Rev. B **62**, (2000).

[307] D. M. Duffy and A. M. Rutherford, J. Phys. Condens. Matter **19**, 16207 (2006).

[308] K. A. Fichthorn and W. H. Weinberg, J. Chem. Phys. **95**, 1090 (1991).

[309] H. Emmerich, Adv. Phys. **57**, 1 (2008).

[310] S. Murphy, S. Daraszewicz, Y. Giret, M. Watkins, A. Shluger, K. Tanimura, and D. Duffy, Phys Rev B. Solid State **92**, 134110 (2015).

[311] R. Darkins, P.-W. Ma, S. T. Murphy, and D. M. Duffy, Phys. Rev. B **98**, 24304 (2018).

[312] E. M. Bringa, R. E. Johnson, and R. M. Papaléo, Phys. Rev. B **65**, 94113 (2002).

[313] N. W. Lima, L. I. Gutierres, R. I. Gonzalez, S. Müller, R. S. Thomaz, E. M. Bringa, and R. M. Papaléo, Phys. Rev. B **94**, 195417 (2016).

[314] Y. Cherednikov, S. Sun, and H. Urbassek, Nucl. Instruments Methods Phys. Res. B **315**, 313 (2013).
[315] Y. Cherednikov, S. Sun, and H. Urbassek, Phys. Rev. B **87**, 245424 (2013).
[316] D. M. Duffy, S. L. Daraszewicz, and J. Mulroue, Nucl. Instruments Methods Phys. Res. Sect. B Beam Interact. with Mater. Atoms **277**, 21 (2012).
[317] H. M. Urbassek, H. Kafemann, and R. E. Johnson, Phys. Rev. B **49**, 786 (1994).
[318] C.-E. Lan, J.-M. Xue, Y.-G. Wang, and Y.-W. Zhang, Chinese Phys. C **37**, 38201 (2013).
[319] A. Rivera, J. Olivares, A. Prada, M. Crespillo, M. Caturla, E. Bringa, J. M. Perlado, and O. Peña-Rodríguez, Sci. Rep. **7**, (2017).
[320] L. I. Gutierres, N. W. Lima, R. S. Thomaz, R. M. Papaléo, and E. M. Bringa, Comput. Mater. Sci. **129**, 98 (2017).
[321] W. Jiang, R. Devanathan, C. J. Sundgren, M. Ishimaru, K. Sato, T. Varga, S. Manandhar, and A. Benyagoub, Acta Mater. **61**, 7904 (2013).
[322] P. A. F. P. Moreira, R. Devanathan, and W. J. Weber, J. Phys. Condens. Matter **22**, 395008 (2010).
[323] R. Devanathan, F. Gao, and C. J. Sundgren, RSC Adv. **3**, 2901 (2013).
[324] J. Pakarinen, M. Backman, F. Djurabekova, and K. Nordlund, Phys. Rev. B **79**, 85426 (2009).
[325] a a Leino, S. L. Daraszewicz, O. H. Pakarinen, K. Nordlund, and F. Djurabekova, EPL (Europhysics Lett. **110**, 16004 (2015).
[326] S. L. Daraszewicz and D. M. Duffy, Nucl. Instruments Methods Phys. Res. Sect. B Beam Interact. with Mater. Atoms **303**, 112 (2013).
[327] D. S. Ivanov and L. V Zhigilei, Phys. Rev. B **68**, 64114 (2003).
[328] A. Caro and M. Victoria, Phys. Rev. A **40**, 2287 (1989).
[329] A. Caro, Radiat. Eff. Defects Solids **126**, 15 (1993).
[330] A. A. Leino, S. L. Daraszewicz, O. H. Pakarinen, F. Djurabekova, K. Nordlund, B. Afra, and P. Kluth, Nucl. Instruments Methods Phys. Res. Sect. B Beam Interact. with Mater. Atoms **326**, 289 (2014).
[331] A. A. Leino, G. D. Samolyuk, R. Sachan, F. Granberg, W. J. Weber, H. Bei, J. Liu, P. Zhai, and Y. Zhang, Acta Mater. **151**, 191 (2018).
[332] O. H. Pakarinen, F. Djurabekova, and K. Nordlund, Nucl. Instruments Methods Phys. Res. Sect. B Beam Interact. with Mater. Atoms **268**, 3163 (2010).
[333] A. Debelle, M. Backman, L. Thomé, K. Nordlund, F. Djurabekova, W. J. Weber, I. Monnet, O. H. Pakarinen, F. Garrido, and F. Paumier, Nucl. Instruments Methods Phys. Res. B **326**, 326 (2014).
[334] M. Backman, M. Toulemonde, O. H. Pakarinen, N. Juslin, F. Djurabekova, K. Nordlund, A. Debelle, and W. J. Weber, Comput. Mater. Sci. **67**, 261 (2013).
[335] Y. Sasajima, N. Ajima, T. Osada, N. Ishikawa, and A. Iwase, Nucl. Instruments Methods Phys. Res. Sect. B Beam Interact. with Mater. Atoms **314**, 202 (2013).
[336] Y. Sasajima, N. Ajima, R. Kaminaga, N. Ishikawa, and A. Iwase, Nucl. Instruments Methods Phys. Res. Sect. B Beam Interact. with Mater. Atoms **440**, 118 (2019).
[337] S. Klaumünzer, Nucl. Instruments Methods Phys. Res. Sect. B Beam Interact. with Mater. Atoms **225**, 136 (2004).
[338] C. Trautmann, S. Klaum??nzer, and H. Trinkaus, Phys. Rev. Lett. **85**, 3648 (2000).
[339] N. Khalfaoui, C. C. Rotaru, S. Bouffard, M. Toulemonde, J. P. Stoquert, F. Haas, C. Trautmann, J. Jensen, and A. Dunlop, Nucl. Instruments Methods Phys. Res. Sect. B

Beam Interact. with Mater. Atoms **240**, 819 (2005).
[340] P. Mota-Santiago, H. Vazquez, T. Bierschenk, F. Kremer, A. Nadzri, D. Schauries, F. Djurabekova, K. Nordlund, C. Trautmann, S. Mudie, M. C. Ridgway, and P. Kluth, Nanotechnology **29**, 144004 (2018).
[341] S. Hooda, K. Avchachov, S. A. Khan, F. Djurabekova, K. Nordlund, B. Satpati, S. Bernstorff, S. Ahlawat, D. Kanjilal, and D. Kabiraj, J. Phys. D. Appl. Phys. **50**, 225302 (2017).
[342] H. Huber, W. Assmann, S. A. Karamian, A. Mücklich, W. Prusseit, E. Gazis, R. Grötzschel, M. Kokkoris, E. Kossionidis, H. D. Mieskes, and R. Vlastou, Nucl. Instruments Methods Phys. Res. Sect. B Beam Interact. with Mater. Atoms **122**, 542 (1997).
[343] J.-H. Zollondz, J. Krauser, A. Weidinger, C. Trautmann, D. Schwen, C. Ronning, H. Hofsaess, and B. Schultrich, Diam. Relat. Mater. **12**, 938 (2003).
[344] D. Schwen, E. Bringa, J. Krauser, A. Weidinger, C. Trautmann, and H. Hofsäss, Appl. Phys. Lett. **101**, 113115 (2012).
[345] A. Debelle, M. Backman, L. Thomé, W. J. Weber, M. Toulemonde, S. Mylonas, A. Boulle, O. H. Pakarinen, N. Juslin, F. Djurabekova, K. Nordlund, F. Garrido, and D. Chaussende, Phys. Rev. B **86**, 100102 (2012).
[346] H. Matzke, in *Radiat. Eff. Solids*, edited by K. E. Sickafus, E. A. Kotomin, and B. P. Uberuaga (Springer Netherlands, Dordrecht, 2007), pp. 401–420.
[347] H. Matzke, P. G. Lucuta, and T. Wiss, Nucl. Instruments Methods Phys. Res. Sect. B Beam Interact. with Mater. Atoms **166**–**167**, 920 (2000).
[348] C. Ronchi, J. Appl. Phys. **44**, 3575 (1973).
[349] C. Ronchi and T. Wiss, J. Appl. Phys. **92**, 5837 (2002).
[350] S. Starikov, High Temp. **53**, 55 (2015).
[351] M. L. Bleiberg, J. Nucl. Mater. **1**, 182 (1959).
[352] A. A. Shoudy, W. E. Mchugh, and M. A. Silliman, *The Effect of Irradiation Temperature and Fission Rate on the Radiation Stability of the Uranium-10 Wt % Molybdenum Alloy* (IAEA, International Atomic Energy Agency (IAEA), 1963).
[353] D. M. Kramer and W. V Johnston, in (1963).
[354] J. Gan, D. D. Keiser, B. D. Miller, A. B. Robinson, D. M. Wachs, and M. K. Meyer, J. Nucl. Mater. **464**, 1 (2015).
[355] Y. Miao, K. Mo, B. Ye, L. Jamison, Z.-G. Mei, J. Gan, B. Miller, J. Madden, J.-S. Park, J. Almer, S. Bhattacharya, Y. S. Kim, G. L. Hofman, and A. M. Yacout, Scr. Mater. **114**, 146 (2016).
[356] B. D. Miller, J. Gan, D. D. Keiser, A. B. Robinson, J. F. Jue, J. W. Madden, and P. G. Medvedev, J. Nucl. Mater. **458**, 115 (2015).
[357] T. Wiss, H. Matzke, C. Trautmann, M. Toulemonde, and S. Klaumünzer, Nucl. Instruments Methods Phys. Res. Sect. B Beam Interact. with Mater. Atoms **122**, 583 (1997).
[358] F. Garrido, L. Vincent, L. Nowicki, G. Sattonnay, and L. Thomé, Nucl. Instruments Methods Phys. Res. Sect. B Beam Interact. with Mater. Atoms **266**, 2842 (2008).
[359] T. Sonoda, M. Kinoshita, N. Ishikawa, M. Sataka, A. Iwase, and K. Yasunaga, Nucl. Instruments Methods Phys. Res. Sect. B Beam Interact. with Mater. Atoms **268**, 3277 (2010).
[360] J. L. Wormald and A. I. Hawari, J. Mater. Res. **30**, 1485 (2015).

[361] H. Matzke, J. Nucl. Mater. **189**, 141 (1992).
[362] C. L. Tracy, J. McLain Pray, M. Lang, D. Popov, C. Park, C. Trautmann, and R. C. Ewing, Nucl. Instruments Methods Phys. Res. Sect. B Beam Interact. with Mater. Atoms **326**, 169 (2014).
[363] K. Yasuda, M. Etoh, K. Sawada, T. Yamamoto, K. Yasunaga, S. Matsumura, and N. Ishikawa, Nucl. Instruments Methods Phys. Res. Sect. B Beam Interact. with Mater. Atoms **314**, 185 (2013).
[364] A. Shelyug, R. I. Palomares, M. Lang, and A. Navrotsky, Phys. Rev. Mater. **2**, 93607 (2018).
[365] T. Sonoda, M. Kinoshita, Y. Chimi, N. Ishikawa, M. Sataka, and A. Iwase, Nucl. Instruments Methods Phys. Res. Sect. B Beam Interact. with Mater. Atoms **250**, 254 (2006).
[366] T. Sonoda, M. Kinoshita, N. Ishikawa, M. Sataka, Y. Chimi, N. Okubo, A. Iwase, and K. Yasunaga, Nucl. Instruments Methods Phys. Res. Sect. B Beam Interact. with Mater. Atoms **266**, 2882 (2008).
[367] W. F. Cureton, R. I. Palomares, C. L. Tracy, E. C. O'Quinn, J. Walters, M. Zdorovets, R. C. Ewing, M. Toulemonde, and M. Lang, J. Nucl. Mater. **525**, 83 (2019).
[368] R. I. Palomares, C. L. Tracy, J. Neuefeind, R. C. Ewing, C. Trautmann, and M. Lang, Nucl. Instruments Methods Phys. Res. Sect. B Beam Interact. with Mater. Atoms **405**, 15 (2017).
[369] R. I. Palomares, C. L. Tracy, F. Zhang, C. Park, D. Popov, C. Trautmann, R. C. Ewing, and M. Lang, J. Appl. Crystallogr. **48**, 711 (2015).
[370] G. S. Was, *Fundamentals of Radiation Materials Science: Metals and Alloys* (Springer Berlin Heidelberg, 2007).
[371] R. C. Ewing, W. J. Weber, and F. W. Clinard, Prog. Nucl. Energy **29**, 63 (1995).
[372] M. Lang, E. C. O'Quinn, J. Shamblin, and J. Neuefeind, MRS Adv. **3**, 1735 (2018).
[373] D. F. Williams and P. F. Britt, in *Technol. Appl. R&D Needs Molten Salt Chem. Innov. Approch. to Accel. Molten Salt React. Dev. Deploy.* (Oak Ridge National Laboratory, 2017).